\documentclass{theoretics}

\usepackage{mathtools}
\usepackage[capitalise,noabbrev,nameinlink]{cleveref}
\usepackage{todonotes}

\usepackage{tikz}
\usetikzlibrary{arrows,automata,calc,positioning,decorations.pathmorphing}

\newcommand{\N}{\mathbb{N}}
\newcommand{\Z}{\mathbb{Z}}

\newcommand{\A}{\mathcal{A}}
\newcommand{\B}{\mathcal{B}}

\newcommand{\M}{\mathcal{M}}
\renewcommand{\O}{\mathcal{O}}
\renewcommand{\P}{\mathcal{P}}
\newcommand{\Q}{\mathcal{Q}}
\newcommand{\BZ}{\mathcal{Z}}

\newcommand{\F}{\mathsf{F}}
\renewcommand{\L}{\mathsf{L}}

\newcommand{\V}{\mathsf{V}}
\newcommand{\X}{\mathsf{X}}
\renewcommand{\r}{\mathsf{r}}

\tikzstyle{state} = [circle,draw, minimum size = 15pt]
\tikzset{
	->,
	>=stealth',
	initial text=$ $
}

\DeclareMathOperator{\last}{\mathsf{last}}
\DeclareMathOperator{\wnd}{\mathsf{wnd}}
\DeclareMathOperator{\mwl}{\mathsf{mwl}}
\DeclareMathOperator{\SW}{\mathsf{SW}}

\DeclareMathOperator{\ps}{ps}
\DeclareMathOperator{\PS}{PS}

\DeclareMathOperator{\dist}{dist}
\DeclareMathOperator{\pdist}{pdist}
\DeclareMathOperator{\cut}{cut}
\DeclareMathOperator{\shift}{shift}
\DeclareMathOperator{\Acc}{Acc}

\newcommand{\NL}{\textsc{NL}}
\newcommand{\PSPACE}{\textsc{Pspace}}
\newcommand{\BPP}{\textsc{BPP}}
\newcommand{\RP}{\textsc{RP}}
\newcommand{\coRP}{\textsc{coRP}}

\newcommand{\Reg}{\mathbf{Reg}}

\newcommand{\Len}{\mathbf{Len}}
\newcommand{\LI}{\mathbf{LI}}
\newcommand{\Triv}{\mathbf{Triv}}
\newcommand{\ST}{\mathbf{ST}}
\newcommand{\SF}{\mathbf{SF}}

\newcommand{\eps}{\varepsilon}

\DeclarePairedDelimiter{\abs}{\lvert}{\rvert}

\renewcommand{\geq}{\geqslant}
\renewcommand{\leq}{\leqslant}
\renewcommand{\ge}{\geq}
\renewcommand{\le}{\leq}

\newcommand{\Omegainf}{\Omega^{\scriptscriptstyle{\infty}}}
\newcommand{\Thetainf}{\Theta^{\scriptscriptstyle{\infty}}}

\crefname{sidefigure}{Figure}{Figures}

\title{Regular Languages in the Sliding Window Model}

\ThCSauthor[MPI]{Moses Ganardi}{ganardi@mpi-sws.org}[0000-0002-0775-7781]
\ThCSauthor[Siegen]{Danny Hucke}{hucke@eti.uni-siegen.de}[]
\ThCSauthor[Siegen]{Markus Lohrey}{lohrey@eti.uni-siegen.de}[0000-0002-4680-7198]
\ThCSauthor[Houston]{Konstantinos Mamouras}{mamouras@rice.edu}[]
\ThCSauthor[ENS]{Tatiana Starikovskaya}{tat.starikovskaya@gmail.com}[]

\ThCSaffil[MPI]{Max Planck Institute for Software Systems (MPI-SWS), Kaiserslautern, Germany}
\ThCSaffil[Siegen]{Universit\"at Siegen, Germany}
\ThCSaffil[Houston]{Rice University, Houston, USA}
\ThCSaffil[ENS]{DI/ENS, PSL Research University, Paris, France}

\ThCSshortnames{M.~Ganardi, D.~Hucke, M.~Lohrey, K.~Mamouras, T.~Starikovskaya}  
\ThCSshorttitle{Regular Languages in the Sliding Window Model}
\ThCSyear{2025}
\ThCSarticlenum{8}
\ThCSreceived{Feb 26, 2024}
\ThCSrevised{Nov 11, 2024}
\ThCSaccepted{Jan 19, 2025}
\ThCSpublished{Mar 6, 2025}
\ThCSdoicreatedtrue
\ThCSkeywords{Regular Languages, Streaming Algorithms, Sliding Window Model}

\begin{document}
\maketitle

\begin{abstract}
We study the space complexity of the following problem:
For a fixed regular language $L$, we receive a stream of symbols and want to test membership of a sliding window of size $n$ in $L$.
For deterministic streaming algorithms we prove a trichotomy theorem,
namely that the (optimal) space complexity is either constant, logarithmic or linear,
measured in the window size $n$.
Additionally, we provide natural language-theoretic characterizations
of the space classes.
We then extend the results to randomized streaming algorithms and we show that in this setting, the space complexity of any regular language is either constant, doubly logarithmic, logarithmic or linear.
Finally, we introduce sliding window testers, which can distinguish
whether a sliding window of size $n$ belongs to the language $L$ or
has Hamming distance $> \epsilon n$ to $L$.
We prove that every regular language has a deterministic (resp., randomized) sliding window tester
that requires only logarithmic (resp., constant) space.
\end{abstract}

\section{Introduction}

\subsection{The sliding window model}

Streaming algorithms process a data stream $a_1 a_2 a_3 \cdots$
of elements $a_i$ from left to right and have at time~$t$ only direct access to the current element $a_t$.
In many streaming applications, elements are outdated after a certain time, i.e.,~they are no longer relevant.
The sliding window model is a simple way to model this.
A sliding window algorithm computes for each time instant~$t$ a value
that only depends on the relevant past (the so-called active window) of $a_1 a_2 \cdots a_t$. 
There are several formalizations of the relevant past. One way to do this is to fix a 
window size~$n$. Then the active window consists at each time instant~$t$ of the last~$n$
elements $a_{t-n+1} a_{t-n+2} \cdots a_t$ (here we assume that $a_i$ is a fixed padding symbol if $i \leq 0$). In the literature this is also called the fixed-size model.
Another sliding window model that can be found in the literature is the variable-size model; see e.g.~\cite{ArasuM04}.
In this model, the arrival of new elements and the expiration of old elements
can happen independently, which means that the window size can vary.\footnote{The reader can also think of a 
queue data structure, where letters can be added to the right and removed at the left, the latter without retrieving the identity of the letter.}
This allows to model for instance time-based windows, where data items arrive at irregular
time instants and the active window contains
all data items that arrive in the last $n$ seconds for a fixed $n$.
The special case of the variable-size model, where old symbols do not expire, is the classical
streaming model.

A general goal in the area of sliding window algorithms is to avoid the explicit storage
of the active window, which would require $\Omega(n)$ space for a window size $n$, and,
instead, to work in considerably smaller space,
e.g.~polylogarithmic space with respect to the window size~$n$.
A detailed introduction into the sliding window model can be found in \cite[Chapter~8]{Aggarwal07}.

The (fixed-size) sliding window model was introduced
in the seminal paper of Datar et al.~\cite{DatarGIM02}
where the authors considered the basic counting problem:
Given a window size $n$ and a stream of bits,
maintain a count of the number of 1’s in the window.
One can easily observe that an exact solution would require $\Theta(n)$ bits.
Intuitively, the reason is that the algorithm cannot see the bit which is
about to expire (the $n$-th most recent bit) without storing it explicitly; this is in fact the main difficulty
in most sliding window algorithms.
However, Datar et al.~show that with $\O(\frac{1}{\epsilon} \cdot \log^2 n)$ bits
one can maintain an approximate count up to a multiplicative factor of $1 \pm \epsilon$.
In \Cref{sec-further-SW} below we briefly discuss further work on sliding window
algorithms.

A foundational problem that has been surprisingly neglected
so far is the language recognition problem over sliding windows:
Given a language $L \subseteq \Sigma^*$ and a stream of symbols over a finite alphabet $\Sigma$,
maintain a data structure which allows to query membership
of the active window in $L$.
In other words, we want to devise a streaming algorithm
which, after every input symbol, either accepts if the current active window belongs to $L$,
or rejects otherwise. This problem finds applications in complex event processing, where
the goal is to detect patterns in data streams. These patterns are usually described in some 
language based on regular expressions; see e.g.~\cite{CuMa12,ZhangDI14} for more details. 
For the standard streaming model, where input symbols do not expire,
some work on language recognition problems has been done; see \Cref{sec-language-recog} below.

\newcommand{\SP}{\textsf{P}}
\newcommand{\SZ}{\textsf{Z}}
\newcommand{\SN}{\textsf{N}}

\begin{example} 
Consider the analysis of the price of a stock in order to identify short-term upward momentum.
The original stream is a time series of stock prices, and it is pre-processed in the following way:
over a sliding window of 5 seconds compute the linear regression of the prices and discretize the slope
into the following values: $\SP_2$ (high positive), $\SP_1$ (low positive), $\SZ$ (zero), $\SN_1$ (low negative),
$\SN_2$ (high negative). This gives rise to a derived stream of symbols in the alphabet $\Sigma = \{\SP_2,\SP_1,\SZ,\SN_1,\SN_2\}$,
over which we describe the {\em upward trend} pattern:
(i) no occurrence of $\SN_2$,
(ii) at most two occurrences of $\SN_1$,
(iii) and any two occurrences of $\SZ$ or $\SN_1$ are separated by at least three positive symbols.
The upward trend pattern can be described as the intersection of the language defined 
 by the following regular expression $e_{ii}$ and the complement of the language
 defined by  $e_{i}$:
\begin{align*}
	e_{i} &= \Sigma^* \cdot \SN_1 \cdot \Sigma^* \cdot \SN_1 \cdot \Sigma^* \cdot \SN_1 \cdot \Sigma^* \\
	e_{ii} &= (\SP_2 + \SP_1)^* \cdot \big( (\SZ + \SN_1) \cdot (\SP_2 + \SP_1)^{[3,\infty)} \big)^* \cdot (\varepsilon + \SZ + \SN_1) \cdot (\SP_2 + \SP_1)^*
\end{align*}
We want to monitor continuously whether the window of the last hour matches the upward trend pattern,
as this is an indicator to buy the stock and ride the upward momentum.
Our results will show that there is a space efficient streaming algorithm for this problem,
which can be synthesized from the regular expressions $e_i$ and $e_{ii}$.
\end{example}

In this paper we focus on querying regular languages over sliding windows.
Unfortunately, there are simple regular languages which require $\Omega(n)$ space in the sliding window model,
i.e.,~one cannot avoid maintaining the entire window explicitly.
However, for certain regular languages, such as for the upward trend pattern from above,
we present sublinear space streaming algorithms.
Before we explain our results in more detail, let us give 
examples of sliding window algorithms for simple regular languages.

\begin{example} 
\label{ex:three-sw}
Let $\Sigma = \{a,b\}$ be the alphabet. In the following examples we refer to the fixed-size sliding window model.
\begin{enumerate}[label=(\roman{*}), ref=(\roman{*})]
\item Let $L = \Sigma^* a$ be the set of all words ending with $a$.
A streaming algorithm can maintain the most recent symbol of the stream in a single bit,
which is also the most recent symbol of the active window.
Hence, the space complexity of $L$ is $\O(1)$ in the sliding window model.
\item Let $L = \Sigma^* a \Sigma^*$ be the set of all words containing $a$.
If $n \in \N$ is the window size,
then a streaming algorithm can maintain the position $1 \le i \le n$ (from right to left)
of the most recent $a$-symbol in the active window or set $i = \infty$ if the active window contains no $a$-symbols.
Let us assume that the initial window is $b^n$
and initialize $i := \infty$.
On input $a$ we set $i := 1$
and on input $b$ we increment $i$
and then set $i := \infty$ if $i > n$.
The algorithm accepts if and only if $i \le n$.
Since the position~$i$ can be stored using $\O(\log n)$ bits,
we have shown that $L$ has space complexity $\O(\log n)$ in the sliding window model.

\item Let $L = a \Sigma^*$ be the set of all words starting with $a$.
We claim that any (deterministic) sliding window algorithm for $L$
and window size $n \in \N$ 
(let us call it $\P_n$) must store at least $n$ bits,
which matches the complexity of the trivial solution where the window is stored explicitly.
More precisely, the claim is that $\P_n$ reaches two distinct memory states
on any two distinct words $x = a_1 \cdots a_n \in \Sigma^n$
and $y = b_1 \cdots b_n \in \Sigma^n$.
Suppose that $a_i \neq b_i$.
Then we simulate $\P_n$ on the streams $x b^{i-1}$ and $y b^{i-1}$, respectively.
The suffixes (or windows) of length $n$ for the two streams
are $a_i \cdots a_n b^{i-1}$ and $b_i \cdots b_n b^{i-1}$.
Since exactly one of the two windows belongs to $L$,
the algorithm $\P_n$ must accept exactly one of the streams $x b^{i-1}$ and $y b^{i-1}$.
In particular, $\P_n$ must reach two distinct memory states on the
words $x b^{i-1}$ and $y b^{i-1}$, and
therefore $\P_n$ must have also reached two distinct memory states on the prefixes $x$ and $y$,
as claimed above.
Therefore, $\P_n$ must have at least $|\Sigma^n| = 2^n$ memory states,
which require $n$ bits of memory. \qedhere
\end{enumerate}
\end{example}

\subsection{Results} \label{sec-results}

Let us now present the main results of this paper.
The precise definitions of all used notions can be found in the main part of the paper.
We denote by $\F_L(n)$ (resp., $\V_L(n)$) the space complexity (measured in bits) of an optimal sliding window algorithm
for the language $L$ in the fixed-size (resp., variable-size) sliding window model.
Here, $n$ denotes the fixed window size for the fixed-size model, whereas for 
the variable-size model $n$ denotes that maximal window
size among all time instants when reading an input stream.

Our first result is a trichotomy theorem for the sliding window model,
stating that the deterministic space complexity is always
either constant, logarithmic, or linear.
This holds for both the fixed- and the variable-size model.
Furthermore, we provide natural characterizations for the three space classes. For this, we need
the following language classes:
\begin{itemize}
\item $\Reg$ is the class of all regular languages.
\item $\Len$ is the class of all regular length languages, i.e.,~regular languages $L\subseteq \Sigma^*$
such that for every $n \geq 0$, either $\Sigma^n \subseteq L$ or $\Sigma^n \cap L = \emptyset$.
\item $\ST$ is the class of all suffix testable languages~\cite[Section~5.3]{Pin97}, i.e.,~finite Boolean combinations of languages of the form $\Sigma^* w$ where $w \in \Sigma^*$
(note that these languages are regular).\footnote{For the results presented in this section, one could equivalently define $\ST$ as the class of all languages $\Sigma^* w$ without taking the Boolean closure.}
\item $\LI$ is the class of all regular left ideals, i.e.,~languages of the form $\Sigma^* L$ for $L\subseteq \Sigma^*$
regular.
\end{itemize}
We emphasize that the three defined language properties
only make sense with respect to an underlying alphabet.
If $\mathbf{L}_1, \dots, \mathbf{L}_n$ are classes of languages over some alphabet $\Sigma$,
then $\langle \mathbf{L}_1, \dots, \mathbf{L}_n \rangle$
denotes the {\em Boolean closure} of the classes $\mathbf{L}_1, \dots, \mathbf{L}_n$, which is
the class of all finite Boolean combinations of languages $L \in \bigcup_{i=1}^n\mathbf{L}_i$.
We also use the following asymptotic notation in our results:
For functions $f,g \colon \N \to \mathbb{R}_{\ge 0}$,
$f(n) = \Omegainf(g(n))$ holds if $f(n) \ge c \cdot g(n)$ for some $c > 0$ and
infinitely many $n \in \N$.
Furthermore, $f(n) = \Thetainf(g(n))$ holds if $f(n) = \O(g(n))$
and $f(n) = \Omegainf(g(n))$. Now we can state our first main result.

\begin{theorem}
	\label{thm:big-thm}
	Let $L \subseteq \Sigma^*$ be regular.
	The space complexity
	$\F_L(n)$ is either $\Theta(1)$, $\Thetainf(\log n)$, or $\Thetainf(n)$. Moreover, we have:
	\begin{alignat*}{2}
	\F_L(n) &= \Theta(1) & \ \iff \ & L \in \langle \ST, \Len \rangle \\
	\F_L(n) &= \Thetainf(\log n) & \ \iff \ & L \in \langle \LI, \Len \rangle \setminus \langle \ST, \Len \rangle \\
	\F_L(n) &= \Thetainf(n) & \ \iff \ & L \in \Reg  \setminus \langle \LI, \Len \rangle 
	\end{alignat*}
	The space complexity 
	$\V_L(n)$ is either $\Theta(1)$, $\Theta(\log n)$, or $\Theta(n)$. Moreover, we have:
	\begin{alignat*}{2}
	\V_L(n) &= \Theta(1) & \ \iff \ & L \in \{\emptyset, \Sigma^*\} \\
	\V_L(n) &= \Theta(\log n) & \ \iff \ & L \in \langle \LI, \Len \rangle \setminus \{\emptyset, \Sigma^*\} \\
	\V_L(n) &= \Theta(n) & \ \iff \ & L \in \Reg  \setminus \langle \LI, \Len \rangle
	\end{alignat*}
	\end{theorem}

\Cref{thm:big-thm} describes which regular patterns can be queried over sliding windows in sublinear space:
Regular left ideals over a sliding window express statements of the form ``recently in the stream some regular event happened''.
Dually, complements of left ideals over a sliding window express statements of the form
``at all recent times in the stream some regular event happened''.

Most papers on streaming algorithms make use of randomness. For many problems, randomized
streaming algorithms are more space efficient than deterministic streaming algorithms; see e.g.~\cite{AlonMS99}
and the remarks at the beginning of \Cref{chap:random}.
So, it is natural to consider
randomized sliding window algorithms for regular languages. 
Our randomized sliding window algorithms have a two-sided or one-sided error of $1/3$ (any constant error probability below $1/2$ would yield the same results).
For a one-sided error we obtain exactly the same space trichotomy for regular languages 
as for deterministic algorithms (\Cref{thm:trichotomy-one-sided}). This changes if we allow a two-sided error.
With $\F^\r_L(n)$ we denote the optimal space complexity of a randomized
sliding window algorithm for $L$ in the fixed size model and with two-sided error. Our second main result says 
that the functions $\F^\r_L(n)$ for $L$ regular fall into four randomized space complexity classes:
constant, doubly logarithmic, logarithmic, and linear space.
A language $L$ is {\em suffix-free} if $xy \in L$ and $x \neq \eps$ implies $y \notin L$.
We denote by $\SF$ the class of all regular suffix-free languages.

\begin{figure}

  \centering
  \includegraphics[scale=1.288]{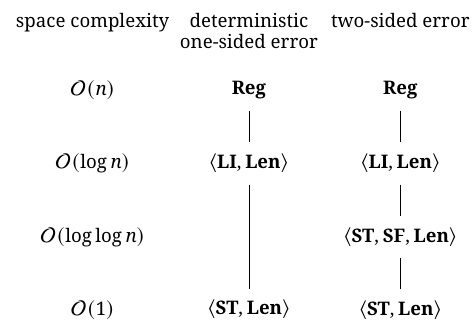}





\caption{The space complexity of regular languages in the fixed-size sliding window model.
$\Reg$: regular languages, $\LI$: regular left ideals, $\ST$: suffix testable languages,
$\SF$: regular suffix-free languages, $\Len$: regular length languages.
The angle brackets $\langle \cdot \rangle$ denote Boolean closure.}
\label{fig:big-picture}
\end{figure}

\begin{theorem} \label{thm:r-char}
	Let $L \subseteq \Sigma^*$ be regular.
	The randomized space complexity $\F^\r_L(n)$ of $L$
	in the fixed-size sliding window model is either
	$\Theta(1)$, $\Thetainf(\log \log n)$, $\Thetainf(\log n)$, or $\Thetainf(n)$.
	Furthermore:
	\begin{alignat*}{2}
        \F^\r_L(n)  &= \Theta(1) & \ \iff \ & L \in \langle \ST, \Len \rangle \\
        \F^\r_L(n) &= \Thetainf(\log \log n) & \ \iff \ & L \in \langle \ST, \SF, \Len \rangle \setminus \langle \ST, \Len \rangle \\
        \F^\r_L(n) &= \Thetainf(\log n) & \ \iff \ & L \in \langle \LI, \Len \rangle \setminus \langle \ST, \SF, \Len \rangle \\
        \F^\r_L(n) &= \Thetainf(n) & \ \iff \ & L \in \Reg \setminus  \langle \LI, \Len \rangle
         \end{alignat*}
\end{theorem}
\Cref{fig:big-picture} compares the deterministic and the randomized space complexity in the fixed-size model (we only show the upper bounds in order
to not overload the figure).
We also consider randomized algorithms in the variable-size model.
In this setting we obtain again the same space trichotomy for regular languages 
as for the deterministic case; see \Cref{lem:rand-length-lb}.

By Theorems~\ref{thm:big-thm} and \ref{thm:r-char}, some (simple) regular languages
(e.g.~$a \{a,b\}^*$) do not admit sublinear space (randomized) algorithms.
This gives the motivation to seek for alternative approaches in order to
achieve efficient algorithms for all regular languages.
We take our inspiration from the property testing model introduced by Goldreich et al.~\cite{GoldreichGR98}.
In this model, the task is to decide (with high probability)
whether the input has a particular property~$P$,
or is ``far'' from any input satisfying~$P$, while querying as few symbols of the input as possible.
Alon et al. prove that every regular language has a property tester
making only $\O(1)$ many queries \cite{AlonKNS00}.
The idea of property testing was also combined with the streaming model,
yielding streaming property testers,
where the objective is not to minimize the number of queries but the required memory
\cite{FeigenbaumKSV02,FrancoisMRS16}.
We define sliding window testers,
which, using as little space as possible,
must accept if the window (of size $n$) belongs to the language~$L$
and must reject if the window has Hamming distance at least $\gamma(n)$ from every
word in $L$. Here $\gamma(n) \leq n$ is a function that is called the Hamming gap of
the sliding window testers. We focus on the fixed-size model.

Two of our main results concerning sliding window testers that we show in \Cref{sec-SW-testing} are the following:
\begin{theorem}\label{thm:testers}
Let $L \subseteq \Sigma^*$ be regular.
\begin{enumerate}[label=(\roman{*}), ref=(\roman{*})]
	\item There exists a deterministic sliding window tester for $L$
	with constant Hamming gap that uses space $\O(\log n)$.
	\item For every $\epsilon > 0$
	there exists a randomized sliding window tester for $L$
	with two-sided error and Hamming gap $\epsilon n$ that uses space $\O(1/\epsilon)$.
\end{enumerate}
\end{theorem}
\Cref{sec-SW-testing} contains additional results that give a rather precise tradeoff between space complexity and the Hamming gap function $\gamma(n)$.
In addition we also study sliding window testers with a one-sided error
and prove optimality for most of our results by providing matching lower bounds. See \Cref{sec main results testing} for a complete discussion of our results for sliding window testers.

\subsection{Related work} \label{sec-related}

This paper builds on four conference papers \cite{GanardiHKLM18,GanardiHL16,GanardiHL18,GanardiHLS19}.
To keep this paper coherent, we decided to omit some of the results from \cite{GanardiHKLM18,GanardiHL16,GanardiHL18,GanardiHLS19}. In this section, we briefly
discuss these results as well as other related work.

\label{sec:omitted}

\subsubsection{Uniform setting}

In all our results we assume a fixed regular language $L$. The space complexity is only measured with respect
to the window size. It is a natural question to ask how the space bounds depend on the size of a finite automaton
(deterministic or nondeterministic) for $L$. This question is considered in \cite{GanardiHKLM18}.
It is shown that,
if $\A$ is a DFA (resp., NFA) with $m$ states for a language $L \in \langle \LI, \Len \rangle$,
then $\V_L(n) = \O(2^m \cdot m \cdot \log n)$ (resp., $\V_L(n)=\O(4^m \cdot \log n)$).
Furthermore, for every 
$k \ge 1$ there exists a language $L_k \subseteq \{0, \ldots, k\}^*$
recognized by a deterministic automaton with $k+3$ states
such that $L_k \in \langle \LI, \Len \rangle$ and 
$F_{L_k}(n) \geq (2^k-1) \cdot (\log n - k)$.
A binary encoding of the words in $L_k$ yields a subexponential lower bound over a fixed alphabet~\cite[Theorem~4.45]{GanardiPhD}.
Further results on the uniform space complexity for languages in $\langle \LI, \Len \rangle$ as well as in $\langle \ST, \Len \rangle$ can be found in \cite[Section~4.3]{GanardiPhD}.

\subsubsection{Membership in the space classes}

In view of \cref{thm:big-thm} it is natural to ask for the complexity of checking whether a given (non)deterministic
finite automaton accepts a language from $\langle \ST, \Len \rangle$ or $\langle \LI, \Len \rangle$, respectively.
Both problems are shown in \cite{GanardiHKLM18} to be $\NL$-complete for deterministic automata and $\PSPACE$-complete for 
nondeterministic automata. We remark that for the class  $\langle \ST, \SF, \Len \rangle$ from 
\cref{thm:r-char} the complexities of the corresponding membership problems are open.

\subsubsection{Other models of randomness}

In \cref{sec-results} we did not specify the underlying model of randomized sliding window algorithms (that is used
in \cref{thm:r-char}) in a precise way. Let us be a bit more specific: we require that a randomized sliding window 
algorithm for a language $L$ running  on an input stream $s$ outputs at every time instant a correct answer on the question whether
the current window belongs to $L$ or not with high probability (say at least $2/3$). This is not the only model of 
randomized sliding window algorithms that can be found in the literature. A stronger model requires that with high
probability the randomized sliding window algorithm outputs at every time instant a correct answer. So the difference
is between ``$\forall$ time instants: $\Pr[\text{answer correct}] \geq 2/3$'' and 
``$\Pr[\text{$\forall$ time instants: answer correct}] \geq 2/3$''. A randomized sliding window algorithm 
that fulfills the latter (stronger) correctness criterion is called strictly correct in \cite{GanardiHLTOCS21}. 
This model is for instance implicitly used in \cite{Ben-BasatEFK16,DatarGIM02}. In \cite{GanardiHLTOCS21} it is shown that every strictly correct 
 randomized sliding window algorithm can be derandomized without increasing the space complexity.
 This result is shown in a very general context for arbitrary approximation problems. 
The proof in \cite{GanardiHLTOCS21} needs input streams of length doubly exponential in the window size for the derandomization.
In contrast, if one restricts to input streams of length polynomial in the window size then strictly correct randomized sliding
window algorithms can be more space efficient than ordinary randomized sliding
window algorithms (as defined in this paper) \cite{GanardiHLTOCS21}. The intermediate case of exponentially long input streams is open.

Finally, we emphasize that our randomized sliding window algorithms are not necessarily \emph{adversarially robust}, 
i.e., an adversary may fool the algorithm by observing the internal memory state and picking the input symbols adaptively.

\subsubsection{Context-free languages} \label{sec-context-free}

It is natural to ask to which extent our results hold for context-free languages. This 
question is considered in \cite{GanardiJL18,Ganardi19}. Let us briefly discuss the results.
In \cite{GanardiJL18} it is shown that if $L$ is a context-free language with $F_L(n) \leq \log n - \omega(1)$
then $L$ must be regular and $F_L(n) = \Theta(1)$. Hence, the gap between constant space and logarithmic
space for regular languages also exists for context-free languages. In contrast, the gap between logarithmic
space and linear space for regular languages does not extend to all context-free languages. In  \cite{GanardiJL18},
the authors construct examples of context-free languages $L$ with $F_L(n) = \Thetainf(n^{1/c})$ and $V_L(n) = \Theta(n^{1/c})$ for every natural
number $c \geq 2$. These languages are not deterministic context-free, but \cite{GanardiJL18} also contains
examples of deterministic one-turn one-counter languages $L$ and $L'$ with $F_L(n) = \Thetainf(\log^2 n)$
and $V_{L'}(n) = \Theta(\log^2 n)$. In \cite{Ganardi19}, the author studies  the space complexity of visibly pushdown languages (a 
language class strictly in-between the regular and deterministic 
context-free languages with good closure and decidability properties \cite{AlurM04}). 
It is shown that for every visibly pushdown language the space complexity in the variable-size sliding window model is either constant, logarithmic or linear in the window size. 
Hence, the space trichotomy that we have seen for regular languages also holds for visibly pushdown languages in the variable-size model.
Whether the visibly pushdown languages also exhibit the space trichotomy in the fixed-size model is open.

\subsubsection{Update times}

In this paper, we only considered the space complexity of sliding window algorithms.
Another important complexity measure is the update time of a sliding window algorithm, i.e.,~the 
worst case time that is spent per incoming symbol for updating the internal data structures.
In \cite{TangwongsanH017}, it is shown that for every regular language $L$ there exists
a deterministic sliding window algorithm (for the fixed-size model) with constant update time.
The underlying machine model of the sliding window algorithm is the RAM model, where basic arithmetic operations on registers
of bit length $ \O(\log n)$ (with $n$ the window size) need constant time. In fact,
the algorithm in \cite{TangwongsanH017} is formulated in a more general context for any associative
aggregation function. The case of a regular language $L$ is obtained by applying the algorithm
from \cite{TangwongsanH017} for the syntactic monoid of~$L$. In \cite{GanardiJLS22} the result of \cite{TangwongsanH017}
is extended to visibly pushdown languages. 

\subsubsection{Further work on sliding windows} \label{sec-further-SW}

We have already mentioned the seminal work of Datar et al.~on the sliding window model
\cite{DatarGIM02}, where the authors considered the problem of estimating the number of ones in the sliding window.
In the same paper, Datar et al.~extend their result for the basic counting problem 
to arbitrary functions which satisfy certain additivity properties,
e.g.~$L_p$-norms for $p \in [1,2]$.
Braverman and Ostrovsky introduced the smooth histogram framework \cite{BravermanO07},
to compute so-called smooth functions over sliding windows, which include all $L_p$-norms and frequency moments.
Further work on computing aggregates, statistics and frequent elements in the sliding window model can be found
in~\cite{ArasuM04,BabcockDMO03,Ben-BasatEF18,Ben-BasatEFK16,BravermanGLWZ18,DatarM02,FeigenbaumKZ04,GibbonsT04,GolabDDLM03}.
The problem of sampling over sliding windows was first studied in \cite{BabcockDM02}
and later improved in \cite{BravermanOZ12}.
As an alternative to sliding windows, Cohen and Strauss consider the problem of maintaining stream aggregates
where the data items are weighted by a decay function \cite{CohenS06}.

\subsubsection{Language recognition in the classical streaming model} \label{sec-language-recog}
 
Whereas language recognition in the sliding window model has been neglected prior to our work, there exists
some work on streaming algorithms for formal languages in the standard setting, where the streaming algorithm
reads an input word $w$ and at the end has to decide whether $w$ belongs to some language. Clearly, for regular
languages, this problem can be solved in constant space. Streaming algorithms for various subclasses of context-free
languages have been studied in \cite{BabuLRV13,FrancoisMRS16,JN14,KrebsLS11,MagniezMN14}. Related to this is the work on querying XML documents
in the streaming model \cite{BarloyMP21,KonradM13,SegoufinV02}.

\subsubsection{Streaming pattern matching} 

Related to our work is the problem of streaming pattern matching, where the goal is to find all occurrences of a pattern (possibly
with some bounded number of mismatches) in a data stream; see e.g.~\cite{DBLP:conf/focs/KociumakaPS21,DBLP:journals/iandc/RadoszewskiS20,DBLP:conf/cpm/GolanKKP20,DBLP:journals/algorithmica/GawrychowskiS22,BreslauerG14,CliffordFPSS15,CliffordFPSS16,CliffordKP19,CliffordS16,GolanKP16,GolanP17,GolanKP18,PoratP09,Starikovskaya17} and search of repetitions in streams~\cite{Ergun:10,stream-periodicity-mismatches,stream-periodicity-wildcards,DBLP:conf/cpm/GawrychowskiRS19,DBLP:journals/algorithmica/GawrychowskiMSU19,DBLP:conf/cpm/MerkurevS19,DBLP:conf/spire/MerkurevS19}.

\subsubsection{Dynamic membership problems for regular languages} 

A sliding window algorithm can be viewed as a dynamic data structure that
maintains a dynamic string $w$ (the window content) under very restricted update operations.
Dynamic membership problems for more general
updates that allow to change the symbol at an arbitrary position have been studied 
in \cite{AmarilliJP21,FrandsenHMRS95,FrandsenMS97}. As in our work, a trichotomy
for the dynamic membership problem of regular languages has been obtained in \cite{AmarilliJP21}
(but the classes appearing the trichotomy in \cite{AmarilliJP21} are different from the classes that
appear in our work).

\subsection{Outline}

The outline of the paper is as follows:
In \Cref{chap:prelim} we give preliminary definitions
and introduce the fixed-size sliding window model
and the variable-size sliding window model.
In \Cref{chp:reg} we study deterministic sliding window algorithms
for regular languages and prove the space trichotomy
and the characterizations of the space classes (\Cref{thm:big-thm}).
In \Cref{chap:random} we turn to randomized sliding window algorithms and
prove the space tetrachotomy (\Cref{thm:r-char}).
Finally, in \Cref{chp:pt} we present deterministic and randomized
sliding window property testers for regular languages (\Cref{thm:testers}).

\section{Preliminaries}

\label{chap:prelim}

\subsection{Words and languages}

An {\em alphabet} $\Sigma$ is a nonempty finite set of {\em symbols}.
A {\em word} over an alphabet $\Sigma$ is a finite sequence
$w = a_1 a_2 \cdots a_n$ of symbols $a_1, \dots, a_n \in \Sigma$.
The {\em length} of $w$ is the number $|w| = n$. 
The {\em empty word} is denoted by $\eps$
whereas the lunate epsilon $\epsilon$ denotes small positive numbers.
The {\em concatenation} of two words $u,v$ is denoted by $u \cdot v$ or $uv$.
The set of all words over $\Sigma$ is denoted by $\Sigma^*$.
A subset $L \subseteq \Sigma^*$ is called a {\em language} over $\Sigma$.

Let $w = a_1 \cdots a_n \in \Sigma^*$ be a word.
Any word of the form $a_1 \cdots a_i$ is a {\em prefix} of $w$,
a word of the form $a_i \cdots a_n$ is a {\em suffix} of $w$,
and a word of the form $a_i \cdots a_j$ is a {\em factor} of $w$.
The concatenation of two languages $K,L$ is $KL = \{ uv \mid u \in K, \, v \in L \}$.
For a language $L$ we define $L^n$ inductively by $L^0 = \{\varepsilon\}$
and $L^{n+1} = L^n L$ for all $n \in \N$.
The {\em Kleene-star} of a language $L$ is the language $L^* = \bigcup_{n \in \N} L^n$.
Furthermore, we define $L^{\le n} = \bigcup_{0 \le k \le n} L^k$
and $L^{< n} = \bigcup_{0 \le k < n} L^k$.

Let $L \subseteq \Sigma^*$ be a language.
We say that $L$ {\em separates} two words $x,y \in \Sigma^*$ with $x \neq y$ if
$|\{x,y\} \cap L| = 1$.
We say that $L$ {\em separates} two languages $K_1,K_2 \subseteq \Sigma^*$ if
$K_1 \subseteq L$ and $K_2 \cap L = \emptyset$, or
$K_2 \subseteq L$ and $K_1 \cap L = \emptyset$.

\subsection{Automata and regular languages} \label{sec-automata}

For good introductions to the theory of formal languages and automata we refer to \cite{Berstel79,HopcroftU79,Kozen97}.

The standard description for regular languages are finite automata.
Let $\Sigma$ be a finite alphabet.
A {\em nondeterministic finite automaton (NFA)} is a tuple 
\[ \A = (Q,\Sigma,I,\Delta,F), \]
where $Q$ is the finite set of {\em states}, $I \subseteq Q$ is the set of initial states,
$\Delta \subseteq Q \times \Sigma \times Q$ is the set of {\em transitions},
and $F \subseteq Q$ is the set of final states.
A {\em run} of $\A$ on a word $w = a_1 \cdots a_n \in \Sigma^*$ is a finite sequence
$\pi = q_0 a_1 q_1 a_2 q_2 \cdots q_{n-1} a_n q_n \in Q (\Sigma Q)^*$
such that $(q_{i-1},a_i,q_i) \in \Delta$ for all $1 \le i \le n$.
We call $\pi$ {\em successful} if $q_0 \in I$ and $q_n \in F$.
The language {\em accepted} by $\A$ is defined as
\[
	\L(\A) = \{ w \in \Sigma^* \mid \text{there exists a successful run of $\A$ on $w$} \}.
\]
A language $L \subseteq \Sigma^*$ is {\em regular} if it is accepted by
some NFA.
The {\em size} $|\A|$ is defined as the number of states.

A {\em \mbox{(left-)}\allowbreak deterministic finite automaton (DFA)}
is an NFA $\A = (Q,\Sigma,I,\Delta,F)$,
where $I = \{q_0\}$ has exactly one initial state $q_0$,
and for all $p \in Q$ and $a \in \Sigma$ there exists exactly one transition $(p,a,q) \in \Delta$.
We view $\Delta$ as a {\em transition function} $\delta \colon Q \times \Sigma \to Q$
and write $\A$ in the format $\A = (Q,\Sigma,q_0,\delta,F)$.
The transition function $\delta$ can be extended
to a right action $\cdot \colon Q \times \Sigma^* \to Q$
of the free monoid $\Sigma^*$ on the state set $Q$
by setting $q \cdot \eps = q$
and defining inductively $q \cdot ua = \delta(q \cdot u,a)$ for all $q \in Q$,
$u \in \Sigma^*$, and $a \in \Sigma$.
We write $\A(w)$ instead of $q_0 \cdot w$.
It is known that any NFA can be turned into an equivalent DFA by the power set construction.

We also consider automata with (possibly) infinitely many states as our formal model for streaming algorithms.
A {\em deterministic automaton} $\A$ has the same format $\A = (Q,\Sigma,q_0,\delta,F)$
as a DFA but we drop the condition that $Q$ must be finite.
We use the notations from the previous paragraph for general deterministic automata as well.

It is well-known that for every regular language $L$ there exists a minimal DFA $\A_L$ for $L$,
which is unique up to isomorphism and whose states are the Myhill-Nerode classes of $L$.
This construction can be carried out for every language $L$ and yields a
deterministic automaton $\A_L$ for $L$ such that for every deterministic automaton $\B$ 
for $L$, we have that $\B(x) = \B(y)$ implies $\A_L(x) = \A_L(y)$ for all $x,y \in \Sigma^*$ \cite[Chapter~III,~Theorem~5.2]{Eilenberg74}.
We call this automaton the {\em minimal deterministic automaton for $L$}.

\subsection{Streaming algorithms} \label{sec streaming general}

A {\em stream} is a finite sequence of elements $a_1 \cdots a_m$,
which arrive element by element from left to right. So, it is just a finite word over some alphabet.
In this paper, the elements $a_i$ are always symbols from a finite alphabet $\Sigma$.
A streaming algorithm reads the symbols of the input stream from left to right.
At time instant $t$ the algorithm only has access to the symbol $a_t$
and the internal storage, which is encoded by a bit string.
The goal of the streaming algorithm is to compute a function $\varphi \colon \Sigma^* \to Y$,
where $\Sigma$ is a finite alphabet and $Y$ is a set of output values.
For the remainder of this paper, we only consider the Boolean case, i.e.,~$Y = \{0,1\}$;
in other words, $\varphi$ is the characteristic function of a language $L$.
Furthermore, we abstract away the actual computation and only analyze
the memory requirement.

Formally, a {\em deterministic streaming algorithm} is the same as 
a deterministic automaton~$\P$ and we say that 
$\P$ is a streaming algorithm for the language $\L(\P)$.
The letter $\P$ stands for {\em program}. If $\P = (M,\Sigma,m_0,\delta,F)$
then the states from $M$ are usually called {\em memory states}.
We require $M \neq \emptyset$ but allow $M$ to be infinite.
The {\em space} of $\P$ (or {\em number of bits used by $\P$}) is given by
$s(\P) = \log |M| \in \mathbb{R}_{\ge 0} \cup \{ \infty \}$.
Here and in the rest of the paper, we denote with $\log$
the logarithm with base two, i.e.,~we measure space in bits.
If $s(\P) = \infty$ we will measure the space restricted to input streams
where some parameter is bounded (namely the window size); see \Cref{sec-variable-size}.

We remark that many streaming algorithms in the literature only produce a single answer
after completely reading the entire stream.
Also, the length of the stream is often known in advance.
However, in the sliding window model we rather assume an input stream of unbounded and unknown length,
and need to compute output values for {\em every} window, i.e.,~at every time instant.

In the following, we introduce the sliding window model in two different variants: the 
fixed-size sliding window model and the variable-size sliding window model.

\subsection{Fixed-size sliding window model} \label{sec-fixed-size}

We fix an arbitrary padding symbol $\Box \in \Sigma$.
Given a stream $x = a_1 a_2 \cdots a_m \in \Sigma^*$ and a {\em window size} $n \in \N$,
we define $\last_n(x) \in \Sigma^n$ by
\[
	\last_n(x) =
	\begin{cases}
		a_{m-n+1} a_{m-n+2} \cdots a_m, & \text{if } n \le m, \\
		\Box^{n-m} a_1 \cdots a_m,         & \text{if } n > m,   
	\end{cases}
\]
which is called the {\em window of size n}, or the {\em active} or {\em current window}.
In other words, $\last_n(x)$ is the suffix of length $n$, padded with $\Box$-symbols on the left.
We view $\Box^n$ as the {\em initial window}; its choice is completely arbitrary.

Let $L \subseteq \Sigma^*$ be a language.
The {\em sliding window problem} $\SW_n(L)$ for $L$ and {\em window size} $n \in \N$
is the language
\[
	\SW_n(L) = \{ x \in \Sigma^* \mid \last_n(x) \in L \}.
\]
Note that for every $L$ and every $n$, $\SW_n(L)$ is regular.
A {\em sliding window algorithm (SW-algorithm)} for $L$ and window size $n \in \N$
is a streaming algorithm for $\SW_n(L)$.
The function $\F_L \colon \N \to \mathbb{R}_{\ge 0}$ is defined by
\begin{equation} \label{def-F_L(n)}
	\F_L(n) = \inf \{ s(\P_n) \mid \P_n \text{ is an SW-algorithm for } L \text{ and window size } n \}.
\end{equation}
It is called the {\em space complexity} of $L$ {\em in the fixed-size sliding window model}.
Note that $\F_L(n) < \infty$ since $\SW_n(L)$ is regular.
A subtle point is that the space complexity $\F_L(n)$ of a language $L$
in general depends on the underlying alphabet.
A simple example is $L = a^*$ which has complexity $\F_L(n) = \O(1)$
over the singleton alphabet $\{a\}$
whereas it has complexity $\F_L(n) = \Theta(\log n)$ over the alphabet $\{a,b\}$
(the latter follows from our results). Here, it is also important that the padding symbol 
$\Box$ belongs to the alphabet $\Sigma$ over which the language $L$ is defined. An alternative
definition would be to take a fresh padding symbol $\Box \notin \Sigma$ and define 
$\last_n(x)$ and $\SW_n(L)$ as above. For instance, for $L = a^*$ we would obtain
$\SW_n(L) = \{ a^i \mid i \geq n\}$, whose minimal DFA has $n+1$ states. Thus, the space complexity
would be $\Omega(\log n)$ instead of $\O(1)$. Note that these differences
only concern the space complexity during the first $n$ steps (until the window is filled up). 
Sliding window algorithms are usually used for streams that are much longer than the window size. So it might be acceptable, if 
during a short initial phase the space complexity is higher than for the rest of the stream.

We draw similarities to circuit complexity, where a language $L \subseteq \{0,1\}^*$ is recognized by a family of circuits
$(\mathcal{C}_n)_{n \in \N}$ in the sense that $\mathcal{C}_n$ recognizes the slice $L \cap \{0,1\}^n$.
Similarly, the sliding window problem $\SW_n(L)$ is solely defined by the slice
$L \cap \Sigma^n$.
If we speak of an SW-algorithm for $L$ and omit the window size $n$,
then this parameter is implicitly universally quantified,
meaning that there exists a family of streaming algorithms $(\P_n)_{n \in \N}$
such that every $\P_n$ is an SW-algorithm for $L$ and window size $n$.

\begin{lemma}
	For any language $L$ we have $\F_L(n) = \O(n)$.
\end{lemma}

\begin{proof}
	A trivial SW-algorithm $\P_n$ for $L$ explicitly
	stores the active window of size $n$ in a queue
	so that the algorithm can always test whether the window belongs to $L$.
	Formally, the state set of $\P_n$ is $\Sigma^n$ and it has transitions of the form
	$(bu,a,ua)$ for $a,b \in \Sigma$, $u \in \Sigma^{n-1}$.
	Viewed as an edge-labeled graph this automaton is also known
	under the name {\em de Bruijn graph} \cite{deBr46}. 
	Since every word $w \in \Sigma^n$ can be encoded with $\O(\log |\Sigma| \cdot n)$ bits and  $|\Sigma|$ is a constant,
	the algorithm uses $\O(n)$ bits.
\end{proof}

Depending on the language $L$ there are more space efficient solutions.
Usually, sliding window algorithms are devised in the following way:
\begin{itemize}
\item Specify some information or property $I(w)$ of the active window $w$
and show that it can be {\em maintained} by a streaming algorithm.
This means that given $I(bu)$ and $a \in \Sigma$ one can compute $I(ua)$.
\item Show that one can decide $w \in L$ from the information $I(w)$.
\end{itemize}
Notice that the complexity function $\F_L(n)$ is not necessarily monotonic.
For instance, let $L$ be the intersection of $a \Sigma^*$ and the set of words with even length. 
By \Cref{ex:three-sw}(iii), we have $\F_L(2n) = \Theta(n)$ but clearly we have $\F_L(2n+1) = \O(1)$
since for odd window sizes the algorithm can always reject.
Therefore, we can only show $\F_L(n) = \Thetainf(n)$ (instead of $\F_L(n) = \Theta(n)$ which is false here),
where $\Thetainf(g(n))$ was defined in the introduction.

Note that the fixed-size sliding window model is a {\em nonuniform} model: for every window size we have
a separate streaming algorithm and these algorithms do not have to follow a common pattern. 
Working with a nonuniform model makes lower bounds stronger.
In contrast, the variable-size sliding window model that we discuss next is a uniform model in the 
sense that there is a single streaming algorithm that works for every window size.
Let us remark that all presented upper bounds for the fixed-size model will be realized by uniform families of algorithms.

\subsection{Variable-size sliding window model} \label{sec-variable-size}

For an alphabet $\Sigma$ we define the extended alphabet $\Sigma_\downarrow = \Sigma \cup \{\downarrow\}$.
In the variable-size model the {\em active window} $\wnd(u) \in \Sigma^*$ for a stream $u \in \Sigma_\downarrow^*$ is defined as follows, where $a \in \Sigma$:
\begin{alignat*}{2}
\wnd(\varepsilon) & = \varepsilon  & \qquad     \wnd(u \! \downarrow) & = \varepsilon, ~ \text{if } \wnd(u) = \varepsilon \\
\wnd(ua) & = \wnd(u) a         &   \wnd(u \! \downarrow) & = v, ~ \text{if } \wnd(u) = av
\end{alignat*}
The symbol $\downarrow$ represents the {\em pop operation}.
We emphasize that a pop operation on an empty window leaves the window empty.
The {\em variable-size sliding window problem} $\SW(L)$ of
a language $L \subseteq \Sigma^*$ is the language
\begin{equation}
\label{def-SW(L)}
\SW(L) = \{ u \in \Sigma_\downarrow^* \mid \wnd(u) \in L \}.
\end{equation}
Note that in general, $\SW(L)$ is not a regular language (even if $L$ is regular).
A {\em variable-size sliding window algorithm (variable-size SW-algorithm)} $\P$ for $L$
is a streaming algorithm for $\SW(L)$.

There are various possible definitions for the space complexity of a variable-size SW-algorithm.
Here, we measure the space complexity as a function in the \emph{maximum} window size over all read prefixes.
This definition enjoys the property that every language $L$ has a variable-size SW-algorithm with \emph{smallest} complexity
among all variable-size SW-algorithms for $L$.
If one would measure the space complexity in the \emph{current} window size instead,
this does not hold anymore, since the memory state encodings of any SW-algorithm can be permuted
to yield an algorithm whose complexity is incomparable to the original one.

To be more formal, for a stream $u = a_1 \cdots a_m \in \Sigma_\downarrow^*$ 
let 
\[\mwl(u) = \max \{ \abs{\wnd(a_1 \cdots a_i)} \mid 0 \le i \le m\}\]
be the {\em maximum window size} of all prefixes of $u$.
If $\P = (M,\Sigma,m_0,\delta,F)$ is a streaming algorithm over $\Sigma_\downarrow$
we define
\begin{equation} \label{defM_n}
	M_n = \{ \P(w) \mid w \in \Sigma_\downarrow^*, \, \mwl(w) = n \}.
\end{equation}
and $M_{\le n} = \bigcup_{0 \le k \le n} M_k$.
The {\em space complexity} of $\P$ in the variable-size sliding window model is 
\[
	v(\P,n) = \log |M_{\le n}| \in \mathbb{R}_{\ge 0} \cup \{\infty\}.
\]
In other words: when we say that the space complexity of a variable-size SW-algorithm
is bounded by $f(n)$, we mean that the algorithm never has to store more than $f(n)$ bits
when it processes a stream $u \in \Sigma_\downarrow^*$ such that for every prefix of $u$
the size of the active window never exceeds $n$. 

Notice that $v(\P,n)$ is a monotonic function.
To prove upper bounds above $\log n$ for the space complexity of $\P$
it suffices to bound $\log |M_n|$ as shown in the following.

\begin{lemma}
	If $s(n) \ge \log n$ is a monotonic function
	and $\log |M_n| = \O(s(n))$ then $v(\P,n) = \O(s(n))$.
\end{lemma}
\begin{proof}
Since $M_{\le n} = M_{0} \cup M_{1} \cup \cdots \cup M_{n}$, we have
\begin{align*}
	\log |M_{\le n}| &= \log \sum_{i=0}^n |M_{i}| \\
	& \le \log \left( (n+1) \cdot \max_{0 \le i \le n} |M_{i}| \right) \\
	& = \log(n+1) + \max_{0 \le i \le n} \log |M_{i}| \\
	& \le \log(n+1) + \max_{0 \le i \le n} \O(s(i)) \\
	& \le \log(n+1) + \O(s(n)) = \O(s(n)),
\end{align*}
which proves the statement.
\end{proof}

\begin{lemma}
	\label{lem:optimal-vs}
	For every language $L \subseteq \Sigma^*$
	there exists a space-optimal variable-size SW-algorithm $\P$,
	i.e.,~$v(\P,n) \leq v(\Q,n)$ for every variable-size SW-algorithm $\Q$ for $L$ and every $n \in \N$.
\end{lemma}

\begin{proof}
	Let $\P$ be the minimal deterministic automaton $\A_{\SW(L)}$ for $\SW(L)$.
	If $\Q$ is any deterministic automaton for $\SW(L)$ then
	$\Q(x) = \Q(y)$ implies $\P(x) = \P(y)$.
	Then, we obtain
	\begin{align*}
		v(\P,n) &= \log |\{ \P(w) \mid w \in \Sigma_\downarrow^*, \, \mwl(w) \le n \}| \\
		&\le \log |\{ \Q(w) \mid w \in \Sigma_\downarrow^*, \, \mwl(w) \le n \}| = v(\Q,n),
	\end{align*}
	which proves the statement.
\end{proof}
One could also define the space complexity $v(\P,n)$ as the number of bits required to encode
a state of $\P$ where the \emph{current} window size is $n$.
It is not difficult to see that \Cref{lem:optimal-vs} fails for this definition.

We define the space complexity of $L$ in the variable-size sliding window model by
$\V_L(n) = v(\P,n)$
where $\P$ is a space-optimal variable-size SW-algorithm for $\SW(L)$
from \Cref{lem:optimal-vs}. It is a monotonic function. 

\begin{lemma}
For any language $L \subseteq \Sigma^*$ and $n \in \N$ we have $\F_L(n) \le \V_L(n)$.
\end{lemma}

\begin{proof}
	If $\P$ is a space-optimal variable-size SW-algorithm for $L$
	then one obtains an SW-algorithm $\P_n$ for window size $n \in \N$ as follows.
	Let us assume $n \ge 1$ (for $n = 0$ we use the trivial SW-algorithm).
	First one simulates $\P$ on the initial window $\Box^n$.
	For every incoming symbol $a \in \Sigma$ we perform a pop operation $\downarrow$ in $\P$,
	followed by inserting $a$.
	Since the maximum window size is bounded by $n$ on any stream,
	the space complexity is bounded by $v(\P,n) = \V_L(n)$.
\end{proof}

The following lemma states that in the variable-size model one must
at least maintain the current window size
if the language is neither empty nor universal.
The issue at hand is performing a pop operation on an empty window.

\begin{lemma}
	\label{lem:length-lb}
	Let $\P$ be a variable-size SW-algorithm
	for a language $\emptyset \subsetneq L \subsetneq \Sigma^*$.
	Then, $\P(x)$ determines\footnotemark~$\abs{\wnd(x)}$ for all $x \in \Sigma_\downarrow^*$
	and therefore $\V_L(n) \ge \log (n+1)$.
	\footnotetext{In other words, for all $x_1,x_2 \in \Sigma_\downarrow^*$,
	if $\P(x_1) = \P(x_2)$ then $\abs{\wnd(x_1)}=\abs{\wnd(x_2)}$.}
\end{lemma}

\begin{proof}
	Let $y \in \Sigma^+$ be a length-minimal nonempty word such that $|\{\eps,y\} \cap L| = 1$.
	Consider streams $x_1, x_2 \in \Sigma_\downarrow^*$ with $\abs{\wnd(x_1)} < \abs{\wnd(x_2)} = m$ and assume $\P(x_1) = \P(x_2)$.
	Then, we also have $\P(x_1 y \!\! \downarrow^m) = \P(x_2 y\!\!\downarrow^m)$.
	But $\wnd(x_2 y \!\! \downarrow^m) = y$ whereas $\wnd(x_1 y \!\!  \downarrow^m)$ is a proper suffix of $y$.
	Now, by the choice of $y$ one these two words belongs to $L$ whereas the other does not, which contradicts
	$\P(x_1 y \!\!  \downarrow^m) = \P(x_2 y \!\! \downarrow^m)$.
	
	For the second statement: if the algorithm reads any stream $a_1 \cdots a_n \in \Sigma^n$
	it must visit $n+1$ pairwise distinct
	memory states and hence $v(\P,n) \ge \log (n+1)$.
\end{proof}

Alternative definitions of the variable-size model are conceivable,
e.g.~one could neglect streams where the popping of an empty window occurs,
or assume that the window size is always known to the algorithm. Then the statement of 
\Cref{lem:length-lb} no longer holds.

\begin{lemma}
	\label{lem:boolean}
	Let $\Sigma$ be a finite alphabet.
	For any function $s(n)$ and $\X \in \{\F,\V\}$, the class
	$\{ L \subseteq \Sigma^* \mid \X_L(n) = \O(s(n)) \}$
	forms a Boolean algebra.
\end{lemma}

\begin{proof}
	Let $L \subseteq \Sigma^*$ be a language.
	Given a SW-algorithm for $L$ for some fixed window size $n$
	or in the variable-size model,
	we can turn it into an algorithm for the complement $\Sigma^* \setminus L$
	by negating its output.
	Clearly, it has the same space complexity as the original algorithm.
	
	Let $L_1,L_2 \subseteq \Sigma^*$ be two languages.
	Let $\P_1,\P_2$ be SW-algorithms for $L_1,L_2$, respectively, either
	for some fixed window size $n$ or in the variable-size model.
	Define $\P$ to be the product automaton of~$\P_1$ and $\P_2$
	which outputs the disjunction of the outputs of the $\P_i$.
	
	In the case of a fixed window size $n$, $\P$ has 
	$2^{s(\P_1)} \cdot 2^{s(\P_2)} = 2^{s(\P_1)+s(\P_2)}$ many states
	and hence $s(\P) = s(\P_1) + s(\P_2)$.
	This implies $\F_{L_1 \cup L_2}(n) \le \F_{L_1}(n) + \F_{L_2}(n)$.
	For the variable-size model, notice that
	$2^{v(\P,n)} \le 2^{v(\P_1,n)} \cdot 2^{v(\P_2,n)} = 2^{v(\P_1,n) + v(\P_2,n)}$.
	Therefore, $\V_{L_1 \cup L_2}(n) \le \V_{L_1}(n) + \V_{L_2}(n)$.
\end{proof}

\begin{remark} \label{ref:internal comp}
Before we start our investigation of the space complexity of regular languages in the sliding window model, we would like to discuss 
an aspect of our definition of the space complexities $\F_L(n)$ and $\V_L(n)$. Both complexity measures do not 
include the space needed for internal computations, i.e.,~the space needed for computing the memory updates of the streaming algorithm. 
Let us explain this in more detail for the variable-size model (the same arguments apply to the fixed-size model). Take
the space-optimal variable-size SW-algorithm $\P$ for a language $L$; see \Cref{lem:optimal-vs}.
The function $\V_L(n)$ measures the number of bits needed to encode the states in the set $M_{\le n}$ (see the line after \eqref{defM_n}).
But the transition function of $\P$ may be difficult to compute. In other words: if we have two memory states $p,q \in M_{\le n}$ 
(both encoded by bit strings of length $\V_L(n)$)
and an $a$-labelled transition from $p$ to $q$ in $\P$
 then additional memory is needed in general in order to compute $q$ from $p$ and $a$.
 This memory is what we mean by the space needed for internal computations. Our definition of $\V_L(n)$ does
 not include this space. One reason for this is that if we would include the space needed for internal computations
 in the total space bound, then it would be difficult to obtain lower bounds that match the upper bounds.
 In particular, techniques based
on communication complexity that we use in the randomized setting (see \Cref{chap:random}) are not able to
take space for internal calculations into account. In our setting,
these techniques only allow to prove lower bounds on the number of memory states of a 
(randomized) streaming algorithm and therefore are not sensitive with respect to the space needed to 
go from one memory state to the next memory state.
\end{remark}

\section{Deterministic sliding window algorithms}

\label{chp:reg}

In this section, we will show that the space complexity of every regular language
in both sliding window models is either constant, logarithmic or linear.
In \Cref{ex:three-sw} we have already seen prototypical languages with these three space complexities,
namely $\Sigma^*a$ (constant), $\Sigma^*a\Sigma^*$ (logarithmic) and~$a\Sigma^*$ (linear) for $\Sigma = \{a,b\}$.
Intuitively, for languages of logarithmic space complexity it suffices to maintain a constant number
of positions in the window.
For languages of constant space complexity it suffices to maintain a constant-length suffix of the window.
Moreover, we describe the languages with logarithmic and constant space complexity
as finite Boolean combinations of simple atomic languages.

\subsection{Right-deterministic finite automata}

It turns out that the appropriate representation of a regular language
for the analysis in the sliding window model are
deterministic finite automata which read the input word, i.e.,~the window, from right to left.
Such automata are called {\em right-deterministic finite automata (rDFA)} in this paper.
The reason why we use rDFAs instead of DFAs can be explained intuitively
for the variable-size sliding window model as follows.
The variable-size model contains operations in both ``directions'':
On the one hand a variable-size window can be extended on the right, and on the other hand the window can be shortened to an arbitrary suffix.
For regular languages the extension to longer windows is ``tame''
because the Myhill--Nerode right congruences have finite index.
Hence, it remains to control the structure of all suffixes with respect to the regular language,
which is best captured by an rDFA for the language.

Formally, a {\em right-deterministic finite automaton (rDFA)} $\B = (Q,\Sigma,F,\delta,q_0)$
consists of a finite state set $Q$, a finite alphabet $\Sigma$,
a set of final states $F \subseteq Q$, a transition function $\delta \colon \Sigma \times Q \to Q$,
and an initial state $q_0 \in Q$.
The transition function $\delta$ extends to a left action $\cdot \colon \Sigma^* \times Q \to Q$
by $\eps \cdot q = q$ and $(aw) \cdot q = \delta(a,w \cdot q)$ for all $a \in \Sigma$,
$w \in \Sigma^*$, $q \in Q$. The language accepted by $\B$ is
$\L(\B) = \{ w \in \Sigma^* \mid w \cdot q_0 \in F \}$.

A {\em run} of $\B$ on a word $w = a_1 \cdots a_n \in \Sigma^*$
from $p_n$ to $p_0$ is a finite sequence
\[ \pi = p_0 a_1 p_1 a_2 p_2 \cdots p_{n-1} a_n p_n \in Q (\Sigma Q)^* \]
such that $p_{i-1} = a_i \cdot p_i$ for all $1 \le i \le n$.
Quite often we write such a run in the following way
\[
\pi : p_0 \xleftarrow{a_1} p_1 \xleftarrow{a_2} p_2 \cdots p_{n-1} \xleftarrow{a_n} p_n .
\]
If the intermediate states $p_1,\ldots, p_{n-1}$ are not important we write this run also as
\[
\pi : p_0 \xleftarrow{a_1 a_2 \cdots a_n} p_n .
\]
The state $p_n$ is also called the \emph{starting state} of the above run $\pi$.
The run $\pi$ is {\em accepting} if $p_0 \in F$ (note that we do not require $p_n = q_0$) and otherwise {\em rejecting}.
Its {\em length} $|\rho|$ is the length $|w|$ of $w$. A run of length zero is called empty;
note that it consists of a single state. A run of length one is also called a {\em transition}.
If $\pi = p_0 a_1 p_1 \cdots a_n p_n$
and $\rho = r_0 b_1 r_1 \cdots b_\ell r_\ell $ are runs such that $p_n = r_0$
then their {\em composition} $\pi \rho$ is defined as
$\pi \rho = p_0 a_1 p_1 \cdots a_n r_0 b_1 r_1 \cdots b_\ell r_\ell$; it is a run on $a_1 \cdots a_n b_1 \cdots b_\ell$.
This definition allows
us to factorize runs in \Cref{subsec:pathsummary}.
We call a run $\pi$ a {\em $P$-run} for a subset $P \subseteq Q$
if all states occurring in $\pi$ are contained in $P$.

A state $q \in Q$ is {\em reachable} from $p \in Q$ if there exists a run from $p$ to $q$, in which case we write $q \preceq_\B p$.
We say that $q$ is {\em reachable} if it is reachable from the initial state $q_0$.
A set of states $P \subseteq Q$ is reachable if all $p \in P$ are reachable.
The reachability relation $\preceq_\B$ is a {\em preorder} on $Q$,
i.e.,~it is reflexive and transitive.
Two states $p,q \in Q$ are {\em strongly connected} if $p \preceq_\B q \preceq_\B p$.
This yields an equivalence relation on $Q$ whose equivalence classes 
 are the {\em strongly connected components (SCCs)} of $\B$.
A subset $P \subseteq Q$ is {\em strongly connected} if it is contained in a single SCC,
i.e.,~all  $p,q \in P$ are strongly connected.

\subsection{Space trichotomy}

In this section, we state two technical results which directly imply \Cref{thm:big-thm}.
Let $\B = (Q,\Sigma,F,\delta,q_0)$ be an rDFA.
A set of states $P \subseteq Q$ is {\em well-behaved} if
for any two $P$-runs $\pi_1, \pi_2$ which start in the same state
and have equal length, either both $\pi_1$ and $\pi_2$ are accepting or both are rejecting.
If every reachable SCC in $\B$ is well-behaved then $\B$ is called {\em well-behaved}.
A state $q \in Q$ is {\em transient} if $x \cdot q \neq q$ for all $x \in \Sigma^+$.
Every transient state in $\B$ forms an SCC of size one (a {\em transient SCC});
however, not every SCC of size one is transient (there can be a loop at the unique state of the SCC).
Let $U(\B) \subseteq Q$ be the set of states $q \in Q$ for which there exists a nontransient state $p \in Q$
such that $q$ is reachable from $p$ and $p$ is reachable from the initial state $q_0$.
Notice that $q \in U(\B)$ if and only if there exist runs of unbounded length from $q_0$ to~$q$
(hence the symbol $U$ for unbounded). Moreover, if $U(\B)$ is well-behaved then $\B$ must be well-behaved.
This follows directly from the above definition and it is also a consequence of \Cref{thm:trichotomy} below.

\begin{figure}[t]
	\centering
        \includegraphics[scale=1.488]{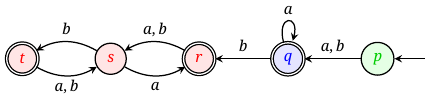}

	\caption{A well-behaved rDFA consisting of three SCCs.}
	\label{fig:well-behaved}
\end{figure}

\begin{example}
Consider the rDFA $\A$ in \Cref{fig:well-behaved}.
It consists of three SCCs, namely
the green SCC $\{p\}$, the blue SCC $\{q\}$ and the red SCC $\{r,s,t\}$.
The red SCC is well-behaved since any run starting in $r$ ends in a final state if and only if its length is even.
The other SCCs are also well-behaved and therefore, the entire automaton is well-behaved.
State $p$ is a transient state and $U(\A) = \{q,r,s,t\}$.
\end{example}

\begin{theorem}
	\label{thm:trichotomy}
	Let $L \subseteq \Sigma^*$ be regular and $\B$ be any rDFA for $L$.
	\begin{enumerate}[label=(\arabic{*}), ref=(\arabic{*})]
		\item If $\B$ is well-behaved then $\V_L(n)=\O(\log n)$ and $\F_L(n)=\O(\log n)$.
		\label{item:tri-1}
		\item If $\B$ is not well-behaved then
		$\V_L(n)=\Omega(n)$ and $\F_L(n)=\Omegainf(n)$.
		\label{item:tri-2}
		\item If $U(\B)$ is well-behaved then $\F_L(n)=\O(1)$.
		\label{item:tri-3}
		\item If $U(\B)$ is not well-behaved then
		$\F_L(n)=\Omegainf(\log n)$.
		\label{item:tri-4}
		\item If $L \in \{\emptyset,\Sigma^*\}$
		then $\V_L(n)=\O(1)$.
		\label{item:tri-5}
		\item If $L \notin \{\emptyset,\Sigma^*\}$
		then $\V_L(n)=\Omega(\log n)$.
		\label{item:tri-6}
	\end{enumerate}
\end{theorem}

\Cref{thm:trichotomy} implies that 
$\F_L(n)$ is either $\Theta(1)$, $\Thetainf(\log n)$, or $\Thetainf(n)$,
and 
$\V_L(n)$ is either $\Theta(1)$, $\Theta(\log n)$, or $\Theta(n)$.
For the characterizations in \Cref{thm:big-thm} it remains to prove:

\begin{theorem}
	\label{thm:characterization}
	Let $L \subseteq \Sigma^*$ be regular.
	\begin{enumerate}[label=(\roman{*}), ref=(\roman{*})]
		\item $\F_L(n) = \O(1)$ $\iff$ $L \in \langle \ST, \Len \rangle$.
		\label{item:const}
		\item $\F_L(n) = \O(\log n)$ $\iff$ $L \in \langle \LI, \Len \rangle$.
		\label{item:log}
	\end{enumerate}	
\end{theorem}
In the rest of \Cref{chp:reg} we prove \Cref{thm:trichotomy} and \Cref{thm:characterization}.
We start with the path summary algorithm, which is our main deterministic SW-algorithm for the variable-size model.

\subsection{The path summary algorithm}

\label{subsec:pathsummary}

In the following, let $\B = (Q,\Sigma,F,\delta,q_0)$ be a right-deterministic finite automaton.
We call a run~$\pi$ {\em internal} if $\pi$ is a $P$-run for some SCC $P$.
The {\em SCC-factorization} of $\pi$ is the unique factorization
$\pi = \pi_k \tau_{k-1} \cdots \tau_2 \pi_2 \tau_1 \pi_1$, where every $\pi_i$ is an internal
(possibly empty) run but cannot be extended to an internal run. The $\tau_i$ are single transitions
(i.e.,~runs from $Q \Sigma Q$) connecting distinct SCCs.
Let $p_k, \dots, p_1 \in Q$ be the starting states of the runs $\pi_k, \dots, \pi_1$.
Then, the {\em path summary} of $\pi$ is defined~as
\[
	\ps(\pi) = (|\pi_k|,p_k) (|\tau_{k-1} \pi_{k-1}|,p_{k-1}) \cdots (|\tau_2 \pi_2|,p_2) (|\tau_1 \pi_1|,p_1),
\]
which is a sequence of pairs from $\N \times Q$.
It specifies the first state that is visited in an SCC,
and the length of the run until reaching the next SCC or the end of the word, respectively.
The leftmost length $|\pi_k|$ can be zero but all other lengths $|\tau_i \pi_i| = 1+|\pi_i|$ are strictly positive.
We define $\pi_{w,q}$ to be the unique run of $\B$ on a word $w \in \Sigma^*$ starting from $q$,
and $\PS_\B(w) = \{\ps(\pi_{w,q}) \mid q \in Q\}$.

\begin{sidefigure}[t]
  \centering
        \includegraphics[scale=1.488]{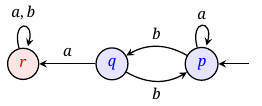}

	\caption{Another rDFA partitioned into two SCCs.}
	\label{fig:path-summary}
\end{sidefigure}

\newcommand{\qr}{\makebox[.7em]{\textcolor{red}{$r$}}}
\newcommand{\qq}{\makebox[.7em]{\textcolor{blue}{$q$}}}
\newcommand{\qp}{\makebox[.7em]{\textcolor{blue}{$p$}}}

\begin{example}
Consider the rDFA $\B$ in \Cref{fig:path-summary}.
For the moment, the final states are irrelevant.
It consists of two SCCs, namely
the blue SCC $\{p,q\}$ and the red SCC $\{r\}$.
All its runs on the word $w = aababb$ are listed here:
\begin{align*}
	\qr \xleftarrow{a} \qr \xleftarrow{a} \qq \xleftarrow{b} \qp \xleftarrow{a} \qp \xleftarrow{b} \qq \xleftarrow{b} \qp \\
	\qr \xleftarrow{a} \qr \xleftarrow{a} \qr \xleftarrow{b} \qr \xleftarrow{a} \qq \xleftarrow{b} \qp \xleftarrow{b} \qq \\
	\qr \xleftarrow{a} \qr \xleftarrow{a} \qr \xleftarrow{b} \qr \xleftarrow{a} \qr \xleftarrow{b} \qr \xleftarrow{b} \qr
\end{align*}
Then, $\PS_\B(w)$ contains the path summaries
$(1,r) (5,p)$, $(3,r) (3,q)$ and $(6,r)$.
\end{example}

The {\em path summary algorithm for $\B$} is a streaming algorithm over $\Sigma_\downarrow$
described in Algorithm~\ref{alg:ps}.
The data structure at time instant $t$ is denoted by $S_t$.
The acceptance condition will be defined later.

\begin{lemma}
\label{lem:ps}
	Algorithm~\ref{alg:ps} correctly maintains $\PS_\B(w)$ for the active window $w \in \Sigma^*$.
\end{lemma}

\begin{proof}
	Initially $\PS_\B(\eps)$ contains the path summary of every empty run from every state,
	which is formally $\{0\} \times Q$.
	
	Assume $S_{t-1} = \PS_\B(w)$ for some window $w \in \Sigma^*$ and that $a \in \Sigma$ is the incoming symbol.
	The claim is that the algorithm computes $S_t = \PS_\B(wa)$ from $S_{t-1}$.
	Suppose that $\pi'$ is a run in $\B$ on~$wa$.
	It can be factorized as $\pi' = \pi \, p_1 a \, p_0$ with $\ps(\pi) \in S_{t-1}$.
	Let $\pi = \pi_k \tau_{k-1}  \pi_{k-1} \cdots \tau_2 \pi_2 \tau_1 \pi_1$ be the SCC-factorization of $\pi$.
	If $p_0$ and $p_1$ are strongly connected then the SCC-factorization
	of $\pi'$ is $\pi' = \pi_k \tau_{k-1}  \pi_{k-1} \cdots \tau_2 \pi_2 \tau_1 \pi_1'$
	where $\pi_1' = \pi_1 \, p_1 a \, p_0$, and otherwise
	$\pi' = \pi_k \tau_{k-1}  \pi_{k-1} \cdots \tau_2 \pi_2 \tau_1 \pi_1 \, p_1 a \, p_0$.
	In this way the algorithm computes $\ps(\pi')$ from $\ps(\pi)$.
	
	Now, consider the case $a = {\downarrow}$.
	We have $w = \eps$ if and only if $\PS_\B(w) = \{0\} \times Q$,
	and in this case the set of path summaries  is unchanged, i.e.,~we set $S_t = S_{t-1}$.
	Otherwise assume $w = bv$ for some $b \in \Sigma$.
	We claim that the algorithm computes $S_t = \PS_\B(v)$ from $S_{t-1}$.
	Suppose that $\pi'$ is a run in $\B$ on $v$ which ends in state $p \in Q$.
	If $q = \delta(b,p)$ in $\B$ then let $\pi = q \, b \, p \, \pi'$. We have $\ps(\pi) \in S_{t-1}$.
	Let $\pi = \pi_k \tau_{k-1}  \pi_{k-1} \cdots \tau_2 \pi_2 \tau_1 \pi_1$ be the SCC-factorization of $\pi$.
	If $|\pi_k| \ge 1$ then $\pi' = \pi_k' \tau_{k-1}  \pi_{k-1} \cdots \tau_2 \pi_2 \tau_1 \pi_1$ 
	is the SCC-factorization of $\pi'$ where $\pi_k = q \, b \, p \, \pi_k'$.
	Otherwise $\pi_k$ is empty and $\tau_{k-1} = q \, b \, p$.
	Therefore, $\pi' = \pi_{k-1} \cdots \tau_2 \pi_2 \tau_1 \pi_1$ is the SCC-factorization of $\pi'$.
	In this way the algorithm computes $\ps(\pi')$ from $\ps(\pi)$.
\end{proof}

\begin{algorithm}[t]
	\KwIn{sequence of operations $a_1 a_2 a_3 \cdots \in \Sigma_\downarrow^\omega$}
	$S_0 = \{0\} \times Q$\;
	\ForEach{$t \ge 1$}
	{
	$S_t = \emptyset$\;
	\If{$a_t \in \Sigma$}{
		\For{$p_0 \in Q$}{
			let $p_1 = a_t \cdot p_0$ and $(\ell_k,p_k) \cdots (\ell_1,p_1) \in S_{t-1}$\;
			\eIf{$p_0$ and $p_1$ are strongly connected\label{line:sc}}{
				add $(\ell_k,p_k) \cdots (\ell_2,p_2) (\ell_1+1,p_0)$ to $S_t$\;
			}{
      			add $(\ell_k,p_k) \cdots (\ell_1,p_1) (1,p_0)$ to $S_t$\;
      		}
		}
 	}
	\If{$a_t = {\downarrow}$}{
		\eIf{$S_{t-1} = \{0\} \times Q$}{
			$S_t = S_{t-1}$\;
		}{
		\For{$(\ell_k,p_k) \cdots (\ell_1,p_1) \in S_{t-1}$}{
			\eIf{$\ell_k \ge 1$}{
				add $(\ell_k-1,p_k) (\ell_{k-1},p_{k-1}) \cdots (\ell_1,p_1)$ to $S_t$\;
			}{
      			add $(\ell_{k-1}-1,p_{k-1}) (\ell_{k-2},p_{k-2})  \cdots (\ell_1,p_1)$ to $S_t$\;
      		}
		}
		}
 	}
	}
	\caption{The path summary algorithm}
	\label{alg:ps}
\end{algorithm}

Observe that Algorithm~\ref{alg:ps} cannot be directly adapted to work for (left-)DFA:
For a pop operation, one would need to remove the first transition from each path summary,
which is generally not possible since a path summary does not store its second state.

\subsection{Proof of Theorem \ref{thm:trichotomy}\ref{item:tri-1}}

Using the path summary algorithm we can prove \Cref{thm:trichotomy}\ref{item:tri-1}:

\begin{proposition}
	\label{prop:wb-log}
	If $\B$ is well-behaved then the regular language $L = \L(\B)$ has space complexity $\V_L(n) = \O(|\B|^2 \cdot \log n)$,
	which is $\O(\log n)$ for a fixed $\B$.
\end{proposition}

\begin{proof}
	Let $\B$ be well-behaved.
	Call a path summary {\em accepting} if it is the path summary of some accepting run.
	The variable-size sliding window algorithm for $L$ is the path summary algorithm for $\B$
	where the algorithm accepts if the path summary starting in $q_0$ is accepting.
	
	For the correctness of the algorithm it suffices to show that
	any run $\pi$ starting in $q_0$ is accepting
	if and only if $\ps(\pi)$ is accepting.
	The direction from left to right is immediate by definition.
	For the other direction, consider the path summary $\ps(\pi) = (\ell_k,p_k) \cdots (\ell_1,p_1)$
	and the SCC-factorization $\pi = \pi_k \tau_{k-1} \cdots \tau_2 \pi_2 \tau_1 \pi_1$.
	Since $\ps(\pi)$ is accepting, there is an accepting run $\pi_k'$
	that starts in $p_k$ and has length $\ell_k$.
	Since $\B$ is well-behaved, the SCC of $p_k$ is well-behaved.
	Therefore, since $\pi'_k$ is accepting and $|\pi_k|=|\pi'_k|=\ell_k$, $\pi_k$ must also be accepting and
	thus $\pi$ is accepting.
	
	We claim that the space complexity of the path summary algorithm is bounded by $\O(|\B|^2 \cdot (\log n + \log |\B|))$.
	Observe that $\PS_\B(w)$ contains $|\B|$ path summaries,
	and a single path summary $\ps(\pi)$ consists of a sequence of at most $|\B|$ states
	and a sequence $(\ell_k, \dots, \ell_1)$ of 
	$k \le |\B|$ numbers up to $|\pi|$.
	Hence, the path summary $\ps(\pi)$ can be encoded using $\O(|\B| \cdot (\log |\B| + \log |\pi|))$ bits, which yields the total
	space complexity $\O(|\B|^2 \cdot (\log n + \log |\B|))$.
	
	To reduce the space complexity to $\O(|\B|^2 \cdot \log n)$ we need to make a case distinction.
	The algorithm maintains the window size $n \in \N$ and the maximal suffix of the window of length up to $|\B|$
	(explicitly) using $\O(\log n + |\B|)$ bits.
	If $n \le |\B|$ then this information suffices to test membership of the window to $L$.
	As soon as $n$ exceeds~$|\B|$ we initialize $\PS_\B(w)$ and use the path summary algorithm
	as described above.
	If $n > |\B|$ then its space complexity is
	$\O(|\B|^2 \cdot (\log n + \log  |\B|)) \subseteq \O(|\B|^2 \cdot \log n)$.
\end{proof}

Observe that the path summary algorithm only stores $\O(\log n)$ bits
where $n$ is the \emph{current} (not the \emph{maximum}) window size.

Before we continue with the proof of the other points from \Cref{thm:trichotomy}, 
we discuss some implementation details for our logspace SW-algorithm.\footnote{These details are not needed for the proof of \Cref{prop:wb-log}
since we abstract from internal computations in our sliding window model; see \Cref{ref:internal comp}.}
To implement the path summary algorithm on a realistic computation model,
we have to be able to efficiently determine whether a path summary is accepting.
Given a number $d \ge 1$,
a set of natural numbers $X \subseteq \N$ is {\em $d$-periodic}
if we have $x \in X$ if and only if $x + d \in X$.

\begin{lemma}
	\label{lem:wb-period}
	Let $P \subseteq Q$ be a well-behaved subset in $\B$ and $p_0 \in P$ be nontransient.
	Then $\mathrm{Acc}(P,p_0) := \{ |\pi| : \pi \text{ is an accepting $P$-run starting in $p_0$ } \}$
	is $d$-periodic for some $d \le |Q|$.
\end{lemma}

\begin{proof}
	Let $\pi_0$ be any nonempty run from $p_0$ to $p_0$, which exists because $p_0$ is nontransient.
	Furthermore, we can choose $\pi_0$ such that its length $d := |\pi_0|$ is at most $|Q|$.
	
	If $\ell \in \mathrm{Acc}(P,p_0)$, then there exists an accepting $P$-run $\pi$ starting in $p_0$
	of length $\ell$.
	Then $\pi \pi_0$ is also an accepting $P$-run
	and we conclude $|\pi\pi_0| = \ell + d \in \mathrm{Acc}(P,p_0)$.

	Now we need to show that $\ell \notin \mathrm{Acc}(P,p_0)$ implies $\ell + d \notin \mathrm{Acc}(P,p_0)$.
	Towards a contradiction assume that $\ell \notin \mathrm{Acc}(P,p_0)$
	and $\ell + d \in \mathrm{Acc}(P,p_0)$,
	i.e.,~there exists an accepting $P$-run $\pi$ starting in $p_0$ of length $\ell+d$.
	Factorize $\pi = \pi_1 \pi_2$ where $|\pi_2| = \ell$.
	Now $\pi_2$ must be rejecting since $\ell \notin \mathrm{Acc}(P,p_0)$.
	But then $\pi_2 \pi_0$ is a rejecting $P$-run of length $\ell+d$,
	which contradicts the well-behavedness of $P$ since $\ell + d \in \mathrm{Acc}(P,p_0)$.
\end{proof}

In the following we describe how to implement the algorithm from \Cref{prop:wb-log}.
We do the following preprocessing on the well-behaved rDFA $\B$.
Using depth-first search we compute all SCCs in $\B$.
For every SCC $P$ we pick a state $p \in P$ and compute the distance
$\mathrm{dist}(p,q)$ from $p$ to all states $q \in P$ using any shortest path algorithm.
Furthermore let $d$ be the minimal length of a nonempty run from $p$ to $p$ itself,
which is the period $d$ from \Cref{lem:wb-period}.
If no such run exists then we store the information that $p$ is transient.
Otherwise we assign to each state $q \in P$ the distance from $p$ modulo $d$.
By traversing an arbitrary $P$-run of length $d$ from $p$ we can compute a bit vector of length $d$
which represents $\mathrm{Acc}(P,p_0)$.
Using this information we can easily answer whether a path summary $(\ell_k,p_k) \cdots (\ell_1,p_1)$ is accepting:
it is accepting if and only if
either $p_k$ is transient, $\ell_k = 0$ and $p_k \in F$,
or $\mathrm{dist}(p_0,p_k) + \ell_k \bmod d$ belongs to $\mathrm{Acc}(P,p_0)$
where $P$ is the SCC of $p_k$ and $p_0$ is the picked state in $P$.

\subsection{Proof of Theorem \ref{thm:trichotomy}\ref{item:tri-2}}
We continue with proving a linear lower bound for rDFA which are not well-behaved.

\begin{lemma}
	\label{lem:linear-lb}
	If $\B$ is not well-behaved then there exist words $u_1, u_2, v_1, v_2, z \in \Sigma^*$
	where $|u_i| = |v_i|$ for $i \in \{1,2\}$
	such that $L = \L(\B)$ separates $u_2\{u_1u_2,v_1v_2\}^*z$ and $v_2\{u_1u_2,v_1v_2\}^*z$.
\end{lemma}

\begin{proof}
	The automaton structure is illustrated in~\cref{fig:wb}.
	Since $\B = (Q,\Sigma,F,\delta,q_0)$ is not well-behaved, 
	there is a reachable SCC $S$ that is not well-behaved. 
	Take a state $p \in S$ and a word $z \in \Sigma^*$ with 
	\[
	p \xleftarrow{z} q_0 .
	\]
	Moreover, since $S$ is strongly connected and not well-behaved
	there are states $q \in S \cap F$, $r \in S \setminus F$
	and nonempty words $u_2, v_2 \in \Sigma^*$ such that $|u_2| = |v_2|$ and
	\[
	 q \xleftarrow{u_2} p \xleftarrow{z} q_0 \quad\text{ and }\quad  r \xleftarrow{v_2} p \xleftarrow{z} q_0 .
	\]
	Finally, since  $S$ is strongly connected, there are words $u_1, v_1 \in \Sigma^*$ such that
	\[
		p \xleftarrow{u_1} q \xleftarrow{u_2} p \xleftarrow{z} q_0 \quad\text{ and }\quad p \xleftarrow{v_1} r \xleftarrow{v_2} p \xleftarrow{z} q_0 .
	\]
	We can ensure that $|u_1| = |v_1|$ and hence also $|u| = |v|$ for $u = u_1 u_2, v = v_1 v_2$.
	If $k = |u|$ and $\ell = |v|$ we replace $u_1$ by $u^{\ell-1} u_1$
	and $v_1$ by $v^{k-1} v_1$ (note that $k > 0$ and $\ell > 0$ since 
	$u$ and $v$ are nonempty), which preserves all properties above.
	Then, $u_2\{u,v\}^*z$ and $v_2\{u,v\}^*z$ are separated by $L$.
\end{proof}
We can now show \Cref{thm:trichotomy}\ref{item:tri-2}:
\begin{proposition}
	\label{prop:linear-lb}
	If $\B$ is not well-behaved then the language $L = \L(\B)$
	satisfies $\F_L(n) = \Omegainf(n)$ and $\V_L(n) = \Omega(n)$.
\end{proposition}
	
\begin{proof}
	Let $u_1, u_2, v_1, v_2, z \in \Sigma^*$ be the words from \Cref{lem:linear-lb}
	and let $u = u_1u_2$ and $v = v_1v_2$.
	Now, consider an SW-algorithm $\P_n$ for $L$ and window size
	$n = |u_2| + |u| \cdot (m-1) + |z|$ for some $m \ge 1$.
	We prove that $\P_n$ has at least $2^m$ many states
	by showing that $\P_n(x) \neq \P_n(y)$ for any $x,y  \in \{ u, v \}^m$ with $x\neq y$.
	Notice that $|\{u,v\}^m| = 2^m$ since $u \neq v$ and $|u| = |v|$.
	
	Read two distinct words $x, y \in \{u,v\}^m$ into two instances of $\P_n$.
	Consider the right most $\{u,v\}$-block where $x$ and $y$ differ.
	Without loss of generality assume $x = x' u s$ and $y = y' v s$ for some $x',y',s \in \{u,v\}^*$ with $|x'|=|y'|$.
	By reading $x' z$ into both instances
	the window of the $x$-instance becomes
	$\last_n(x x' z) = u_2 s x' z$
	and the window of the $y$-instance becomes
	$\last_n(y x' z) = v_2 s x' z$.
	By \Cref{lem:linear-lb} the two windows are separated by $L$,
	and therefore the algorithm $\P_n$ must accept one of the streams $x x' z$ and $y x' z$, and reject the other.
	In conclusion $\P_n(x) \neq \P_n(y)$ and hence $\P_n$ must use at least $m = \Omega(n)$ bits. This holds for infinitely
	many $n$, namely all $n$ of the form $|u_2|+|z| + |u| \cdot (m-1)$ for some $m \ge 1$.
	
	The argument above shows that there exist numbers $c,d \in \N$
	such that for all $m \ge 1$ we have $\V_L(cm+d) \ge \F_L(cm+d) = \Omega(m)$.
	If $n \geq d$ then $m = \lfloor (n-d)/c \rfloor = \Omega(n)$ satisfies $cm+d \le n$.
	Therefore, $\V_L(n) \ge \V_L(cm+d) = \Omega(m)$ by monotonicity of $\V_L$ and hence $\V_L(n) = \Omega(n)$.
\end{proof}

\begin{sidefigure}
  \centering
      \includegraphics[scale=1.288]{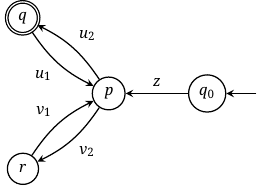}

	\caption{Forbidden pattern for well-behaved rDFAs where $|u_1| = |v_1|$ and $|u_2| = |v_2|$.}
	\label{fig:wb}
\end{sidefigure}

From \Cref{prop:wb-log} and \Cref{prop:linear-lb} we obtain:
\begin{corollary}
	\label{cor:log-wb}
	Let $\X \in \{ \F,\V\}$.
	A regular language $L \subseteq \Sigma^*$ satisfies $\X_L(n) = \O(\log n)$
	if and only if $L$ is recognized by a well-behaved rDFA.
\end{corollary}

\subsection{Proof of Theorem \ref{thm:trichotomy}\ref{item:tri-3}--\ref{item:tri-6}}

\label{subsec:const}

Next, we study which regular languages have sublogarithmic complexity.
Recall that in the variable-size model any such language must be empty or universal
because the algorithm must at least 
maintain the current window size by \Cref{lem:length-lb}.

\begin{corollary}
	\label{cor:v-triv}
	The empty language $L = \emptyset$ and the universal language $L = \Sigma^*$ satisfy $\V_L(n) = \O(1)$.
	All other languages satisfy $\V_L(n) = \Omega(\log n)$.
\end{corollary}

This proves points \ref{item:tri-5} and \ref{item:tri-6}
in \Cref{thm:trichotomy}.
Now, we can turn to the fixed-size model
and prove the points \ref{item:tri-3} and \ref{item:tri-4}. Point~\ref{item:tri-3} follows from:

\begin{proposition}
	\label{prop:r-wb}
	If $U(\B)$ is well-behaved then $L = \L(\B)$ has space complexity $\F_L(n) = \O(|\B|)$,
	which is $\O(1)$ when $\B$ is fixed.
\end{proposition}

\begin{proof}
	Let $k = |\B|$.
	The SW-algorithm $\P_n$ for $\SW_n(L)$ maintains $\last_{k}(x)$ for an input stream $x \in \Sigma^*$
	using $\O(k)$ bits.
	If $n \leq k$ then $\last_n(x)$ is a suffix of $\last_{k}(x)$ and hence $\P_n$ can determine
	whether $\last_n(x) \in L$.
	If $n > k$ then $\last_{k}(x)$ is a suffix of $\last_n(x)$,
	say $\last_n(x) = s  \last_{k}(x)$.
	We can decide if $\last_n(x) \in L$ as follows:
	Consider the run of $\B$ on $\last_n(x)$ starting from the initial state:
	\[ r \xleftarrow{s} q \xleftarrow{\last_{k}(x)} q_0 . \]
	By the choice of $k$ some state $p \in Q$ must occur twice in the run $q \xleftarrow{\last_{k}(x)} q_0$.
	Therefore, $p$ is nontransient and all states in the run $r \xleftarrow{s} q$ belong to $U(\B)$.
	Since $U(\B)$ is well-behaved, $r$ is final
	if and only if some run of length $|s| = n - k$ starting in $q$ is accepting.
	This information can be precomputed for each state $q$ in the fixed-size model.
\end{proof}
For the lower bound in \Cref{thm:trichotomy}\ref{item:tri-4} we need the following lemma:
\begin{lemma}
	\label{lem:nonwb-loop}
	If $U(\B)$ is not well-behaved
	then there exist words $x, y, z \in \Sigma^*$
	where $|x| = |y|$ such that
	$L = \L(\B)$ separates $xy^*z$ and $y^*z$.
\end{lemma}

\begin{proof}
	Since $U(\B)$ is not well-behaved, there are $U(\B)$-runs $\pi$ and $\rho$
	from the same starting state $q \in U(\B)$ such that $|\pi| = |\rho|$
	and exactly one of the runs $\pi$ and $\rho$ is accepting.
	By definition of $U(\B)$ the state $q$ is reachable from a nontransient state $p$ via some run $\sigma$
	such that $p$ is reachable from the initial state $q_0$, say $p \xleftarrow{z_0} q_0$.
	We can replace $\pi$ by $\pi \sigma$ and $\rho$ by $\rho \sigma$
	preserving the properties of being $U(\B)$-runs and $|\pi| = |\rho|$.
	Assume that $\pi$ and $\rho$ are runs on words $v \in \Sigma^*$ and $w \in \Sigma^*$, respectively.
	Since $p$ is nontransient,
	we can construct runs from $p$ to $p$ of unbounded lengths.
	Consider such a run $p \xleftarrow{u} p$ of length $|u| \ge |v| = |w|$.
	Then, $L$ separates $v u^* z_0$ and $w u^* z_0$.
	Factorize $u = u_1 u_2$ so that $|u_2| = |v| = |w|$.
	Notice that all words in $u_2 u^*z_0$ reach the same state in $\B$
	and hence $u_2 u^*z_0$ is either contained in $L$ or disjoint from $L$.
	Then, $L$ separates either $u_2 u^* z_0$ and $v u^* z_0$,
	or $u_2 u^* z_0$ and $w u^* z_0$.
	Hence, $L$ also separates $(u_2 u_1)^*  u_2 z_0$ from either $v u_1 (u_2 u_1)^*  u_2 z_0$ or from $w u_1 (u_2 u_1)^*  u_2 z_0$.
	This yields the words $z= u_2 z_0$, $y=u_2 u_1$ and $x=v u_1$ or $x=w u_1$ with the claimed properties.
\end{proof}

\begin{proposition}
	\label{prop:logn-lb}
	If $U(\B)$ is not well-behaved then $L = \L(\B)$
	satisfies $\F_L(n) \geq \log n - \O(1)$ for infinitely many $n$.
	In particular, $\F_L(n) = \Omegainf(\log n)$.
\end{proposition}

\begin{proof}
	Let $x,y,z \in \Sigma^*$ be the words from \Cref{lem:nonwb-loop}.
	Consider an SW-algorithm $\P_n$ for $L$ and window size $n = |x| + |y| \cdot m + |z|$ for some $m \ge 1$.
	We prove that $\P_n$ has at least $m$ many states
	by showing that $\P_n(x y^i) \neq \P_n(x y^j)$ for any $1 \le i < j \le m$.
	Let $1 \le i < j \le m$.
	Then, we have
	\[
		\last_n(x y^i y^{m-i} z) = \last_n(x y^m z) = x y^m z
	\]
	and
	\[
		\last_n(x y^j y^{m-i} z) = \last_n(x y^{m+j-i} z) = y^{m+1} z.
	\]
	Since exactly one of the words $x y^m z$ and $y^{m+1} z$ belongs to $L$,
	it also holds that exactly one of the streams $x y^i y^{m-i} z$ and $x y^j y^{m-i} z$
	is accepted by $\P_n$.
	This proves that $\P_n$ must reach different memory states on inputs $x y^i$ and $x y^j$.
	In conclusion $\P_n$ must use $\log m \ge \log n - \O(1)$ bits, and this holds for infinitely many $n$.
\end{proof}

\subsection{Characterization of constant space}

\label{sec:char-const}

Next, we prove that a regular language $L$ has constant space complexity $\F_L(n)$
if and only if it is a Boolean combination of suffix testable languages
and regular length languages (\Cref{thm:characterization}\ref{item:const}).

The language $L$ is called {\em $k$-suffix testable}
if for all $x,y \in \Sigma^*$ and $z \in \Sigma^k$ we have
$xz \in L$ if and only if $yz \in L$.
Equivalently, $L$ is a Boolean combination of languages of the form $\Sigma^* w$ where $w \in \Sigma^{\le k}$.
Clearly, a language is suffix testable if and only if
it is $k$-suffix testable for some $k \in \N$.
Let us remark that the class of suffix testable languages corresponds to the variety
$\mathbf{D}$ of definite monoids~\cite{Straubing85}.
Clearly, every finite language
is suffix testable: If $k$ is the maximum length of a word in $L \subseteq \Sigma^*$ then $L$ is $(k+1)$-suffix testable
since $L = \bigcup_{w \in L} \{w\}$ and $\{w\} = \Sigma^* w \setminus \bigcup_{a \in \Sigma} \Sigma^*aw$. 

We will utilize a distance notion between states in a DFA,
which is also studied in~\cite{GawrychowskiJ09}.
The {\em symmetric difference} of two sets $A$ and $B$ is $A \mathop{\triangle} B = (A \cup B) \setminus (A \cap B)$.
We define the distance $d(K,L)$ of two languages $K,L\subseteq\Sigma^*$ by
\[
	d(K,L) = \begin{cases}
	\sup_{u \in K \mathop{\triangle} L} |u| + 1, & \text{if } K \neq L, \\
	0, & \text{if } K = L.
	\end{cases}
\]
Notice that $d(K,L)<\infty$ if and only if $K\mathop{\triangle} L$ is finite.
For a DFA $\A = (Q,\Sigma,q_0,\delta,F)$ and a state $p\in Q$, we define $\A_p = (Q,\Sigma,p,\delta,F)$.
Moreover, for two states $p,q\in Q$, we define the distance $d(p,q)=d(\L(\A_p),\L(\A_q))$.
If we have two runs $p \xrightarrow{u} p'$ and $q \xrightarrow{u} q'$ where $p' \in F$,
$q' \notin F$ and $|u| \ge |Q|^2$ then some state pair occurs twice in the runs
and we can pump the runs to unbounded lengths.
Therefore, $d(p,q) < \infty$ implies $d(p,q) \le |Q|^2$.
In fact $d(p,q) < \infty$ implies $d(p,q) \le |Q|$ by \cite[Lemma~1]{GawrychowskiJ09}.
	
\begin{lemma}
	\label{lem:testable-distance}
	Let $L \subseteq \Sigma^*$ be regular and $\A = (Q,\Sigma,q_0,\delta,F)$ be its minimal DFA. 
	We have:
	\begin{enumerate}[label=(\roman{*}), ref=(\roman{*})]
		\item for all $p,q \in Q$, $d(p,q) \le k$ if and only if $\forall z \in \Sigma^k : p \cdot z = q \cdot z$,
		\label{item:st1}
		\item $L$ is $k$-suffix testable if and only if $d(p,q) \le k$ for all $p,q \in Q$,
		\label{item:st2}
		\item if there exists $k \ge 0$ such that $L$ is $k$-suffix testable, then $L$ is $|Q|$-suffix testable.
		\label{item:st3}
	\end{enumerate}
\end{lemma}

\begin{proof}
	The proof of \ref{item:st1} is an easy induction.
	If $k=0$, the statement is $d(p,q) = 0$ if and only if $p = q$,
	which is true because $\A$ is minimal. For the induction step, we have
	$d(p,q) \le k+1$ if and only if $d(\delta(p,a),\delta(q,a)) \le k$
	for all $a \in \Sigma$ if and only if $\delta(p,a) \cdot z = \delta(q,a) \cdot z$ for all $z \in \Sigma^{k}$
	if and only if $p \cdot z = q \cdot z$
	for all $z \in \Sigma^{k+1}$.
					
	For \ref{item:st2}, assume that $L$ is $k$-suffix testable and consider two states $p = \A(x)$ and $q= \A(y)$.
	If $z \in \L(\A_p) \triangle \L(\A_q)$, then $|z| < k$ because $xz \in L$ if and only if $yz \notin L$
	and $L$ is $k$-suffix testable.
						
	Now, assume that $d(p,q) \le k$ for all $p,q \in Q$ and consider $x,y \in \Sigma^*$, $z \in \Sigma^k$.
	Since we have $d(\A(x),\A(y)) \le k$, \ref{item:st1} implies $\A(xz) = \A(yz)$, and in particular $xz \in L$ if and only if $yz \in L$.
	Therefore, $L$ is $k$-suffix testable.

	Point \ref{item:st3} follows from \ref{item:st2} and from the above cited \cite[Lemma~1]{GawrychowskiJ09}.
\end{proof}

\begin{lemma}\label{thm:suffix-language}
	For any $L \subseteq \Sigma^*$ and $n \ge 0$,
	the language $\SW_n(L)$ is $2^{\F_L(n)}$-suffix testable.
\end{lemma}
\begin{proof}
	Let $\P_n$ be an SW-algorithm for $L$ and window size $n$
	with space complexity $\F_L(n)$.
	Therefore, $\P_n$ has at most $2^{\F_L(n)}$ states.
	The definition of $\SW_n(L)$ directly implies that $\SW_n(L)$ is $n$-suffix testable.
	By \Cref{lem:testable-distance}\ref{item:st3} $\SW_n(L)$ is $2^{\F_L(n)}$-suffix testable.
\end{proof}
Note that \Cref{thm:suffix-language} holds for arbitrary languages and not only for regular languages.

\begin{proof}[Proof of Theorem \ref{thm:characterization}\ref{item:const}]
	First, let $L \subseteq \Sigma^*$ be a regular language with $\F_L(n) = \O(1)$
	and let $k = \max_{n \in \N} 2^{\F_L(n)}$.
	By \Cref{thm:suffix-language} the language $\SW_n(L)$ is $k$-suffix testable for all $n \ge 0$.
	We can express $L$ as the Boolean combination
	\[
		L = (L \cap \Sigma^{\le k-1}) \cup \bigcup_{z \in \Sigma^k} (Lz^{-1}) \, z = (L \cap \Sigma^{\le k-1}) \cup \bigcup_{z \in \Sigma^k} ((Lz^{-1}) \, \Sigma^k \cap \Sigma^* z)
	\]
	where the right quotient $Lz^{-1} = \{ x \in \Sigma^* \mid xz \in L \}$ is regular \cite[Chapter~3,~Example~5.7]{Berstel79}.
	The set $L \cap \Sigma^{\le k-1}$ is finite and hence suffix testable. 
	It remains to show that each $Lz^{-1}$ for $z \in \Sigma^k$  is a length language.
	Consider two words $x,y \in \Sigma^*$ of the same length $|x|=|y|=n$.
	Since $|xz|=|yz| = n+k$ and $\SW_{n+k}(L)$ is $k$-suffix testable, we have $xz \in L$ if and only if $yz \in L$,
	and hence $x \in Lz^{-1}$ if and only if $y \in Lz^{-1}$.
			
	For the other direction note that
	(i) if $L$ is a length language or a suffix testable language then clearly
	$\F_L(n) = \O(1)$, and
	(ii) $\{ L \subseteq \Sigma^* \mid \F_L(n) = \O(1) \}$ is closed under Boolean operations by \Cref{lem:boolean}.
	This proves the theorem.
\end{proof}

\subsection{Characterization of logarithmic space}

\label{sec:char-log}

Recall from \Cref{thm:trichotomy} that well-behaved \mbox{rDFAs} precisely define those regular languages
with logarithmic space complexity $\F_L(n)$ or equivalently $\V_L(n)$.
In the following, we will show that
well-behaved rDFAs recognize precisely the finite Boolean combinations of
regular left ideals and regular length languages,
which therefore are precisely the regular languages with logarithmic space complexity (\Cref{thm:characterization}\ref{item:log}).
Let us start with the easy direction:

\begin{proposition}
	Every language $L \in \langle \LI, \Len \rangle$ is recognized by a well-behaved rDFA.
\end{proposition}

\begin{proof}
	Let $\B$ be an rDFA for $L$.
	If $L$ is a length language then for all reachable states 
	$q$ and all runs $\pi,\pi'$ starting from $q$
	with $|\pi| = |\pi'|$ we have:
	$\pi$ is accepting if and only if $\pi'$ is accepting.
	If $L$ is a left ideal, then whenever a final state $p$ is reachable,
	and $q$ is reachable from $p$, then $q$ is also final.
	Hence, for every reachable SCC $P$ in $\B$ either all states of $P$ are final or all states
	of $P$ are nonfinal. In particular, $\B$ is well-behaved.

	It remains to show that the class of languages $L \subseteq \Sigma^*$ recognized by well-behaved rDFAs
	is closed under Boolean operations.
	If $\B$ is well-behaved then the complement automaton $\overline \B$ is also well-behaved.
	Given two well-behaved rDFAs $\B_1,\B_2$,
	we claim that the product automaton $\B_1 \times \B_2$ recognizing the intersection language is also well-behaved.
	Suppose that $\B_i = (Q_i,\Sigma,F_i,\delta_i,q_{0,i})$ for $i \in \{1,2\}$.
	The product automaton for the intersection language is defined by
	\[
	\B_1 \times \B_2 = (Q_1 \times Q_2,\Sigma,F_1 \times F_2,\delta,(q_{0,1},q_{0,2}))
	\]
	where $\delta(a,(q_1,q_2)) = (\delta_1(a,q_1),\delta_2(a,q_2))$ for all $q_1 \in Q_1, q_2 \in Q_2$ and $a \in \Sigma$.
	Consider an SCC $S$ of $\B_1 \times \B_2$ which is reachable from the initial state and let $(p_1,p_2), (q_1,q_2), (r_1,r_2) \in S$
	such that
	\[
		(q_1,q_2) \xleftarrow{u} (p_1,p_2) \text{ and } (r_1,r_2) \xleftarrow{v} (p_1,p_2)
	\]
	for some words $u,v \in \Sigma^*$ with $|u| = |v|$.
	Since for $i \in \{1,2\}$ we have $q_i \xleftarrow{u} p_i$ and $r_i \xleftarrow{v} p_i$, and $\{p_i,r_i,q_i\}$
	is contained in an SCC of $\B_i$ (which is also reachable from the initial state $q_{0,i}$), we have
	\begin{eqnarray*}
            (q_1,q_2) \text{ is final} & \iff & q_1 \text{ and } q_2 \text{ are final} \\
            & \iff & r_1 \text{ and } r_2 \text{ are final} \\
            & \iff & (r_1,r_2) \text{ is final},
         \end{eqnarray*}
	and therefore $\B_1 \times \B_2$ is well-behaved.
\end{proof}

It remains to prove that every well-behaved rDFA recognizes a finite Boolean combination
of regular left ideals and regular length languages.
With a right-deterministic finite automaton $\B = (Q,\Sigma,F,\delta,q_0)$
we associate the directed graph $(Q,E)$
with edge set $E = \{ (p,a \cdot p) \mid p \in Q, a \in \Sigma\}$.
The {\em period} $g(G)$ of a directed graph $G$ is the greatest common divisor of all cycle lengths in $G$.
If $G$ is acyclic we define the period to be $\infty$.
We will apply the following lemma from Alon et al.~\cite{AlonKNS00} to the nontransient SCCs of $\B$.

\begin{lemma}[\cite{AlonKNS00}] \label{lemma-alon}
Let $G = (V, E)$ be a strongly connected directed graph with $E \neq \emptyset$ and finite period~$g$. 
Then there exist a partition $V = \bigcup_{i=0}^{g-1} V_i$ and a constant $m(G) \leq  3|V|^2$ with the following properties:
\begin{itemize}
\item For every $0 \leq i,j \leq g-1$ and for every $u \in V_i$, $v \in V_j$ the length of every directed path from $u$ to 
$v$ in $G$ is congruent to $j-i$ modulo $g$.
\item For every $0 \leq i,j \leq g-1$, for every $u \in V_i$, $v \in V_j$ and every 
integer $r \geq m(G)$, if $r$ is congruent to $j-i$ modulo $g$, then there exists a directed path from $u$ to $v$ in $G$ of length $r$.
\end{itemize}
\end{lemma}

\begin{lemma}[uniform period]
\label{lem:uni-per}
For every regular language there exists an rDFA $\B$ recognizing $L$ and a number~$g$
such that every nontransient SCC $C$ in $\B$ has period $g(C) = g$.
\end{lemma}

\begin{proof}
Let $\B = (Q,\Sigma,F,\delta,q_0)$ be any rDFA for $L$.
Let $g$ be the product of all periods $g(C)$ over all nontransient SCCs $C$ in $\B$.
In the following, we compute in the additive group $\Z_g = \{0,\ldots,g-1\}$.
We define
\[
	\B \times \Z_g = (Q \times \Z_g, \Sigma, F \times \Z_g, \delta', (q_0,0)),
\]
where for all $(p,i) \in Q \times \Z_g$ and $a \in \Sigma$ we set
\[
	\delta'(a,(p,i)) = \begin{cases}
	(\delta(a,p),i+1), & \text{if } p \text{ and } \delta(a,p) \text{ are strongly connected}, \\
	(\delta(a,p),0), & \text{otherwise.}
	\end{cases}
\]
Clearly, $\L(\B \times \Z_g) = \L(\B)$.
We show that every nontransient SCC of $\B \times \Z_g$ has period $g$.
Let $D$ be a nontransient  SCC of $\B \times \Z_g$.
Clearly, every cycle length in $D$ is a multiple of $g$.
Take any state $(q,i) \in D$ and let $C$ be the SCC of $q$ in $\B$.
Since $D$ is nontransient, there exists a cycle in $\B \times \Z_g$
containing $(q,i)$,
which induces a cycle in $\B$ containing $q$.
This implies that $C$ is nontransient.
Hence, we can apply \Cref{lemma-alon} and obtain a cycle of length $k \cdot g(C)$ in $C$
for every sufficiently large $k \in \N$ ($k \geq m(C)$ suffices).
Since $g$ is a multiple of $g(C)$, $C$ also contains a cycle of length 
$k \cdot g$ for every sufficiently large $k$. But every such cycle induces
a cycle of the same length $k \cdot g$ in $D$.
Hence, there exists $k \in \N$ such that 
$D$ contains cycles of length $k \cdot g$ and $(k+1) \cdot g$.
It follows that the period of  $D$ divides $\gcd(k \cdot g, (k+1) \cdot g) = g$.
This proves that the period of $D$ is exactly~$g$.
\end{proof}

\begin{proof}[Proof of Theorem \ref{thm:characterization}\ref{item:log}]
It remains to show the direction from left to right.
Consider a well-behaved rDFA $\B = (Q,\Sigma,F,\delta,q_0)$
for a regular language $L \subseteq \Sigma^*$.
We prove that $L$ is a finite Boolean combination of regular left ideals
and regular length languages.
By \Cref{lem:uni-per} we can ensure that all nontransient SCCs in $\B$
have the same period $g$.
This new rDFA $\B$ is also well-behaved since in fact any rDFA for $L$ must be well-behaved; this follows from 
\Cref{prop:linear-lb} and \Cref{cor:log-wb}.
Alternatively, one can verify that the transformation from \Cref{lem:uni-per} preserves the well-behavedness of $\B$.

A {\em path description} $P$ is a sequence
\begin{equation}
	\label{eq:path-desc}
	C_k,(q_k,a_{k-1},p_{k-1}),C_{k-1}, \dots ,(q_3,a_2,p_2),C_2,(q_2,a_1,p_1),C_1,q_1
\end{equation}
where $C_k, \dots, C_1$ are pairwise distinct SCCs of $\B$,
$q_1 = q_0$,
$(q_{i+1},a_i,p_i)$ is a transition in $\B$ for all $1 \le i \le k-1$,
$p_i, q_i \in C_i$ for all $1 \le i \le k-1$, and $q_k \in C_k$.
A run $\pi$ in $\B$ {\em respects} the path description $P$ if the SCC-factorization of $\pi$ is
$\pi = \pi_k \tau_{k-1} \cdots \tau_2 \pi_2 \tau_1 \pi_1$,
$\pi_i$ is a $C_i$-internal run from $q_i$ to $p_i$ for all $1 \le i \le k-1$,
$\tau_i = q_{i+1} a_i p_i$ for all $1 \le i \le k-1$,
and $\pi_k$ is a $C_k$-internal run starting in $q_k$.
Let $L_P$ be the set of words $w \in \Sigma^*$ such that the unique run of $\B$ on $w$
starting in $q_0$ respects the path description $P$.
We can write $L = \bigcup_P (L_P \cap L)$ where $P$ ranges over all path descriptions.
Notice that the number of path descriptions is finite.

Let us fix a path description $P$ as in \eqref{eq:path-desc}. We prove that $L_P \cap L$ is a finite Boolean combination of regular left ideals and regular length languages.
First, we claim that $L_P$ is a finite Boolean combination of regular left ideals.
Let $\Delta = \{ (a \cdot p,a,p) \mid p \in Q, a \in \Sigma \}$ be the set of all transition triples
and let $\Delta_P \subseteq \Delta$ be the set of transition triples contained in any of the SCCs $C_k, \dots, C_1$
together with the transition triples $(q_{i+1},a_i,p_i)$ for $1 \le i \le k-1$.
A word $w \in \Sigma^*$ then belongs to $L_P$ if and only if
$w$ belongs to the regular left ideal $\Sigma^* L_P$ and the run of $\B$ on $w$ starting in $q_0$
does not use any transitions from $\Delta \setminus \Delta_P$.
It is easy to construct for every  $\tau \in \Delta \setminus \Delta_P$
an rDFA $\mathcal{D}_\tau$ which accepts all words $w$ such that the run of $\B$ on $w$ starting in $q_0$
uses the transition $\tau$. Clearly, this language is a left ideal.
In total we have $L_P = \Sigma^* L_P \setminus \bigcup_{\tau \in \Delta \setminus \Delta_P} \L(\mathcal{D}_\tau)$,
which proves the claim.

If $C_k$ is a transient SCC then $L_P \cap L$ is either empty or $L_P$, and we are done.
For the rest of the proof we assume that $C_k$ is nontransient. Recall that all nontransient SCCs in $\B$ have period $g$.
Furthermore, $C_k$ is well-behaved since it is reachable from $q_0$ according to the path description $P$.
Let $C_k = \bigcup_{i=0}^{g-1} V_i$ be the partition from \Cref{lemma-alon}.
We claim that $F \cap C_k$ is a union of some of the~$V_i$'s.
Towards a contradiction assume that there exist states $p,q \in V_i$ where $p \in F$ and $q \notin F$.
Let $r \ge m(C)$ be any number divisible by $g$.
Then, by \Cref{lemma-alon} there exist runs from $p$ to $p$ and from $p$ to $q$,
both of length $r$. This contradicts the fact that $C_k$ is well-behaved.

Let $\pi, \pi'$ be two runs of $\B$ starting from $q_0$
which respect $P$.
We claim that $|\pi| \equiv |\pi'| \pmod{g}$ if and only if $\pi$ and $\pi'$ end in the same part $V_i$ of $C_k$.
Consider the SCC-factorizations
$\pi = \pi_k \tau_{k-1} \pi_{k-1} \cdots \tau_2 \pi_2 \tau_1 \pi_1$ and
$\pi' = \pi'_k \tau_{k-1} \pi'_{k-1} \cdots \tau_2 \pi'_2 \tau_1 \pi'_1$.
For all $1 \le i \le k-1$ the subruns $\pi_i$ and $\pi'_i$ start in $q_i$ and end in $p_i$.
If $C_i$ is nontransient then $|\pi_i| \equiv |\pi_i'| \pmod{g}$ by \Cref{lemma-alon},
and otherwise $|\pi_i| = |\pi_i'| = 0$.
This implies $|\tau_{k-1} \pi_{k-1} \cdots \tau_2 \pi_2 \tau_1 \pi_1| \equiv |\tau_{k-1} \pi'_{k-1} \cdots \tau_2 \pi'_2 \tau_1 \pi'_1| \pmod{g}$.
Also by \Cref{lemma-alon} we know that $|\pi_k| \equiv |\pi'_k| \pmod{g}$ if and only if
$\pi$ and $\pi'$ end in the same part~$V_i$.
This proves the claim.

It follows that we can write $L_P \cap L = L_P \cap K$ where
\[ K = \{ w \in \Sigma^* \mid \exists r \in R \colon |w| \equiv r \pmod{g} \} \]
for some $R \subseteq \{0, \dots, g-1\}$.
Since $K$ is a regular length language, we have proved the claim that $L$ is a Boolean combination
of regular left ideals and regular length languages.
This concludes the proof of \Cref{thm:characterization}\ref{item:log}.
\end{proof}

\section{Randomized sliding window algorithms}

\label{chap:random}

Most of the work in the context of streaming uses randomness and/or approximation
to design space- and time-efficient algorithms.
For example, the AMS-algorithm \cite{AlonMS99} approximates the number of distinct elements
in a stream with high probability in $\O(\log m)$ space where $m$ is the size of the universe.
Furthermore, it is proved that any deterministic approximation algorithm and any randomized
exact algorithm must use $\Omega(n)$ space \cite{AlonMS99}.
On the other hand, the exponential histogram algorithm by Datar et al.~\cite{DatarGIM02} for approximating
the number of $1$'s in a sliding window is a
{\em deterministic} sliding window approximation algorithm that uses $\O(\frac{1}{\epsilon} \log^2 n)$ bits.
It is proven in \cite{DatarGIM02} that $\Omega(\frac{1}{\epsilon} \log^2 n)$ bits are necessary
even for randomized (Monte Carlo or Las Vegas) sliding window algorithms.

In this section, we will study if and how randomness helps for testing membership to regular languages
over sliding windows.
The main result of this section is a space tetrachotomy in the fixed-size sliding window model,
stating that every regular language has optimal space complexity $\Theta(1)$, $\Thetainf(\log \log n)$,
$\Thetainf(\log n)$ or $\Thetainf(n)$ if the streaming algorithms are randomized with two-sided error.

\subsection{Randomized streaming algorithms}

In the following, we will introduce probabilistic automata \cite{Paz71,Rabin63}
as a model of randomized streaming algorithms. With $[0,1]$ we denote the set of all real numbers $r$ with $0 \leq r \leq 1$.
A {\em probabilistic automaton} $\P = (Q,\Sigma,\iota,\rho,F)$
consists of a nonempty countable set of states $Q$,
a finite alphabet $\Sigma$,
an initial state distribution $\iota \colon Q \to [0,1]$,
a transition probability function $\rho \colon Q \times \Sigma \times Q \to [0,1]$,
and a set of final states $F \subseteq Q$,
such that
\begin{enumerate}[label=(\roman{*}), ref=(\roman{*})]
\item $\sum_{q \in Q} \iota(q) = 1$,
\item $\sum_{q \in Q} \rho(p,a,q) = 1$ for all $p \in Q$, $a \in \Sigma$.
\end{enumerate}
If $Q$ is infinite then this means of course that the above infinite sums converge to 1. 
This implies that these sums are absolutely convergent 
(all $\iota(q)$ and $\rho(p,a,q)$ are non-negative) and therefore the order of summation
is not relevant.

If $\iota$ and $\rho$ map into $\{0,1\}$, then $\P$ can be viewed as a deterministic automaton.
We will refer to probabilistic automata also as \emph{randomized streaming algorithms}. 

For a word $w \in \Sigma^*$ and a state $q$ we define the probability $\P(w,q)$ that $\P$ after reading the word $w$ arrives
in state $q$ inductively over the length of $w$ as follows, where $q \in Q$, $v \in \Sigma^*$ and $a \in \Sigma$:
\begin{itemize}
\item $\P(\varepsilon,q) = \iota(q)$ and
\item $\P(va,q) = \sum_{p \in Q} \P(v,p) \cdot \rho(p,a,q)$.
\end{itemize}
Then the probability that $\P$ accepts the word $w$ is 
\[ \Pr[\P \text{ accepts } w] = \sum_{q \in F} \P(w,q).\] 
and the probability
that $\P$ rejects the word $w$ is 
\[ \Pr[\P \text{ rejects } w] = \sum_{q \in Q \setminus F} \P(w,q).\]
The {\em space} of $\P$ (or {\em number of bits used by $\P$}) is given by
$s(\P) = \log |Q| \in \mathbb{R}_{\ge 0} \cup \{ \infty \}$.
We say that $\P$ is a randomized streaming algorithm for $L \subseteq \Sigma^*$
with error probability $0 \le \lambda \le 1$ if
\begin{itemize}
\item $\Pr[\P \text{ accepts } w]  \ge 1 - \lambda$ for all $w \in L$,
\item $\Pr[\P \text{ rejects } w]  \ge 1 - \lambda$  for all $w \notin L$.
\end{itemize}
The error probability $\lambda$ is also called a {\em two-sided error}.
If we omit $\lambda$ we choose $\lambda = 1/3$.

For a randomized streaming algorithm $\P = (Q,\Sigma,\iota,\rho,F)$ and a number $k \ge 1$
let $\P^{(k)}$ be the randomized streaming algorithm which simulates $k$ instances of $\P$
in parallel with independent random choices and outputs the majority vote.
Formally the states of $\P^{(k)}$ are multisets of size $k$ over $Q$ (using multisets instead of 
ordered tuples will yield a better space bound in \Cref{sec-bernoulli}).
Therefore, $s(\P^{(k)}) \le k \cdot s(\P)$.

\begin{lemma}[probability amplification]
	\label{lem:amplification}
	For all $0 < \lambda' < \lambda < 1/2$ there exists a number
	$k = \O(\log \left( \frac{1}{\lambda'} \right) \cdot \left(\frac{1}{2}-\lambda\right)^{-2})$
	such that
	for all randomized streaming algorithms $\P$ and all $w \in \Sigma^*$ we have:
	\begin{enumerate}[label=(\roman{*}), ref=(\roman{*})]
	\item If $\Pr[\P \text{ accepts } w]  \ge 1-\lambda$ then
	$\Pr[\P^{(k)} \text{ accepts } w] \ge 1-\lambda'$.
	\item If $\Pr[\P \text{ rejects } w]  \ge 1-\lambda$ then
	$\Pr[\P^{(k)} \text{ rejects } w] \le 1-\lambda'$.
	\end{enumerate}
\end{lemma}

\begin{proof}
	We will choose $k$ later.
	Let $X_1, \dots, X_k$ be independent Bernoulli random variables
	with $\Pr[X_i = 0] = \lambda$ and $\Pr[X_i = 1] = 1-\lambda$.
	Let $X = \sum_{i=1}^k X_i$ with expectation $\mu = k (1-\lambda)$.
	Suppose that $\P$ accepts $w$ with probability $\ge 1-\lambda$,
	i.e.,~$\P$ rejects $w$ with probability at most $\lambda$.
	Then, $\P^{(k)}$ rejects $w$ with probability at most $\Pr[X \le k/2]$.
	By the Chernoff bound \cite[Theorem~4.5]{MitzenmacherU17},
	for any $0 < \delta < 1$ we have
	\begin{equation}
		\label{eq:chernoff}
		\Pr[X \le (1-\delta)\mu] \le \exp\left(-\frac{\mu\delta^2}{2}\right).
	\end{equation}
	By choosing $\delta = 1 - \frac{1}{2(1-\lambda)}$ we get $(1-\delta)\mu = k/2$.
	Then, \eqref{eq:chernoff} gives the following estimate:
	\begin{align*}
		 \Pr[X \le k/2] 
		&\le \exp\left(-\frac{k(1-\lambda)(1 - \frac{1}{2(1-\lambda)})^2}{2}\right) \\
		&= \exp\left(-\frac{k}{2}\cdot \frac{(\frac{1}{2}-\lambda)^2}{1-\lambda} \right) \\
		&\le \exp\left(-\frac{k}{2}\cdot \left(\frac{1}{2}-\lambda\right)^2 \right).
	\end{align*}
	By choosing
	\[
		k \ge 2 \cdot \ln \left( \frac{1}{\lambda'} \right) \cdot \left(\frac{1}{2}-\lambda\right)^{-2}
	\]
	we can bound the probability that $\P^{(k)}$ rejects $w$ by $\lambda'$.
	Statement (ii) can be shown analogously.
\end{proof}

\subsection{Space tetrachotomy}
A {\em randomized sliding window algorithm} for a language $L$ and window size $n$
is a randomized streaming algorithm for $\SW_n(L)$.
The {\em randomized space complexity} $\F^\r_L(n)$ of $L$ in the fixed-size sliding window model
is the minimal space complexity $s(\P_n)$ of a randomized sliding window algorithm
$\P_n$ for $L$ and window size $n$. For this to be well-defined it is important that we require the error
probability to be at most $1/3$.

Before we investigate randomized streaming algorithms in more detail, let us first comment on the fact that in our 
definition of randomized SW-algorithms we allow arbitrary (even irrational) probabilities in the state transitions.
On the other hand, in all correctness proofs for our randomized SW-algorithms we only need the fact that
the probabilities are from a certain  interval $I \subseteq [0,1]$. Therefore, if $d$ is the length of the interval $I$, we can always choose a probability $p \in I$ with $\O(\log_2(1/d))$ bits such that the algorithm still achieves an error probability of at most $1/3$.
However, the size of the interval $I$ may depend on the window size $n$; more precisely it may shrink when $n$ grows.
In particular, the number of bits needed to write down the probabilities used in $\P_n$ (the algorithm for window size $n$)
may grow with $n$. One might argue that these bits should also enter the definition of the space used by the algorithm.
The reason why we do not take these bits into account is the same as why we do not consider the space  
 for internal calculations; see \Cref{ref:internal comp}.
 Assume for instance that we need to implement a randomized branching with probabilities $p$ and $1-p$ and 
 let $m$ be the number of bits of $p$. Let us moreover assume that we have a randomized machine model that
 apart from deterministic commands can only toss fair coins.
  It is not difficult to see that one can implement a biased coin with probabilities $p$ and $1-p$ using 
  $m$ many fair coins. For this, one also needs some additional
 control structure for which $\O(\log m)$ bits are needed (basically to store the program counter). But this is internal space that we do not take into account in our definition of space (as justified in \Cref{ref:internal comp}).\footnote{The reader may
 view this control structure as additional $\varepsilon$-transitions in a probabilistic automaton that are taken with probability
 $1/2$.}

Clearly we have $\F^\r_L(n) \le \F_L(n)$.
Furthermore, we prove that randomness can reduce
the space complexity at most exponentially:

\begin{lemma}
\label{lem:rabin}
For any language $L$ we have $\F_L(n) = 2^{\O(F^\r_L(n))}$.
\end{lemma}

\begin{proof}
	Rabin proved that any probabilistic finite automaton with a so-called isolated cut-point
	can be made deterministic with an exponential size increase \cite{Rabin63}.
	Let $\P = (Q,\Sigma,\iota,\rho,F)$ be a probabilistic finite automaton with $m$ states.
	Suppose that $\lambda \in [0,1]$ is an {\em isolated cut-point} with radius $\delta > 0$,
	i.e.,~$\abs{\Pr[\P \text{ accepts } w] - \lambda} \ge \delta$ for all $w \in \Sigma^*$.
	Then, $L = \{ w \in \Sigma^* \mid \Pr[\P \text{ accepts } w] \ge \lambda \}$
	is recognized by a DFA $\A$ with at most $(1 + m/\delta)^{m-1} = 2^{\O(m \log m)}$ states
	\cite[Theorem~3]{Rabin63}.

	Now, let $\P_n$ be a minimal probabilistic finite automaton for $\SW_n(L)$ with $m$ states and error probability $\leq 1/3$.
	Since $\P_n$ has $1/2$ as an isolated cutpoint with radius $1/2-1/3 = 1/6$,
	there exists an equivalent DFA $\Q_n$ with $|\Q_n| \le 2^{\O(m \log m)}$ states.
	The statement follows from
	$\F_L(n) \le \log |\Q_n| = \O(m \log m) = \O(2^{\F^\r_L(n)} \cdot \F^\r_L(n))$,
	which is bounded by $2^{\O(\F^\r_L(n))}$.
\end{proof}

In this section, we will prove \Cref{thm:r-char},
which is a tetrachotomy for the randomized space complexity of regular languages
in the fixed-size sliding window model.
Let us rephrase \Cref{thm:r-char} and split it into three upper bounds and three lower bounds.

\begin{theorem}\label{thm:tetrachotomy}
	Let $L \subseteq \Sigma^*$ be a regular language.
	\begin{enumerate}[label=(\arabic{*}), ref=(\arabic{*})]
	\item \label{point-O(1)} If $L \in \langle \ST, \Len \rangle$
	then $\F^\r_L(n) = \O(1)$.
	\item \label{lower-loglog} If $L \notin \langle \ST, \Len \rangle$
	then $\F^\r_L(n) = \Omegainf(\log \log n)$.
	\item \label{upper-loglog} If $L \in \langle \ST, \SF, \Len \rangle$
	then $\F^\r_L(n) = \O(\log \log n)$.
	\item \label{lower-log} If $L \notin \langle \ST, \SF, \Len \rangle$
	then $\F^\r_L(n) = \Omegainf(\log n)$.
	\item \label{point-O(log)} If $L \in \langle \LI, \Len \rangle$
	then $\F^\r_L(n) = \O(\log n)$.
	\item \label{lower-lin} If $L \notin \langle \LI, \Len \rangle$
	then $\F^\r_L(n) = \Omegainf(n)$.
	\end{enumerate}
\end{theorem}

Points \ref{point-O(1)} and \ref{point-O(log)} already hold in the deterministic setting,
see \Cref{thm:characterization}.
In the next sections we prove points \ref{lower-loglog}, \ref{upper-loglog}, \ref{lower-log},
and \ref{lower-lin}.

We first transfer \Cref{lem:boolean} to the fixed-size model in the randomized setting:
\begin{lemma}\label{thm:randombool}
        Let $\Sigma$ be a finite alphabet.
	For any function $s(n)$, the class $\{ L \subseteq \Sigma^* \mid \F^\r_L(n) = \O(s(n)) \}$
	forms a Boolean algebra.
\end{lemma}

\begin{proof}
       Let $L \subseteq \Sigma^*$ be a language and $n \in \N$ a window size.
	If $\P_n$ is a randomized SW-algorithm for $L$ and window size $n$ then $\overline{\P}_n$ is a randomized SW-algorithm for $\Sigma^* \setminus L$ and window size $n$,
	where $\overline{\P}_n$ simulates $\P_n$ and returns the negated output.
	Let $\P_n$ and $\Q_n$ be randomized SW-algorithms for $K$ and $L$, respectively, and window size $n$.
	By \Cref{lem:amplification} we can reduce their error probabilities to 1/6 with a constant space increase.
	Then, the algorithm which simulates $\P_n$ and $\Q_n$ in parallel and returns the disjunction of the outputs
	is a randomized SW-algorithm for $K \cup L$ and window size $n$.
	Its error probability is at most 1/3 by the union bound.
\end{proof}

\subsection{The Bernoulli counter} \label{sec-bernoulli}

The crucial algorithmic tool for the proof of \cref{thm:tetrachotomy}\ref{upper-loglog} is 
a simple probabilistic counter. It is inspired by the approximate counter by Morris \cite{Flajolet85,Morris78a},
which uses $\O(\log \log n)$ bits.
For our purposes, it suffices to detect whether the counter has exceeded a certain threshold,
which can be accomplished using only $\O(1)$ bits.

Formally, a {\em probabilistic counter} is a probabilistic automaton 
\[  \BZ = (C,\{\mathtt{inc}\},\iota,\rho,F) \]
over the unary alphabet $\{\mathtt{inc}\}$. 
States in $F$ are called {\em high} and states in $C \setminus F$ are called {\em low}.
We make the restriction that there is a low state $c_0 \in C$ such that
$\iota(c_0) = 1$ (and hence $\iota(c)=0$ for all $c \in C \setminus \{c_0\}$); thus
$\BZ$ has a unique initial state $c_0$ (which must be low) and we write $\BZ = (C,\{\mathtt{inc}\},c_0,\rho,F)$.
This restriction is not really important (and can in fact be achieved for every 
probabilistic automaton by adding a new state), but it will simplify our constructions.

In the following we write $\BZ(k,c)$ for $\BZ(\mathtt{inc}^k, c)$ ($k \geq 0$, $c \in C$), which is the probability
that $\BZ$ arrives in state $c$ after $k$ increments. Moreover, $\BZ^{\mathsf{hi}}(k)$ is the probability
that $\BZ$ is in a high state after $k$ increments (this is the same as $\Pr[\BZ \text{ accepts } \mathtt{inc}^k]$).  
Given numbers $0 \le \ell < h$ we say that $\BZ$ is an
{\em $(h,\ell)$-counter with error probability $\lambda < \frac{1}{2}$}
if for all $k \in \N$ we have:
\begin{itemize}
	\item If $k \le \ell$, then $\BZ^{\mathsf{hi}}(k) \le \lambda$.
	\item If $k \ge h$, then $\BZ^{\mathsf{hi}}(k)  \ge 1-\lambda$.
\end{itemize}
In other words, a probabilistic counter can distinguish with high probability 
between values below $\ell$ and values above $h$
but does not give any guarantees for counter values strictly between $\ell$ and $h$.
A {\em Bernoulli counter} is a probabilistic counter 
$\BZ_p$ that is parameterized by a probability $0 < p < 1$ and that
has the state set $\{0,1\}$, where 0 is a low state and 1 is a high state.
Initially the counter is in the state $x = 0$.
On every increment we set $x = 1$ with probability $p$;
the state remains unchanged with probability $1-p$.
We have
\[
 \BZ_p^{\mathsf{hi}}(k) = \BZ_p(k,1) = 1-(1-p)^k.
\]
Let us first show the following claim.

\begin{lemma} \label{claim-bernoulli}
	For all $h>0$, $0 < \xi < 1$ and $0 <\ell \le (1-\xi) h$
	there exists $0 < p < 1$ such that $\BZ_p$ is an $(h,\ell)$-counter
	with error probability $1/2 - \xi/8$. 
\end{lemma}

\begin{proof}
We need to choose $0<p<1$ such that
\begin{enumerate}[label=(\roman*)]
\item $1-(1-p)^{(1-\xi) h} \le 1/2-\xi/8$, or equivalently, $1/2+\xi/8 \le (1-p)^{(1-\xi)h}$, and
\item $(1-p)^{h} \le 1/2-\xi/8$, or  equivalently, $(1-p)^{(1-\xi)h} \le (1/2-\xi/8)^{1-\xi}$.
\end{enumerate}
It suffices to show
\begin{equation}
\label{eq:pow}
\frac{1}{2}+\frac{\xi}{8} \le \left( \frac{1}{2}-\frac{\xi}{8} \right)^{1-\xi}.
\end{equation}
Then one can take for instance $p = 1-(1/2-\xi/8)^{1/h} \in (0,1)$, which satisfies (ii). Moreover, (i) is satisfied due to \eqref{eq:pow}.

Taking logarithms shows that \eqref{eq:pow} is equivalent to 
\[ 
\ln(4+\xi) - \ln 8 \le (1-\xi) \cdot (\ln(4-\xi) - \ln 8),\] 
which can be rearranged to $\ln(4+\xi) \le \ln(4-\xi) + \xi (\ln 8 - \ln(4-\xi))$. Since $\ln 8 - \ln(4-\xi) \ge \ln 8 - \ln 4 = \ln 2$, it suffices to prove
\begin{equation}
	\label{eq:log}
	\ln(4+\xi) \le \ln(4-\xi) + \xi \ln 2.
\end{equation}
One can verify $3\ln 2 \approx 2.0794 \ge 2$. We have
\begin{eqnarray*}
	4+\xi &\le& 4 + (3\ln 2 - 1) \xi \\
	&=& 4 + (4\ln 2 - 1) \xi - \xi \ln 2 \\
	&\le& 4 + (4\ln 2 - 1) \xi - \xi^2 \ln 2 \\
	&=& (4-\xi)(\xi \ln 2 + 1).
\end{eqnarray*}
By taking logarithms and plugging in $\ln x \le x-1$ for all $x > 0$, we obtain
\[
	\ln(4+\xi) \le \ln(4-\xi) + \ln(\xi \ln 2 + 1) \le \ln(4-\xi) + \xi \ln 2.
\]
This proves \eqref{eq:log} and hence \eqref{eq:pow}, and thus the lemma.
\end{proof}

\begin{proposition}
\label{prop:bernoulli}
For all $h>0$, $0 < \xi < 1$, $0 < \ell \le (1-\xi) h$ and $0 < \lambda' < 1/2$ 
there exists an $(h,\ell)$-counter $\BZ$ with error probability $\lambda'$
which uses $\O(\log \log (1/\lambda') + \log (1/\xi))$ bits.
\end{proposition}

\begin{proof}
Take the $(h,\ell)$-counter $\BZ_p$ from \Cref{claim-bernoulli}, whose error
probability is $\lambda := 1/2 - \xi/8$. To $\BZ_p$  we apply 
\Cref{lem:amplification},
which states that we need to run $k = \O(\log (\frac{1}{\lambda'}) \cdot \frac{1}{\xi^2})$ independent copies
to reduce the error probability to $\lambda'$.
The states of $\BZ^{(k)}_p$ are multisets over $\{0,1\}$ of size $k$,
which can be encoded with $\O(\log k) = \O(\log \log \frac{1}{\lambda'} + \log \frac{1}{\xi})$ bits
by specifying the number of 1-bits in the multiset. Note that the unique initial state of $\BZ^{(k)}_p$
is the multiset with $k$ occurrences of $0$.
\end{proof}

\subsection{Suffix-free languages} \label{sec-random-suffix-free}

In this section, we prove \Cref{thm:tetrachotomy}\ref{upper-loglog}.
Since languages in $\ST \cup \Len$ have constant space (deterministic) SW-algorithms
it suffices by  \Cref{thm:randombool} to show:
\begin{theorem}
	\label{thm:sf-loglog}
	If $L$ is regular and suffix-free then $\F^\r_L(n) = \O(\log \log n)$.
\end{theorem}

Fix a regular suffix-free language $L \subseteq \Sigma^*$ and
let $\B = (Q,\Sigma,F,\delta,q_0)$ be an rDFA for $L$ where all states are reachable.
Excluding the trivial case $L = \emptyset$, we assume that $\B$ contains at least one final state.
Furthermore, since $L$ is suffix-free, any run in $\B$ contains at most one final state.
Therefore, we can assume that $F$ contains exactly one final state $q_F$, and all outgoing transitions
from $q_F$ lead to a sink state.
For a stream $w \in \Sigma^*$ define the function $\ell_w \colon Q \to \N \cup \{ \infty \}$ by
\begin{equation}
 \ell_w(q) = \inf \{  k \in \N \mid \last_k(w) \cdot q = q_F \}, \label{def-l_q(w)}
\end{equation}
where we set $\inf(\emptyset) = \infty$ (note that $\{  k \in \N \mid \last_k(w) \cdot q = q_F \}$ is either empty or a singleton set).
Notice that $\last_n(w) \in L$ if and only if $\ell_{w}(q_0) = n$. Also, it holds $\ell_w(q_F) = 0$ for every $w \in \Sigma^*$.
A {\em deterministic} streaming algorithm can maintain the function $\ell_w$ where $w \in \Sigma^*$ is
the stream prefix read so far:
If a symbol $a \in \Sigma$ is read, we can determine
\begin{equation}
	\label{eq:maintain-ellw}
	\ell_{wa}(q) = \begin{cases}
	0, & \text{if } q = q_F, \\
	1 + \ell_w(a \cdot q), & \text{otherwise,}
	\end{cases}
\end{equation}
where $1 + \infty = \infty$.
Storing $\ell_w(q)$ may require up to $\log |w|$ bits. Therefore, if an SW-algorithm for window size $n$
wants to store all $\ell_w(q)$ for $q \in Q$ ($w$ is the input stream and not just the sliding window), then the space is not bounded in the window size. 
The solution is to use probabilistic counters with suitable threshold values $\ell$ and $h$.

Let $n \in \N$ be a window size.
The randomized sliding window algorithm $\P_n$ for $L$ consists of two parts:
a constant-space threshold algorithm $\mathcal{T}_n$,
which rejects with high probability whenever $\ell_{w}(q_0) \geq 2n$,
and a modulo counting algorithm $\M_n$, which maintains $\ell_w$ modulo a random prime number
with $\O(\log \log n)$ bits.

\begin{lemma}[threshold counting] \label{lemma-threshold}
	There exists a randomized streaming algorithm $\mathcal{T}_n$ with $\O(1)$ bits
	such that for all $w \in \Sigma^*$ we have:
	\begin{itemize}
		\item $\Pr[\mathcal{T}_n \text{ accepts } w] \ge 2/3$, if $\ell_w(q_0) \le n$, and
		\item $\Pr[\mathcal{T}_n \text{ rejects } w] \ge 2/3$, if $\ell_w(q_0) \ge 2n$.
	\end{itemize}
\end{lemma}

\begin{proof}
By \Cref{prop:bernoulli} there is a 
$(2n,n)$-counter $\BZ = (C,\{\mathtt{inc}\},c_0,\rho,F)$ with error probability 1/3 which uses $\O(1)$ space.
Let  $c_{\infty} \in F$ be an arbitrary high state.
The algorithm $\mathcal{T}_n$ maintains for every $q \in Q$ an instance $\BZ_q$ of the $(2n,n)$-counter $\BZ$.
The input alphabet of $\BZ_q$ is $\Sigma$ (instead of $\{\mathtt{inc}\}$) and
the probability $\BZ_q(w,c)$ of reaching $c \in C$ after reading the word $w \in \Sigma^*$ will satisfy
\begin{equation} \label{eq-c(q)}
\BZ_q(w,c) = \BZ(\ell_w(q),c),
\end{equation}
where we set $\BZ(\infty,c_\infty) = 1$ (and $\BZ(\infty,c)=0$ for all states $c \neq c_\infty$)
We initialize $\BZ_q$ in order to get \eqref{eq-c(q)} for $w = \eps$.
To this end, we distinguish whether state $q$ has finite or infinite value $\ell_\eps(q)$.
Notice that $\ell_\eps(q)$ is finite if and only if the final state $q_F$ can be reached from state $q$
by only reading the padding symbol $\Box$.
If $\ell_\eps(q) < \infty$, then we initialize $\BZ_q$ in its initial state $c_0$ and then execute
$\ell_\eps(q)$ increments.
If $\ell_\eps(q) = \infty$, we set $\BZ_q$ to state $c_\infty$ (with probability one).
Given an input symbol $a \in \Sigma$, we compute the new states of the counters
$\BZ_q$ as follows: Assume that $c_q$ is the current state of $\BZ_q$.
First we set $\BZ_{q_F}$ to the initial state $c_0$.
This ensures \eqref{eq-c(q)} for $q_F$ since $\ell_{wa}(q_F) = 0$ and $\BZ(0,c_0) = 1$.
For $q \in Q \setminus \{q_F\}$ we set the new state of $\BZ_q$ with probability $\rho(c_{a \cdot q}, \mathtt{inc}, c)$
to $c$. This ensures again~\eqref{eq-c(q)}:
\begin{align*}
	\BZ_q(wa,c) &= \sum_{c' \in C} \BZ_{a \cdot q}(w,c') \cdot \rho(c',\mathtt{inc},c) \\
	&= \sum_{c' \in C} \BZ(\ell_w(a \cdot q),c') \cdot \rho(c',\mathtt{inc},c) \\
	& = \BZ(\ell_w(a \cdot q)+1,c) = \BZ(\ell_{wa}(q),c).
\end{align*}
The algorithm $\mathcal{T}_n$ accepts the word $w$ if and only if $\BZ_{q_0}$ is in a low state after reading $w$.
Note that this happens with probability $1-\BZ_{q_0}^{\mathsf{hi}}(w) = 1- \BZ^{\mathsf{hi}}(\ell_w(q_0))$ ($\BZ^{\mathsf{hi}}(k)$ is the probability
that $\BZ$ is in a high state after $k$ increments). 
Correctness follows from the fact that $\BZ$ is a $(2n,n)$-counter with error probability 1/3: 
\[ 
\Pr[\mathcal{T}_n \text{ accepts } w] = 1-\BZ^{\mathsf{hi}}(\ell_w(q_0)) 
\begin{cases}
\ge 2/3 \text{ if } \ell_w(q_0) \le n, \\
\le 1/3 \text{ if }  \ell_w(q_0) \ge 2n.
\end{cases}
\]
This proves the lemma.
\end{proof}
Also note that the randomized SW-algorithm from the previous proof 
uses several probabilistic counters $\BZ_q$ (one for each state $q \in Q$) and they all have the same parameters $\ell = n$ and $h = 2n$.
For each new input symbol, a subset of these counters have to be incremented. These increments are not needed to be independent. 
Hence, in each step, only the random bits for incrementing a single $(2n,n)$-counter are needed. These random bits can be used for all $\BZ_q$ that have to be incremented.

We now come to the modulo counting algorithm, for which we use the following simple fact on prime numbers.

\begin{lemma} \label{lemma-primes}
There is a constant $c$ such that for every large enough $m \in \mathbb{N}$ and all $0 \leq a,b \le m$ with $a \neq b$ the following holds: If the prime number $p$ is picked uniformly
at random among all prime numbers that are no greater than $c \log m \log \log m$, then $\Pr[a \equiv b \pmod p] \leq 1/3$.
\end{lemma}

\begin{proof}
Let $p_i$ be the $i$-th prime number.
It is known that 
$p_i < i \cdot  (\ln i + \ln\ln i)$ for $i \geq 6$ \cite[3.13]{RosserS62}. 
Fix an $m$ and
let $k$ be the first natural number such that $\prod_{i=1}^k p_i \ge m$. Since  
$\prod_{i=1}^k p_i \geq 2^k$, we have $k \le \log m$ and hence $p_{3k} \leq
3 \log m \cdot (\ln (3 \log m) + \ln\ln(3 \log m)) \leq c \log m \log \log m$ for some constant $c$
and all large enough $m$.

Since $-m \le a - b \le m$ and any product of at least $k+1$ pairwise distinct primes
exceeds $m$, the integer $a - b \neq 0$ has at most $k$ prime factors. Hence, for a randomly chosen prime
$p \in \{p_1, \dots, p_{3k}\}$ we have $\Pr[a \equiv b \pmod p] \leq 1/3$.  
\end{proof}

\begin{lemma}[modulo counting]
	\label{lem:modulocounting}
	There exists a randomized streaming algorithm $\M_n$ with $\O(\log \log n)$ bits
	such that for all $w \in \Sigma^*$ we have:
	\begin{itemize}
		\item $\Pr[\M_n \text{ accepts } w] = 1$, if $\ell_w(q_0) = n$, and
		\item $\Pr[\M_n \text{ rejects } w] \ge 2/3$, if $\ell_w(q_0) < 2n$ and $\ell_w(q_0) \neq n$.
	\end{itemize}
\end{lemma}

\begin{proof}
Let $c$ be the constant from \Cref{lemma-primes} which is applied with $m = 2n$.
The algorithm $\M_n$ initially picks a random prime $p   \leq c \log(2n) \log \log(2n)$
which is stored throughout the run using $\O(\log \log n)$ bits.
Then, after reading  $w \in \Sigma^*$, $\M_n$ stores for every $q \in Q$ a bit telling whether
$\ell_w(q) < \infty$ and, if the latter holds,
the value $\ell_w(q) \bmod p$ using in total $\O(|Q| \cdot \log \log n)$ bits.
These numbers can be maintained according to \eqref{eq:maintain-ellw}.
The algorithm accepts if and only if $\ell_w(q_0) \equiv n \pmod p$.

If $\ell_w(q_0) = n$ then the algorithm always accepts.
Now, assume $\ell_w(q_0) < 2n$ and $\ell_w(q_0) \neq n$.
Then \Cref{lemma-primes}  with $a = \ell_w(q_0)$ and $b = n$ yields
$\Pr[\ell_w(q_0) \equiv n \pmod p] \leq 1/3$.
Therefore, $\M_n$ rejects with probability at least~$2/3$.
\end{proof}
It is worth mentioning that in the above modulo counting algorithm
the errors after reading different prefixes of an input stream $w$ are not independent.
If for instance $w = uu$ with $|u|$ the window size, then the algorithm will make an
 error after reading $u$ if and only if it makes an error after reading $uu$.
 This is of course due to the fact that the only random
choice is made at the very beginning. After this random choice, the algorithm proceeds deterministically.

By combining the algorithms from \Cref{lemma-threshold} and  \Cref{lem:modulocounting} we can prove \Cref{thm:sf-loglog}.
The algorithm $\P_n$ is the conjunction of the threshold algorithm $\mathcal{T}_n$
and the modulo counting algorithm~$\M_n$.
Recall that $\last_n(w) \in L$ if and only if $\ell_{w}(q_0) = n$.
If $\ell_{w}(q_0) = n$ then $\mathcal{T}_n$ accepts with probability at least $2/3$ and
$\M_n$ accepts with probability 1; hence $\P_n$ accepts with probability at least $2/3$.
If $\ell_{w}(q_0) \neq n$ then $\M_n$ or $\mathcal{T}_n$ rejects with probability at least $2/3$.
Hence, $\P_n$ rejects with probability at least~$2/3$.

\subsection{Lower bounds} \label{sec-CC}

In this section, we prove the lower bounds from \Cref{thm:tetrachotomy}.
Point \ref{lower-loglog} from \Cref{thm:tetrachotomy} follows easily from
the relation $\F_L(n) = 2^{\O(\F^\r_L(n))}$ (\Cref{lem:rabin}).
Since every language $L \in \Reg \setminus \langle \ST, \Len \rangle$
satisfies $\F_L(n) = \Omegainf(\log n)$
(\Cref{thm:trichotomy} and \ref{thm:characterization}),
it also satisfies $\F^\r_L(n) = \Omegainf(\log \log n)$.

For \ref{lower-log} and \ref{lower-lin}
we apply known lower bounds from communication complexity
by deriving a randomized communication protocol from 
a randomized SW-algorithm.
This is in fact a standard technique for obtaining lower bounds for streaming algorithms; see e.g.~\cite{Roughgarden16}.

We present the necessary background from communication complexity; see \cite{KushilevitzN97}
for a detailed introduction.  
We only need the one-way setting where Alice sends a single message to Bob.
Consider a function $f \colon X \times Y \to \{0,1\}$ for some finite sets $X$ and $Y$.
A {\em randomized one-way (communication) protocol} $P=(a,b)$ consists of functions
$a \colon X\times R_a \to \{0,1\}^*$ and $b \colon \{0,1\}^* \times Y \times R_b \to\{0,1\}$, where
$R_a$ and $R_b$ are finite sets of random choices of Alice and Bob, respectively.
The {\em cost} of $P$ is the maximum number of bits transmitted by Alice, i.e.
\[
	\mathrm{cost}(P) = \max_{x \in X, r_a \in R_a} |a(x,r_a)|. 
\]
Moreover, probability distributions are given on $R_a$ and $R_b$.
Alice computes from her input $x \in X$ and a random choice $r_a \in R_a$
the value $a(x,r_a)$ and sends it to Bob.
Using this value, his input $y \in Y$ and a random choice $r_b \in R_b$
he outputs $b(a(x,r_a),y,r_b)$. 
The random choices $r_a \in R_a, r_b \in R_b$ are chosen independently of each other.
The protocol $P$ {\em computes} $f$ if 
for all $(x,y) \in X \times Y$ we have
\begin{equation} \label{eq-rcc}
\Pr_{r_a \in R_a, r_b \in R_b}[P(x,y) \neq f(x,y)] \le \frac{1}{3},
\end{equation}
where $P(x,y)$ is the random variable $b(a(x,r_a),y,r_b)$.
The {\em randomized one-way communication complexity} of $f$ is the minimal cost
among all one-way randomized protocols that compute $f$ (with an arbitrary number of random bits).
The choice of the constant $1/3$ in \eqref{eq-rcc} is arbitrary in the sense that changing the constant
to any $\lambda < 1/2$ only changes the communication complexity
by a constant (depending on $\lambda$),
see \cite[p.~30]{KushilevitzN97}.
We will use established lower bounds on the randomized one-way communication complexity of some functions.

\begin{theorem}[{\cite[Theorem~3.7 and 3.8]{KremerNR99}}] \label{thm:coco}
Let $n \in \N$.
\begin{itemize}
\item The {\em index function}
\[ 
	\mathrm{IDX}_n \colon \{0,1\}^n\times \{1,\dots,n\} \to \{0,1\} 
\]
with $\mathrm{IDX}_n(a_1 \cdots a_n,i) = a_i$
has randomized one-way communication complexity $\Theta(n)$.
\item The {\em greater-than function}
\begin{align*}
	\mathrm{GT}_n \colon \{1,\dots,n\}\times \{1,\dots,n\} \to \{0,1\}
\end{align*}
with $\mathrm{GT}_n(i,j) = 1$ if and only if $i > j$
has randomized one-way communication complexity $\Theta(\log n)$.
\end{itemize}
\end{theorem}
The upper bounds from these statements also hold for the deterministic one-way communication complexity
as witnessed by the trivial deterministic protocols.
We also define the {\em equality function} 
\[ 
\mathrm{EQ}_n \colon \{1, \dots, n\} \times \{1,\dots,n\} \to \{0,1\}
\]
by $\mathrm{EQ}_n(i,j) = 1$ if and only if $i = j$.
Its randomized one-way communication complexity is $\Theta(\log \log n)$
whereas its deterministic one-way communication complexity is $\Theta(\log n)$ \cite{KushilevitzN97}.

We start with the proof of \ref{lower-lin} from \Cref{thm:tetrachotomy}, which extends our linear
space lower bound from the deterministic setting to the randomized setting.

\begin{proposition}
\label{thm:rand-lin}
If $L \in \Reg \setminus \langle \LI, \Len \rangle$
then $\F^\r_L(n) = \Omegainf(n)$.
\end{proposition}

\begin{proof}
By \Cref{thm:characterization}\ref{item:log} any rDFA for $L$ is not well-behaved 
and by \Cref{lem:linear-lb} there exist words $u = u_1u_2, v = v_1v_2, z \in \Sigma^*$
such that $|u_1| = |v_1|$, $|u_2| = |v_2|$ and $L$ separates $u_2\{u,v\}^*z$ and $v_2\{u,v\}^*z$.
Let $\eta \colon \{0,1\}^* \to \{u,v\}^*$ be the injective homomorphism defined by
$\eta(0) = u$ and $\eta(1) = v$.

Now, consider a randomized SW-algorithm $\P_n$ for $L$ and window size
$n = |u_2| + |u| \cdot m + |z|$ for some $m \ge 1$.
We describe a randomized one-way communication protocol for $\mathrm{IDX}_m$.

Let $\alpha = \alpha_1 \cdots \alpha_m \in\{0,1\}^m$ be Alice's input and $i\in\{1,\dots,m\}$ be Bob's input.
Alice reads $\eta(\alpha)$ into $\P_n$ (using here random choices in order to select the outgoing
transitions in $\P_n$) and sends the memory state using $\O(s(\P_n))$ bits to Bob.
Continuing from the received state, Bob reads $u^i z$ into $\P_n$.
Then, the active window is
\[
	\last_n(\eta(\alpha) u^i z) = s \, \eta(\alpha_{i+1} \cdots \alpha_m) u^i z \in \{u_2,v_2\}\{u,v\}^*z
\]
where $s = u_2$ if $\alpha_i = 0$ and $s = v_2$ if $\alpha_i = 1$.
Hence, from the output of $\P_n$ Bob can determine whether $\alpha_i = 1$.
The cost of the protocol is bounded by $\O(s(\P_n))$
and must be at least $\Omega(m) = \Omega(n)$ by \Cref{thm:coco}.
We conclude that $s(\P_n) = \Omega(n)$ for infinitely many $n$
and therefore $\F^\r_L(n) = \Omegainf(n)$.
\end{proof}

Next, we prove point \ref{lower-log} from \Cref{thm:tetrachotomy}.
For that, we need the following automaton property, where 
$\B = (Q,\Sigma,F,\delta,q_0)$ is an rDFA. 

A pair $(p,q)\in Q\times Q$ of states is called {\em synchronized}
if there exist words $x,y,z\in\Sigma^*$ with $|x|=|y|=|z| \geq 1$ such that
\[ 
q \xleftarrow{x} q  \xleftarrow{y} p \xleftarrow{z} p .
\]
A pair $(p,q)\in Q\times Q$ is called {\em reachable} if $p$ and $q$ are reachable from $q_0$
and $(p,q)$ is called {\em $F$-consistent} if either $\{p,q\} \cap F = \emptyset$
or $\{p,q\} \subseteq F$.
We remark that synchronized state pairs have no connection to the notion of synchronizing words.

Our main technical result for synchronized pairs is the following:

\begin{lemma}\label{lem:sync-pair}
Assume that every reachable synchronized pair in $\B$ is $F$-consistent.
Then, $\L(\B)$ belongs to $\langle \ST, \SF, \Len \rangle$.
\end{lemma}

For the proof of \Cref{lem:sync-pair} we need two lemmas.

\begin{lemma}
	\label{lem:factorial-length}
	A state pair $(p,q)$ is synchronized
	if and only if $p$ and $q$ are nontransient
	and there exists a nonempty run from $p$ to $q$ whose 
	length is a multiple of $|Q|!$.
\end{lemma}

\begin{proof}
        First assume that $(p,q)$ is synchronized.
	Let $x,y,z\in\Sigma^+$ with $|x|=|y|=|z|=k$ such that
	$q \xleftarrow{x} q \xleftarrow{y} p \xleftarrow{z} p$.
	Then, $p$ and $q$ are nontransient and we have 
	\[  q \xleftarrow{x^{|Q|!-1}y} p, \]
	where $x^{|Q|!-1}y$ has length $(|Q|!-1) \cdot k + k = |Q|! \cdot k$.

	Conversely, assume that $p$ and $q$ are nontransient
	and there exists a nonempty run $q \xleftarrow{y} p$ whose length is divided by $|Q|!$.
	Since the states $p$ and $q$ are nontransient,
	there are words $x$ and $z$ of length at most $|Q|$ with
	$q \xleftarrow{x} q$ and $p \xleftarrow{z} p$.
	These words can be pumped up to have length $|y|$.
\end{proof}

Let $Q = T \cup N$ be the partition of the state set into the set $T$ of transient states and the set $N$ of nontransient states.
A function $\beta \colon \N \to \{0,1\}$ is {\em $k$-periodic} if $\beta(i) = \beta(i+k)$ for all $i \in \N$.

\begin{lemma}
	\label{lem:periodic}
	Assume that every reachable synchronized pair in $\B$ is $F$-consistent.
	Then, for every word $v \in \Sigma^*$ of length at least $|Q|! \cdot (|T|+1)$
	there exists a $|Q|!$-periodic function $\beta_v \colon \N \to \{0,1\}$ such that the following holds:
	If $w \in \Sigma^* v$ and $w \cdot q_0 \in N$, then we have
	$w \in \L(\B)$ if and only if $\beta(|w|) = 1$.
\end{lemma}

\begin{proof}
	Let $v = a_k \cdots a_2 a_1$  with
	$k \ge |Q|! \cdot (|T|+1)$, and consider the run
	\begin{equation}
		\label{eq:prefix-run}
		q_k \xleftarrow{a_k} \cdots \xleftarrow{a_2} q_1 \xleftarrow{a_1} q_0
	\end{equation}
	of $\B$ on $v$.
	Clearly, each transient state can occur at most once in the run.
	First notice that for each $0 \le i \le |Q|!-1$ at least one of the states in
	\[
		Q_i = \{ q_{i+j|Q|!} \mid 0 \le j \le |T| \}
	\]
	is nontransient because otherwise the set would contain $|T|+1$ pairwise distinct transient states.
	Furthermore, we claim that the nontransient states in $Q_i$ are either all final or all nonfinal:
	Take two nontransient states $q_{i+j_1|Q|!}$ and $q_{i+j_2|Q|!}$ with $j_1 < j_2$.
	Since we have a run of length $(j_2-j_1)|Q|!$ from $q_{i+j_1|Q|!}$ to $q_{i+j_2|Q|!}$,
	these two states form a synchronized pair by \Cref{lem:factorial-length}, which by assumption must
	be $F$-consistent.
	
	Now, define $\beta_v \colon \N \to \{0,1\}$ by
	\[
		\beta_v(m) = \begin{cases}
		1, & \text{if the states in } Q_{m \bmod |Q|!} \cap N \text{ are final}, \\
		0, & \text{if the states in } Q_{m \bmod |Q|!} \cap N \text{ are nonfinal},
		\end{cases}
	\]
	which is well-defined by the remarks above.
	Clearly $\beta_v$ is $|Q|!$-periodic.
	
	Let $w = a_m \cdots a_2 a_1 \in \Sigma^* v$ be a word of length $m \ge k$.
	The run of $\B$ on $w$ starting from the initial state prolongs the run in \eqref{eq:prefix-run}:
	\[
		q_m \xleftarrow{a_m} \cdots \xleftarrow{a_{k+2}} q_{k+1} \xleftarrow{a_{k+1}} q_k \xleftarrow{a_k} \cdots \xleftarrow{a_2} q_1 \xleftarrow{a_1} q_0
	\]
	Assume that $q_m \in N$.
	As argued above, there is a position $0 \le i < k$
	such that $i \equiv m \pmod{|Q|!}$ and $q_i \in N$.
	Therefore, there exists a nonempty run from $q_i$ to $q_m$ whose length is a multiple of $|Q|!$.
	Hence, $(q_i,q_m)$ is a synchronized pair by \Cref{lem:factorial-length},
	which is $F$-consistent by assumption.
	Therefore, $w \in L$ if and only if $q_m \in F$ if and only if  $q_i \in F$ if and only if  $\beta_v(|w|) = 1$.
\end{proof}
We can now prove \Cref{lem:sync-pair}.

\begin{proof}[Proof of lemma \ref{lem:sync-pair}]
Given a subset $P \subseteq Q$ let $\L(\B,P) := \L(Q,\Sigma,P,\delta,q_0)$.
Let $F_N = N \cap F$ and $F_T = T \cap F$.
We disjointly decompose $L$ into
\[
	L = \L(\B, F_N) \cup \bigcup_{q \in F_T} \L(\B, \{q\}).
\]
First observe that $\L(\B, \{q\}) \in \SF$ for all $q \in F_T$
because a transient state $q$ can occur at most once in a run of $\B$.

It remains to show  that  $\L(\B, F_N)$ belongs to $\langle \ST,\SF,\Len \rangle$.
Using the threshold $k=|Q|! \cdot (|T|+1)$,
we distinguish between words of length at most $k-1$ and words of length at least $k$,
and group the latter set by their suffixes of length $k$:
\[
	\L(\B, F_N) = (\L(\B, F_N) \cap \Sigma^{\le k-1})\cup\bigcup_{v\in\Sigma^{k}}(\L(\B, F_N) \cap \Sigma^* v).
\]
The first part $\L(\B, F_N) \cap \Sigma^{\le k-1}$ is finite and thus suffix testable.
To finish the proof, we will show that $\L(\B, F_N) \cap \Sigma^*v \in \langle \ST, \SF, \Len \rangle$
for each $v \in \Sigma^k$.
Let $v \in \Sigma^k$ and let $\beta_v \colon \N \to \{0,1\}$ be the $|Q|!$-periodic function from \Cref{lem:periodic}.
The lemma implies that
\[
	\L(\B, F_N) \cap \Sigma^*v = (\Sigma^*v \cap \{ w \in \Sigma^* \mid \beta(|w|) = 1 \}) \setminus \L(\B, T).
\]
The language $\{ w \in \Sigma^* \mid \beta(|w|) = 1 \}$ is 
a regular length language, $\Sigma^* v$ is suffix testable and $\L(\B, T)$ is a finite union of
regular suffix-free languages.
\end{proof}

The following lemma is an immediate consequence of \Cref{lem:sync-pair}.

\begin{lemma}\label{lemma:logprop}
If $L \in \Reg \setminus \langle \ST, \SF, \Len \rangle$
then there exist $u,x,y,z\in\Sigma^*$ with $|x|=|y|=|z| \ge 1$ such that
$L$ separates $x^*yz^*u$ and $z^*u$.
\end{lemma}
Now, we can finally prove point \ref{lower-log} from \Cref{thm:tetrachotomy}.

\begin{proposition}\label{lowerlognrandom}
If $L \in \Reg \setminus \langle \ST, \SF, \Len \rangle$
then $\F^\r_L(n) = \Omegainf(\log n)$.
\end{proposition}

\begin{proof}
Consider the words $u,x,y,z\in\Sigma^*$ described in \Cref{lemma:logprop}.
Let $n= |z| \cdot m+|u|$ for some $m \ge 1$
and let $\P_n$ be a randomized SW-algorithm for $L$.
We describe a randomized one-way protocol for $\mathrm{GT}_m$:
Let $1 \le i \le m$ be the input of Alice and $1 \le j \le m$ be the input of Bob.
Alice starts with reading $x^m y z^{m-i}$ into $\P_n$. Then she sends the reached state to Bob using $\O(s(\P_n))$ bits.
Bob then continues the run of $\P_n$ from the transmitted state with the word $z^j u$.
Hence, $\P_n$ is simulated on the word $w := x^m y z^{m-i} z^j u = x^m y z^{m-i+j} u$.
We have
\[
	\last_n(w) = \begin{cases}
	x^{i-1-j} y z^{m-i+j} u, & \text{if } i > j, \\
	z^m u, & \text{if } i \le j.
	\end{cases}
\]
By \Cref{lemma:logprop}, $\last_n(w)$ belongs to $L$ in exactly one of the two cases
$i>j$ and $i \leq j$. Hence, Bob can distinguish these two cases with probability at least $2/3$.
It follows that the protocol computes $\mathrm{GT}_m$ and its cost is bounded by $s(\P_n)$.
By \Cref{thm:coco} we can conclude that $s(\P_n) = \Omega(\log m) = \Omega(\log n)$,
and therefore $\F^\r_L(n) = \Omegainf(\log n)$.
\end{proof}

\subsection{Sliding window algorithms with one-sided error} \label{sec-one-sided}

So far, we have only considered randomized SW-algorithms with two-sided error (analogously to the 
complexity class \BPP).
Randomized SW-algorithms with one-sided error (analogously to the 
classes \RP~and \coRP) can be motivated by applications where all ``yes''-outputs or
all ``no''-outputs, respectively, have to be correct. 
We distinguish between true-biased and false-biased algorithms.
A {\em true-biased (randomized) streaming algorithm} $\P$ for a language $L$ 
satisfies the following properties:
\begin{itemize}
	\item If $w \in L$ then $\Pr[\P \text{ accepts } w] \ge 2/3$.
	\item If $w \notin L$ then $\Pr[\P \text{ rejects } w] = 1$.
\end{itemize}
A {\em false-biased (randomized) streaming algorithm} $\P$ for a language $L$ satisfies the following properties:
\begin{itemize}
	\item If $w \in L$ then $\Pr[\P \text{ accepts } w]  = 1$.
	\item If $w \notin L$ then $\Pr[\P \text{ rejects } w] \ge 2/3$.
\end{itemize}
Let $F^\mathsf{0}_L(n)$ (resp., $F^\mathsf{1}_L(n)$) be the minimal space complexity $s(\P_n)$
of any true-biased (resp., false-biased) SW-algorithm $\P_n$ for $L$ and window size $n$.
We have the relations $F^\r_L(n) \le F^i_L(n) \le F_L(n)$ for $i \in \{ \mathsf{0}, \mathsf{1} \}$,
and $F^\mathsf{0}_L(n) = F^\mathsf{1}_{\Sigma^* \setminus L}(n)$.

For $F^\mathsf{0}_L(n)$ and $F^\mathsf{1}_L(n)$ a statement analogous to \Cref{thm:randombool} does not hold, i.e., 
the classes  $\{ L \subseteq \Sigma^* \mid \F^\mathsf{i}_L(n) = \O(s(n)) \}$ for $\mathsf{i} \in \{0,1\}$ and a function $s(n)$ do not
form a Boolean algebra. To see this, consider the language $L= \{ \$ w \# w : w \in \{0,1\}^*\}$. It is easy to see
that $F^\mathsf{1}_{L}(n) = \O(\log n)$. On the other hand, every true-biased randomized (in fact, every nondeterministic)
communication protocol for $\mathrm{EQ}_n$ (over the domain $\{1, \dots, n\}$) has cost $\Omega(\log n)$ \cite[Chapter~5]{Roughgarden16}. This implies
$F^\mathsf{1}_{\Sigma^* \setminus L}(n) = F^\mathsf{0}_{L}(n)  = \Omegainf(n)$, where $\Sigma= \{ 0,1,\$, \#\}$.

We show that for all regular languages
SW-algorithms with one-sided error have no advantage over their deterministic counterparts:
\begin{theorem}[one-sided error]
\label{thm:trichotomy-one-sided}
Let $L$ be regular.
\begin{enumerate}[label=(\roman{*}), ref=(\roman{*})]
	\item If $L \in \langle \ST, \Len \rangle$ then
	$F^\mathsf{0}_L(n)$ and $F^\mathsf{1}_L(n)$ are $\O(1)$. \label{onesided-1}
	\item If $L \notin \langle \ST, \Len \rangle$ then
	$F^\mathsf{0}_L(n)$ and $F^\mathsf{1}_L(n)$ are $\Omegainf(\log n)$. \label{onesided-2}
	\item If $L \in \langle \LI, \Len \rangle$ then
	$F^\mathsf{0}_L(n)$ and $F^\mathsf{1}_L(n)$ are $\O(\log n)$. \label{onesided-3}
	\item If $L \notin \langle \LI, \Len \rangle$ then
	$F^\mathsf{0}_L(n)$ and $F^\mathsf{1}_L(n)$ are $\Omegainf(n)$. \label{onesided-4}
\end{enumerate}
\end{theorem}
The upper bounds in \ref{onesided-1} and \ref{onesided-3} already hold
for deterministic SW-algorithms (\Cref{thm:big-thm}).
Moreover, the lower bound in \ref{onesided-4} already holds for SW-algorithms with two-sided error
(\Cref{thm:tetrachotomy}\ref{lower-lin}).
It remains to prove point \ref{onesided-2} of the theorem.\footnote{Note that \Cref{thm:trichotomy-one-sided}\eqref{onesided-2}
generalizes the lower bound $F_L(n) = \Omegainf(\log n)$ for languages $L  \in \Reg \setminus \langle \ST, \Len \rangle$; see \Cref{thm:big-thm}.}
In fact we show that any {\em nondeterministic} SW-algorithm
for a regular language $L \notin \langle \ST, \Len \rangle$
requires space $\Omegainf(\log n)$ (this generalizes the lower bound in the second equivalence of \Cref{thm:big-thm}).
A nondeterministic SW-algorithm for a language $L$ and window size $n$ is an NFA $\P_n$
with $\L(\P_n) = \SW_n(L)$, and its space complexity is $s(\P_n) = \log |\P_n|$.
If we have a true-biased randomized SW-algorithm for $L$ we can turn it into a
nondeterministic SW-algorithm by keeping only those transitions with nonzero probabilities
and making all states $q$ initial which have a positive initial probability $\iota(q) > 0$.
Therefore, it suffices to show the following statement:

\begin{proposition} \label{prop-lower-sqrt}
	Let $L \in \Reg \setminus \langle \ST, \Len \rangle$.
	Then, for infinitely many $n$ every nondeterministic SW-algorithm $\P_n$ for $L$ has $\Omega(\sqrt{n})$ many states.
\end{proposition}

For the proof of \Cref{prop-lower-sqrt} we need the following lemma.

\begin{lemma} \label{lemma-NFA-lower}
Let $L \subseteq a^*$ and $n \in \N$ such that
$L$ separates $\{a^n\}$ and $\{a^k \mid k > n\}$.
Then, every NFA for~$L$ has at least $\sqrt{n}$ many states.
\end{lemma}

\begin{proof}
The easy case is $a^n \in L$ and $a^k \notin L$ for all $k > n$.
If an NFA for $L$ has at most $n$ states then any successful run on $a^n$
must have a state repetition.
By pumping one can construct a successful run on $a^k$ for some $k > n$, which is a contradiction.

Now, assume $a^n \notin L$ and $a^k \in L$ for all $k > n$.
The proof is essentially the same as for \cite[Lemma~6]{JiraskovaM14}, where the statement of the lemma is shown for $L = a^* \setminus \{a^n\}$.
Let us give the proof for completeness. 
It is known that every unary NFA has an equivalent NFA in so-called Chrobak normal form.  A unary NFA in Chrobak normal
form consists of a simple path (called the initial path in the following) whose starting state is the unique initial state of the NFA. 
From  the last state of the initial path, edges go to a collection of 
disjoint cycles. In \cite{Gawrychowski11} it is shown that an $m$-state unary NFA has an equivalent NFA in Chrobak normal form whose
initial path consists of $m^2 - m$ states. Now, assume that $L$ is accepted by an NFA with $m$ states and let $\A$ 
be the equivalent NFA in Chrobak normal form, whose initial path consists of $m^2-m$ states. If $n \geq m^2-m$ then all states
that are reached in $\A$ from the initial state via $a^n$ belong to a cycle and every cycle contains such a state. Since
$a^n \notin L$, all these states are rejecting. Hence, $a^{n+ x \cdot d} \notin L$ for all $x \geq 0$, where $d$ is the product
of all cycle lengths. This contradicts the fact that $a^k \in L$ for all $k > n$. Hence, we must have $n < m^2-m$ and therefore
$m > \sqrt{n}$.
\end{proof}

\begin{proof}[Proof of Proposition \ref{prop-lower-sqrt}]
Let $L \in \Reg \setminus \langle \ST, \Len \rangle$.
By \Cref{lem:nonwb-loop} and the results from \Cref{sec:char-const}
there are words $x,y,z \in \Sigma^*$ such that $|x|=|y|$ and
$L$ separates $x y^* z$ and $y^* z$.
Note that we must have $x \neq y$.

Fix $m \geq 0$ and consider the window size $n  =  |x| + m|y| + |z|$.
Let $\P_n = (Q,\Sigma,I,\Delta,F)$ be a nondeterministic SW-algorithm for $L$ and window size $n$,
i.e.,~it is an NFA for $\SW_n(L)$.
Notice that $\P_n$ separates $\{x y^m z\}$ and $\{ x y^k z \mid k > m\}$.
We define an NFA $\A$ over the unary alphabet $\{a\}$ as follows:
\begin{itemize}
\item The state set of $\A$ is $Q$.
\item The set of initial states of $\A$ is $\{ q \in Q \mid \exists p \in I \colon p \xrightarrow{ x } q \text{ in } \P_n \}$.
\item The set of final states of $\A$  is $\{ p \in Q \mid \exists q \in F \colon p \xrightarrow{ z } q \text{ in } \P_n \}$.
\item The set of transitions of $\A$ is $\{ (p,a,q) \mid p \xrightarrow{y} q \text{ in } \P_n \}$.
\end{itemize}
It recognizes the language $\L(\A) = \{ a^k \mid x y^k z \in \SW_n(L) \}$,
and therefore $\L(\A)$ separates $\{a^m\}$ and $\{ a^k \mid k > m \}$.
By \Cref{lemma-NFA-lower}, $\A$ has at least $\sqrt{m} = \Omega(\sqrt{n})$ states.
Hence, also the number of states of $\P_n$ is in $\Omega(\sqrt{n})$.
\end{proof}

 \Cref{prop-lower-sqrt} implies $F^\mathsf{0}_L(n) \geq 1/2 \log n - \O(1)$ on infinitely many $n$ 
 for all $L \in \Reg \setminus \langle \ST, \Len \rangle$.
Since $\Reg \setminus \langle \ST, \Len \rangle$ is closed under complement,
this implies $F^\mathsf{1}_L(n)=F^\mathsf{0}_{\Sigma^* \setminus L}(n) \geq 1/2 \log n - \O(1)$ 
on infinitely many $n$ for all $L \in \Reg \setminus \langle \ST, \Len \rangle$.

\subsection{Randomized variable-size model} \label{sec-randomized-variable}

In this section, we briefly look at randomized algorithms in the variable-size model.
First we transfer the definitions from \Cref{sec-variable-size} in a straightforward way.
A {\em randomized variable-size sliding window algorithm}~$\P$ for $L \subseteq \Sigma^*$
is a randomized streaming algorithm for $\SW(L)$ (defined in \eqref{def-SW(L)} on page~\pageref{def-SW(L)}).
Its {\em space complexity} is $v(\P,n) = \log |M_{\le n}| \in \N \cup \{\infty\}$
where $M_{\le n}$ contains all memory states in $\P$ which are reachable with nonzero probability in $\P$
on inputs $w \in \Sigma_\downarrow^*$ with $\mwl(w) \le n$.
Since the variable-size sliding window model subsumes the fixed-size model,
we have $\F^\r_L(n) \le v(\P,n)$ for every randomized variable-size
sliding window algorithm $\P$ for $L$.

Again we raise the question if randomness can improve the space complexity in the variable-size model.
We claim that, in contrast to the fixed-size model, randomness does not allow more space efficient
algorithms in the variable-size setting. 
Clearly, all upper bounds for the deterministic variable-size setting transfer to the randomized
variable-size setting,
i.e.,~languages in $\langle \LI, \Len \rangle$ have $\O(\log n)$ space complexity,
and empty and universal languages have $\O(1)$ space complexity.
For every regular language $L$ which is not contained in $\langle \LI, \Len \rangle$
we proved a linear lower bound on $\F^\r_L(n)$ (\Cref{thm:rand-lin}),
which is also a lower bound on the space complexity of any randomized variable-size sliding window algorithm for $L$.
It remains to look at languages $\emptyset \subsetneq L \subsetneq \Sigma^*$,
for which we have proved a logarithmic lower bound
in the deterministic setting (\Cref{lem:length-lb}).

\begin{lemma}
	\label{lem:rand-length-lb}
	If $\P$ is a randomized variable-size SW-algorithm 
	for a language $\emptyset \subsetneq L \subsetneq \Sigma^*$
	then $v(\P,n) = \Omega(\log n)$.
\end{lemma}

\begin{proof}
	Let $\emptyset \subsetneq L \subsetneq \Sigma^*$ be a language.
	There must be a length-minimal nonempty word
	$a_1 \cdots a_k \in \Sigma^+$ such that $|\{\eps,a_1 \cdots a_k\} \cap L| = 1$ and we fix such a word
	$a_1 \cdots a_k$.
	By minimality we also have $|\{a_1 \cdots a_k,a_2 \cdots a_k\} \cap L| = 1$.
	Let~$\P$ be a randomized variable-size SW-algorithm for $L$.
	By \Cref{lem:amplification} we can assume that the error probability of $\P$ is at most 1/6,
	which increases its space complexity $v(\P,n)$ by a constant factor.
	
	For every $n \in \N$ we construct a protocol for $\mathrm{GT}_n$
	with cost $\O(v(\P,n))$. With \Cref{thm:coco} this implies that $v(\P,n) = \Omega(\log n)$.
	Let $1 \le i \le n$ be the input of Alice and $1 \le j \le n$ be the input of Bob.
	Alice starts two instances of $\P$ (using independent random bits)
	and reads $a_1^i$ into both of them.
	She sends the memory states to Bob
	using $\O(v(\P,i)) \le \O(v(\P,n))$ bits.
	Bob then continues from both states, and reads $\downarrow^j a_2 \cdots a_k$ into the first instance
	and $\downarrow^{j+1} a_1 a_2 \cdots a_k$ into the second instance.
	Let $y_1, y_2 \in \{0,1\}$ be the outputs of the two instances of $\P$.
	With high probability, namely $(1-1/6)^2 \ge 2/3$,
	both answers are correct, i.e.
	\begin{align*}
		y_1 = 1 \ & \iff  \ \wnd(a_1^i \downarrow^j a_2 \cdots a_k) \in L \\
	\intertext{and}
		y_2 = 1 \ & \iff \  \wnd(a_1^i \downarrow^{j+1} a_1 a_2 \cdots a_k) \in L.
	\end{align*}
	Bob returns true, i.e.,~he claims $i > j$, if and only if $y_1 = y_2$.
	
	Let us prove the correctness. If $i > j$ then
	\[
		\wnd(a_1^i \downarrow^j a_2 \cdots a_k) = a_1^{i-j} a_2 \cdots a_k = \wnd(a_1^i \downarrow^{j+1} a_1 a_2 \cdots a_k)
	\]
	and hence Bob returns true with probability at least 2/3.
	If $i \le j$ then
	\[
		\wnd(a_1^i \downarrow^j a_2 \cdots a_k) = a_2 \cdots a_k
	\]
	and
	\[
		\wnd(a_1^i \downarrow^{j+1} a_1 a_2 \cdots a_k) = a_1a_2 \cdots a_k.
	\]
	By assumption, exactly one of the words $a_1 \cdots a_k$, $a_2 \cdots a_k$ belongs to $L$,
	and therefore Bob returns false with probability at least 2/3.
\end{proof}

The lower bound from \Cref{lem:rand-length-lb} also holds for variable-size SW-algorithms with one-sided error
since they are more restricted than algorithms with two-sided error.
In fact, \Cref{lem:rand-length-lb} also holds for nondeterministic and co-nondeterministic SW-algorithms
since the (co-)nondeterministic communication complexity of $\mathrm{GT}_n$ is $\Theta(\log n)$~\cite[Chapter~5]{Roughgarden16}.

\section{Property testing in the sliding window model} \label{sec-SW-testing}

\label{chp:pt}
In all settings discussed so far, there are some regular languages for which testing membership in the sliding window model requires linear space.
To be more specific, for any language $L\in\Reg\setminus\langle \LI, \Len \rangle$ it requires linear space to test membership even for randomized sliding window algorithms with two-sided error.
In order to achieve space-efficient sliding window algorithms for all regular languages, we have
to allow randomized sliding window algorithms that are allowed to err with
unbounded probability on some specific inputs.
We formalize this in the context of the property testing framework. More precisely,
we introduce in this section {\em sliding window (property) testers},
which must accept if the active window belongs to a language $L$
and reject if it has large Hamming distance from $L$.

 For words that are not in $L$ but that have
small Hamming distance from $L$ the algorithm is allowed to give any answer.
We consider deterministic sliding window property testers
and randomized sliding window property testers.

While at first sight the only connection between property testers and sliding window property testers is that we must accept the input if it satisfies a property $P$ and reject if it is far from satisfying $P$, there is, in fact, a deeper link.
In particular, the property tester for regular languages due to Alon et al.~\cite{AlonKNS00}
combined with an optimal sampling algorithm for sliding windows~\cite{BravermanOZ12} immediately yields $\O(\log n)$-space, two-sided error sliding window property testers with Hamming gap $\gamma(n) = \epsilon n$ for all regular languages. We will improve on this observation. 

\subsection{Sliding window testers}

The {\em Hamming distance} between two words
$u = a_1 \cdots a_n$ and $v = b_1 \cdots b_n$ of equal length
is the number of positions where $u$ and $v$ differ, i.e.,~$\dist(u,v) = |\{ i \mid  a_i \neq b_i\}|$.
If $|u| \neq |v|$ we set $\dist(u,v) = \infty$.
The distance of a word $u$ to a language $L$ is defined as
\[
\dist(u,L) = \inf \{ \dist(u,v) \mid v \in L \} \in \N \cup \{\infty\}.
\]
Additionally, we define the \emph{prefix distance} between equal-length words
$u = a_1 \cdots a_n$ and $v = b_1 \cdots b_n$
by $\pdist(u,v) = \min \{ i \in \{0,\dots,n\} \mid a_{i+1} \cdots a_n = b_{i+1} \cdots b_n\}$.
For instance, we have $\pdist(abbaca, abcaca) = 3$ and $\pdist(abccca, abcaca) = 4$ (whereas the Hamming distance
in both cases is $1$).
Clearly, we have $\dist(u, v) \le \pdist(u,v)$.
The algorithms presented in this section satisfy the stronger property
that windows whose prefix distance to the language $L$ is large are rejected by the algorithm.

In this section, $\gamma$ is always a function $\gamma \colon \N \to \mathbb{R}_{\geq 0}$ such that $\gamma(n) < n$ for all $n$.
A {\em deterministic sliding window (property) tester} with Hamming gap $\gamma(n)$ for a language $L \subseteq \Sigma^*$
and window size $n$  is a deterministic streaming algorithm $\P_n$
over the alphabet $\Sigma$ with the following properties:
\begin{itemize}
\item If $\last_n(w) \in L$, then $w \in \L(\P_n)$.
\item If $\dist(\last_n(w),L) > \gamma(n)$, then $w \notin \L(\P_n)$.
\end{itemize}
If neither of the two cases hold, the behavior of $\P_n$ can be arbitrary. Recall that $s(\P_n)$ is the space used by $\P_n$ (see \Cref{sec streaming general}).
A {\em randomized sliding window tester} with Hamming gap $\gamma(n)$ for a language $L \subseteq \Sigma^*$
and window size $n$  is a randomized streaming algorithm $\P_n$
over the alphabet $\Sigma$ with the following properties.
It has {\em two-sided error} if for all $w \in \Sigma^*$ we have:
\begin{itemize}
\item If $\last_n(w) \in L$, then $\Pr[\P_n \text{ accepts } w] \ge 2/3$.
\item If $\dist(\last_n(w),L) > \gamma(n)$, then $\Pr[\P_n \text{ rejects } w] \ge 2/3$.
\end{itemize}
It is {\em true-biased} if for all $w \in \Sigma^*$ we have:
\begin{itemize}
\item If $\last_n(w) \in L$, then $\Pr[\P_n \text{ accepts } w] \ge 2/3$.
\item If $\dist(\last_n(w),L) > \gamma(n)$, then $\Pr[\P_n \text{ rejects } w] = 1$.
\end{itemize}
It is {\em false-biased} if for all $w \in \Sigma^*$ we have:
\begin{itemize}
\item If $\last_n(w) \in L$, then $\Pr[\P_n \text{ accepts } w] = 1$.
\item If $\dist(\last_n(w),L) > \gamma(n)$, then $\Pr[\P_n \text{ rejects } w] \ge 2/3$.
\end{itemize}
True-biased and false-biased algorithms are algorithms with {\em one-sided error}.
Again, the success probability $2/3$ is an arbitrary choice in light of \Cref{lem:amplification}.

Intuitively, the Hamming gap function $\gamma$ should be a small function.
Typical choices for  $\gamma(n)$ are $\epsilon n$ for some constant $\epsilon$ 
($0 < \epsilon < 1$) or $\gamma(n) = c$ for a constant $c$. We will also consider Hamming gap functions that
are between these two cases.
The case $\gamma(n) = 0$ for all $n$ corresponds to exact membership testing to $L$, which was studied in 
the previous sections.

Let us also remark that we only consider the fixed-size sliding window model in this section. One might also consider
variable-size sliding window testers, where the size of the window can grow and shrink. 
We leave this for future work. Moreover, we only consider the Hamming distance in this paper.
One might also consider other distances on words, like for instance edit distance. We believe
that Hamming distance is the most basic distance measure on strings. The upper bounds stated below also apply
to edit distance (since the edit distance is always bounded by the Hamming distance). Whether our 
lower bounds can be extended to edit distance remains open.

\subsection{Main results of this section} \label{sec main results testing}

Let us now state and discuss the main results of this section. We start with our upper bounds:

\begin{theorem}
\label{theorem:deterministic_ub}
For every regular language $L$ and window size $n$ there exists a deterministic sliding window tester $\P_n$ 
with Hamming gap $\O(1)$ and $s(\P_n) = \O(\log n)$.
\end{theorem}
We will later see that allowing a larger (but not too large) Hamming gap in \Cref{theorem:deterministic_ub} does not allow a better space bound.
This changes if we allow randomized sliding window testers with a two-sided error:

\begin{theorem}
\label{theorem:two-sided-general-HG}
For every regular language $L$ there is a constant $c$ such that the following holds:
If the function $\gamma(n)$ and the window size $n$ satisfy $\gamma(n) \ge c$ then there is a randomized sliding window tester $\P_n$ for $L$ and window size $n$ with a two-sided error,
Hamming gap $\gamma(n)$, and $s(\P_n) = \O(\log(n/\gamma(n)))$.
\end{theorem}
From \Cref{theorem:two-sided-general-HG} we will easily obtain the following corollary:
\begin{corollary}
\label{theorem:two-sided}
For every regular language $L$, every window size $n$ and every $0 < \epsilon < 1$
there exists a randomized sliding window tester $\P_n$
with two-sided error, Hamming gap $\epsilon n$ and $s(\P_n) = \O(1/\epsilon)$.
\end{corollary}

The upper bounds in \Cref{theorem:deterministic_ub} and \Cref{theorem:two-sided-general-HG} hold for all regular languages.
We will also identify subclasses for which these upper bounds can be improved. Recall the definition of suffix-free languages from 
\Cref{sec-results}. Another important language class in the context of sliding-window testers is the class of {\em trivial} languages.
A language $L \subseteq \Sigma^*$ is {\em $\gamma(n)$-trivial} (for a function $\gamma(n)<n$)
if for all $n \in \N$ with $L \cap \Sigma^n \neq \emptyset$
and all $w \in \Sigma^n$ we have $\dist(w,L) \le \gamma(n)$.
If $L$ is $\O(1)$-trivial we say that $L$ is {\em trivial}. 
Examples of trivial languages include all length languages, all suffix (resp., prefix) testable languages (in particular, $L = a \{a,b\}^*$ for which 
$\F_L(n) = \Theta(n)$ holds),
and also the set  of all words over $\{a,b\}$ which contain an even number of $a$'s. 
Note that Alon et al.~\cite{AlonKNS00} call a language $L$ trivial
if $L$ is $o(n)$-trivial according to our definition,
i.e.,~$\gamma(n)$-trivial for some function $\gamma(n) = o(n)$.
In fact, we will prove that both definitions coincide for regular languages (\Cref{triv-alon}).
With $\Triv$ we denote the set of all  regular trivial languages.

We can achieve a Hamming gap of $\gamma(n)$ simply with a deterministic sliding window tester that accepts or rejects all input words depending on the input length.
Moreover, the tester has only one state and hence
uses space $\log(1) = 0$. It turns out that
for finite unions of regular trivial languages and regular suffix-free languages, we can obtain a doubly logarithmic space
 bound if we allow false-biased randomized sliding window testers. Let us write $\bigcup(\Triv, \SF)$ for the class
 of all finite unions of regular trivial languages and regular suffix-free languages.

\begin{theorem}
	\label{thm:false-biased}
	For every $L \in \bigcup(\Triv, \SF)$ and window size $n$ there exists a false-biased randomized sliding window tester $\P_n$
	with Hamming gap $\O(1)$ and $s(\P_n) = \O(\log \log n)$.
\end{theorem}

Let us now discuss our lower bounds. It turns out that the above upper bounds are sharp in most cases.
First of all, the logarithmic space bound in \Cref{theorem:deterministic_ub} cannot be improved
whenever $L$ is a regular nontrivial language. This holds even for randomized true-biased algorithms and a Hamming gap $\epsilon n$ (assuming $\epsilon<1$ is not too big).
Similarly, the doubly logarithmic space bound in \Cref{thm:false-biased} cannot be improved.

\begin{figure}
\centering
 \includegraphics[scale=1.288]{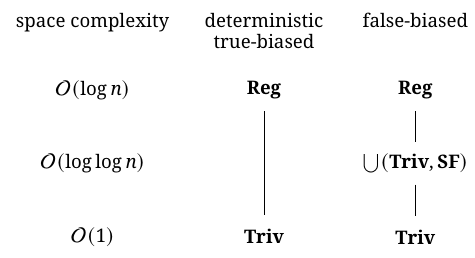}

         


\caption{The space complexity of regular languages with respect to deterministic, true-biased and false-biased sliding window testers.
As in \Cref{fig:big-picture}, only upper bounds are shown, and they hold for every Hamming gap function $\gamma(n)$ provided that $\gamma(n) \geq c$ for a constant $c$ that depends on the language. All upper bounds can be matched with lower bounds that hold for every $\gamma(n) \leq \epsilon n$ for a constant $\epsilon$ that depends on the language.}
\label{fig:big-picture-tester}
\end{figure}

\begin{theorem}
\label{theorem:deterministic_lb}
For every language $L \in \Reg \setminus \Triv$ there exist $\epsilon > 0$ and infinitely many window sizes $n \in \N$
for which every  true-biased (resp., false-biased) randomized sliding window tester for $L$ with Hamming gap $\epsilon n$
uses space at least $\log n - \O(1)$ (resp. $\log \log n - \O(1))$.
\end{theorem}
Moreover, also for false-biased randomized sliding window testers the 
 logarithmic space bound from \Cref{theorem:deterministic_ub} cannot be improved
whenever $L \in \Reg \setminus  \bigcup(\Triv, \SF)$:

\begin{theorem}
\label{thm:lb-false-biased}
If $L \in \Reg \setminus  \bigcup(\Triv, \SF)$ then there exist $\epsilon > 0$ and infinitely many window sizes $n \in \N$
for which every false-biased randomized sliding window tester for $L$ with Hamming gap $\epsilon n$
uses at least $\log n - \O(1)$ space. 
\end{theorem}
The above results provide matching upper and lower space bounds for deterministic, true-biased and false-biased sliding window testers; see also \Cref{fig:big-picture-tester}.
Moreover, the upper bounds hold for a constant Hamming gap (Theorems~\ref{theorem:deterministic_ub} and \ref{thm:false-biased}) 
whereas the lower bounds hold for Hamming gap $\epsilon n$ as long as $\epsilon$ is larger than a language-dependent constant (Theorems~\ref{theorem:deterministic_lb} and \ref{thm:lb-false-biased}).
Thus, in the deterministic, true-biased and false-biased settings, the space complexity is quite insensitive to the choice of the Hamming gap function $\gamma(n)$. 

For randomized sliding window testers with a two-sided error, the situation is different. We have already discussed \Cref{theorem:two-sided-general-HG}, 
where the Hamming gap $\gamma(n)$ is reflected in the space bound. It turns out that the upper bound in \Cref{theorem:two-sided-general-HG} is tight whenever $L$ is not a finite union of regular trivial languages and regular suffix-free languages:

\begin{theorem} \label{lemma-lb-log(n)-log(gamma(n))}
If $L \in \Reg \setminus  \bigcup(\Triv, \SF)$ then there exist $\epsilon > 0$ and infinitely many window sizes $n \in \N$ for which
every randomized sliding window tester with two-sided error for $L$ and Hamming gap $\gamma(n) \le \epsilon n$ needs space $\Omega(\log(n/\gamma(n)))$.
\end{theorem}
If $L \in \bigcup(\Triv, \SF)$ then the lower bound from \Cref{lemma-lb-log(n)-log(gamma(n))} does 
not hold in general, since we have
an upper bound of $\O(\log \log n)$ from \Cref{thm:false-biased}.\footnote{Note that if $\gamma(n) = \O(n/\log n)$ then the lower bound 
$\Omegainf(\log(n/\gamma(n)))$ becomes $\Omegainf(\log \log n)$.}
 We do not know whether there is a matching lower bound of $\Omegainf(\log \log n)$
for nontrivial languages. Currently, we can only show a slightly weaker lower bound in this case:

\begin{theorem} \label{lemma-loglog(n-gamma)}
If $L \in \Reg \setminus \Triv$ then there exist $\epsilon > 0$ and infinitely many window sizes $n \in \N$ for which
every randomized sliding window tester with two-sided error for $L$ and Hamming gap $\gamma(n) \le \epsilon n$ needs space $\Omega(\log\log(n/\gamma(n)))$.
\end{theorem}
Note that whenever $\gamma(n) = \O(n^c)$ for some $c < 1$ then the lower bound 
$\Omegainf(\log\log(n/\gamma(n)))$ from \Cref{lemma-loglog(n-gamma)} becomes 
$\Omegainf(\log \log n)$, which matches the upper bound from \Cref{thm:false-biased}.
It is left open to classify the space complexity for languages in $\bigcup(\Triv, \SF) \setminus \Triv$, e.g.\ $L = ab^*$,
for sublinear Hamming gaps $\gamma(n)$ which are $\Omega(n^c)$ for all $c < 1$, e.g.\ $\gamma(n) =  n/\log n$.

Let us also remark that \Cref{lem:rabin} does not generalize to sliding window testers (with the obvious generalization of the space complexities
$\F_L(n)$ and $\F_L^r(n)$ to sliding window testers). In the proof of \Cref{lem:rabin} we used the fact that a randomized sliding window algorithm
for a language $L$ and a window size $n$ is a probabilistic finite automaton with the isolated cut-point $1/2$. This is not necessarily true for 
randomized sliding window testers. If $w$ is a word such that neither  $\last_n(w) \in L$ nor $\dist(\last_n(w),L) > \gamma(n)$ holds then 
$\Pr[\P_n \text{ accepts } w] = 1/2$ is possible. Indeed, the generalization of \Cref{lem:rabin} to sliding window testers would
contradict \Cref{theorem:two-sided} together with \Cref{theorem:deterministic_lb}.

\subsection{Upper bounds}

In this section we prove Theorems~\ref{theorem:deterministic_ub}, \ref{theorem:two-sided-general-HG} and \ref{thm:false-biased}.

\subsubsection{Deterministic sliding window testers}

In this section, we prove \Cref{theorem:deterministic_ub}: every regular language has a deterministic sliding window tester
with constant Hamming gap which uses $\O(\log n)$ space.
It is based on the path summary algorithm from \Cref{subsec:pathsummary}.
In the following, we fix a regular language $L$ and an rDFA $\B = (Q,\Sigma,F,\delta,q_0)$ for $L$.
By \Cref{lem:uni-per} we can assume that every nontransient SCC
of $\B$ has the same period $g \geq 1$.

\begin{figure}[t]
  \centering
   \includegraphics[scale=1.288]{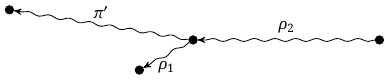}
  
\caption{The run~$\pi = \pi' \rho_2$ {\em $t$-simulates} the run $\rho = \rho_1 \rho_2$. We have $|\rho_1| \leq t$.}\label{fig:simulate-rin}
\end{figure}

For a state $q \in Q$ we define
$\Acc(q) = \{ n \in \N \mid \exists w \in \Sigma^n \colon w \cdot q \in F \}$.
The following lemma is the main tool to prove correctness of our sliding window testers.
It states that if a word of length $n$ is accepted from state $p$ and $\rho$ is any internal run (see \Cref{subsec:pathsummary} and page~\pageref{subsec:pathsummary})
of length at most $n$ starting from state $p$, then after removing a bounded length 
run at the end of $\rho$, $\rho$ can be extended 
to an accepting run of length $n$.
Formally, a run~$\pi$ {\em $t$-simulates} a run $\rho$
if one can factorize $\rho = \rho_1 \rho_2$ and $\pi = \pi' \rho_2$ where $|\rho_1| \le t$ for runs $\rho_1, \rho_2$, and $\pi'$; see also Figure~\ref{fig:simulate-rin}.
Note that this implies that $\rho$ and $\pi$ start in the same state. Also note that runs go from right to left (we work with an rDFA), so $\rho_1$ (resp., $\pi'$) is the final part of $\rho$
(resp., $\pi$). 

\begin{lemma}
\label{lem:sim}
There exists a number $t \in \N$ (which only depends on $\B$) such that for every internal run $\rho$ 
starting from a state $p$ and every $n \in \Acc(p)$ with $n \geq |\rho|$, there exists an accepting run $\pi$ of length $n$
which $t$-simulates $\rho$.
\end{lemma}
Note that the run $\pi$ in this lemma is not necessarily internal.

Based on \Cref{lem:sim} we can prove \Cref{theorem:deterministic_ub}. Afterwards we prove~\Cref{lem:sim}.

\begin{proof}[Proof of Theorem  \ref{theorem:deterministic_ub}]
Let $t$ be the constant from \Cref{lem:sim}.
We present a deterministic sliding window tester with constant Hamming gap $t$ which uses $\O(\log n)$ space.
Let $n \in \N$ be the window size.
By \Cref{lem:ps} we can maintain the set of all path summaries
$\PS_\B(w) = \{ \ps(\pi_{w,q}) \mid q \in Q \}$ for the active window $w \in \Sigma^n$,
using $\O(\log n)$ bits.
In fact, the path summary algorithm works for variable-size windows but we do not need this here.

It remains to define the acceptance condition.
Consider the SCC-factoriza\-tion of $\pi_{w,q_0}$,
say 
\[  \pi_{w,q_0} = \pi_m \tau_{m-1} \pi_{m-1} \cdots \tau_1 \pi_1 \]
and its path summary $(\ell_m,q_m) \cdots (\ell_1,q_1)$.
The algorithm accepts if and only if this path summary is accepting, i.e.,~$\ell_m = |\pi_m| \in \Acc(q_m)$.
If $w \in L$ then clearly $|\pi_m| \in \Acc(q_m)$.
On the other hand, if $|\pi_m| \in \Acc(q_m)$
then the internal run $\pi_m$ can be $t$-simulated by an accepting run $\pi_m'$ of equal length by \Cref{lem:sim}.
The run $\pi_m' \tau_{m-1} \pi_{m-1} \cdots \tau_1 \pi_1$ is accepting
and witnesses that $\pdist(w,L) \le t$. We get $\dist(w, L) \leq \pdist(w,L) \leq t$.
\end{proof}

To prove \Cref{lem:sim} we need to analyze the sets $\Acc(q)$ first.
For $a \in \N$ and $X \subseteq \N$ we use the standard notation
$X + a = \{a+x \mid x \in X\}$.
A set $X \subseteq \N$ is {\em eventually $d$-periodic},
where $d \ge 1$ is an integer,
if there exists a {\em threshold} $t \in \N$ such that for all $x \ge t$
we have $x \in X$ if and only if $x + d \in X$.
If $X$ is eventually $d$-periodic for some $d \ge 1$, then $X$ is {\em eventually periodic}.

\begin{lemma}
\label{lem:acc-per}
For every $q \in Q$ the set $\Acc(q)$ is eventually $g$-periodic.
\end{lemma}

\begin{proof}
	It suffices to show that for all $0 \le r \le g-1$
	the set $S_r = \{ i \in \N \mid r + i \cdot g \in \Acc(q) \}$
	is either finite or co-finite.
	Consider a remainder $0 \le r \le g-1$ where $S_r$ is infinite.
	We need to show that $S_r$ is indeed co-finite.
	Let $i \in S_r$ with $i \ge |Q|$,
	i.e.,~there exists an accepting run $\pi$ from $q$ of length $r + i \cdot g$.
	Since $\pi$ has length at least $|Q|$, it must traverse a state $p$ in a nontransient SCC~$C$.
	Choose $j_0$ such that $j_0 \cdot g \ge m(C)$
	where $m(C)$ is the reachability constant from \Cref{lemma-alon}.
	By \Cref{lemma-alon} for all $j \ge j_0$
	there exists a cycle from $p$ to $p$ of length $j \cdot g$.
	Therefore, we can extend $\pi$ to a longer accepting run by $j \cdot g$ symbols
	for any $j \ge j_0$.
	This proves that $x \in S_r$ for every $x \geq i + j_0$
	and that $S_r$ is co-finite.
\end{proof}

Two sets $X,Y \subseteq \N$ are {\em equal up to a threshold $t \in \N$}, in symbol $X =_t Y$, if for all $x \geq t$:
$x \in X$ if and only if $x \in Y$. Two sets $X,Y \subseteq \N$ are {\em almost equal} if they are equal up to some threshold $t \in \N$. 

\begin{lemma}
	\label{lem:ae}
	A set $X \subseteq \N$ is eventually $d$-periodic if and only if $X$ and $X+d$ are almost equal.
\end{lemma}
\begin{proof}
Let $t \in \N$ be such that for all $x \ge t$ we have $x \in X$ if and only if $x + d \in X$. Then, $X$ and $X + d$ are equal up to threshold $t+d$. Conversely, if $X =_t X+d$, then for all $x \ge t$ we have $x + d \in X$ if and only if $x + d \in X+d$, which is true if and only if $x \in X$.
\end{proof}

If the graph $G = (V,E)$ is strongly connected with $E \neq \emptyset$ and finite period $g$,
and $V_0, \ldots, V_{g-1}$ satisfy the properties from \Cref{lemma-alon}, then
we define the {\em shift} from $u \in V_i$ to $v \in V_j$
by
\begin{equation}\label{shift}
\shift(u,v) = (j-i) \bmod g \in \{0, \dots, g-1\}.
\end{equation}
Notice that $\shift(u,v)$ could be defined without referring to the partition $\bigcup_{i=0}^{g-1} V_i$ 
since the length of any path from $u$ to $v$ is congruent to $\shift(u,v)$ modulo $g$
by \Cref{lemma-alon}.  Also, note that $\shift(u,v) + \shift(v,u) \equiv 0 \pmod g$.

\begin{lemma}
\label{lem:con-run}
Let $C$ be a nontransient SCC in $\B$, $p,q \in C$ and $s = \shift(p,q)$.
Then, $\Acc(p)$ and $\Acc(q) + s$ are almost equal.
\end{lemma}

\begin{proof}
	Let $k \in \N$ such that $k \cdot g \ge m(C)$ where $m(C)$ is the constant from \Cref{lemma-alon}.
	By \Cref{lemma-alon} there exists a run from $p$ to $q$ of length $s + k \cdot g$,
	and a run from $q$ to $p$ of length $(k+1) \cdot g - s$ (the latter number is congruent 
	to $\shift(q,p)$ modulo $g$).
	By prolonging accepting runs we obtain
	\[
		\Acc(q) + s + k \cdot g \subseteq \Acc(p) \mbox{ and } \Acc(p) + (k+1) \cdot g - s \subseteq \Acc(q).
	\]
	Adding $s + k \cdot g$ to both sides of the last inclusion yields
	\[
		\Acc(p) + (2k+1) \cdot g \subseteq \Acc(q) + s + k \cdot g \subseteq \Acc(p).
	\]
	By \Cref{lem:acc-per} and \Cref{lem:ae} the three sets above are almost equal.
	Also, $\Acc(q) + s + k \cdot g$ is almost equal to $\Acc(q) + s$
	by \Cref{lem:acc-per} and \Cref{lem:ae}. Since almost equality is a transitive relation,
	this proves the statement.
\end{proof}

\begin{corollary}
\label{cor:gt}
There exists a threshold $t \in \N$ such that
\begin{enumerate}[label=(\roman{*}), ref=(\roman{*})]
	\item $\Acc(q) =_t \Acc(q) + g$ for all $q \in Q$, and \label{item:gt-i}
	\item $\Acc(p) =_t \Acc(q) + \shift(p,q)$ for all nontransient SCCs $C$ and all $p,q \in C$.
	\label{item:gt-ii}
\end{enumerate}
\end{corollary}

Let us fix the threshold $t$ from \Cref{cor:gt} in the following.
We can now prove \Cref{lem:sim}.

\begin{proof}[Proof of Lemma \ref{lem:sim}]
Let $\rho$ be an internal run starting from $p$ with $|\rho| \leq n \in \Acc(p)$.
We have to find an accepting run $\pi$ of length $n$ starting from $p$ and factorizations 
$\rho = \rho_1 \rho_2$ and $\pi = \pi' \rho_2$ with $|\rho_1| \le t$.

If $|\rho| \le t$, then we can choose for $\pi$ any accepting run from $p$ of length $n \in \Acc(p)$.
Otherwise, if $|\rho| > t$, then the internal run $\rho$ is nonempty, which implies that the 
SCC $C$ containing $p$ is nontransient.
Moreover, writing $\rho = \rho_1 \rho_2$ where $|\rho_1| = t$, it is the case
that $\rho_2$ leads from $p$ to some state $q$ of the same SCC.
Set $s := \shift(q,p)$, which satisfies $s + |\rho_2| \equiv 0 \pmod g$
by the properties in \Cref{lemma-alon} (see also the discussion before \Cref{lem:con-run}).
Since $\Acc(q) =_t \Acc(p) + s$ by \Cref{cor:gt}\ref{item:gt-ii}, $n > t$ and $n \in \Acc(p)$,
we have $n + s \in \Acc(q)$.
Finally, since $n + s \equiv n-|\rho_2| \pmod g$ and $n-|\rho_2| = n-|\rho|+t \ge t$,
we know $n-|\rho_2| \in \Acc(q)$ by \Cref{cor:gt}\ref{item:gt-i}.
This yields an accepting run $\pi'$ from $q$ of length $n-|\rho_2|$.
Then, $\rho$ is $t$-simulated by $\pi = \pi' \rho_2$.
\end{proof}

\subsubsection{Sliding window testers with two-sided error}
\label{sec:two-sided}

In this section, we will prove \Cref{theorem:two-sided-general-HG}.
We will construct for every regular language a randomized sliding window tester
with two-sided error and Hamming gap $\gamma(n)$ that uses $\O(\log(n/\gamma(n)))$ bits
assuming the window size $n$ satisfies $\gamma(n) \geq c$ for a suitably chosen constant.
We still assume that the regular language $L$
is recognized by an rDFA $\B = (Q,\Sigma,F,\delta,q_0)$
whose nontransient SCCs have uniform period $g \ge 1$.
Furthermore, we again use the constant $t$ from \Cref{cor:gt}.

We will set the constant $c$ from \Cref{theorem:two-sided-general-HG} to $c = 4(t+1)$.
Let us fix a window size $n$ such that $\gamma(n) \ge 4(t+1)$.
We define the parameters $h = n - t$ and $\ell = n - \gamma(n) + t + 1$, which satisfy
\begin{align}
\begin{split}
	\label{eq:hl-estimate}
	\frac{\ell}{h} &= \frac{n - \gamma(n) + t + 1}{n - t} \le \frac{n - \gamma(n) + \frac{1}{4}\gamma(n)}{n - \frac{1}{4} \gamma(n)}  \\
	&= \frac{n - \frac{1}{4}\gamma(n) - \frac{1}{2}\gamma(n)}{n - \frac{1}{4} \gamma(n)} \leq 1 -   \frac{\gamma(n)}{2n}.
\end{split}
\end{align}
Let $\BZ = (C,\{\mathtt{inc}\},c_0,\rho,F)$ be the $(h,\ell)$-counter with error probability $1/(3|Q|)$ from \Cref{prop:bernoulli},
which uses $\O(\log\log |Q| + \log(n/\gamma(n))) = \O(\log(n/\gamma(n)))$ space by \eqref{eq:hl-estimate} (as usual, we consider
$|Q|$ as a constant).
The counter $\BZ$ is used to define so-called compact summaries of runs. 

A {\em compact summary} $\kappa = (q_m, r_m, c_m) \cdots (q_2, r_2, c_2) (q_1, r_1, c_1)$ is a sequence of triples,
where each triple $(q_i, r_i, c_i)$ consists of a state $q_i \in Q$, a remainder $0 \le r_i \le g-1$,
and a state $c_i \in C$ of the counter $\BZ$. The state $c_1$ of the counter is always its initial state $c_0$ (and hence low)
and $r_1 = 0$.
We say that $\kappa$ {\em represents} a run $\pi$ if
the SCC-factorization of $\pi$ has the form $\pi_m \tau_{m-1} \pi_{m-1} \cdots \tau_1 \pi_1$,
and the following properties hold for all $1 \le i \le m$:
\begin{enumerate}[label=(C\arabic{*}), ref=(C\arabic{*})]
\item $\pi_i$ starts in $q_i$; \label{C1}
\item $r_i = |\tau_{i-1} \pi_{i-1} \cdots \tau_1 \pi_1| \bmod {g}$; \label{C2}
\item if $|\tau_{i-1} \pi_{i-1} \cdots \tau_1 \pi_1| \le n - \gamma(n) + t + 1$
then $c_i$ is a low state;\label{C3}
\item if $|\tau_{i-1} \pi_{i-1} \cdots \tau_1 \pi_1| \ge n-t$ then $c_i$ is a high state. \label{C4}
\end{enumerate}
Note that $\kappa$ does not restrict $\pi_m$ except that the latter must be an internal run
starting in $q_m$.

\begin{figure}[t]
  \centering
   \includegraphics[scale=1.288]{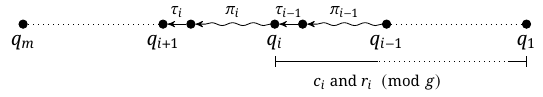}
  


  
 
\caption{A compact summary of a run $\pi$.}\label{fig:compact}
\end{figure}

The idea of a compact summary is visualized in \Cref{fig:compact}.
If $m > |Q|$ then the above compact summary cannot represent a run.
Therefore, we can assume that $m \leq |Q|$.
For every triple $(q_i, r_i, c_i)$, the entries $q_i$ and $r_i$ only depend on the rDFA $\B$,
and hence can be stored with $\O(1)$ bits.
Each state $c_i$ of the probabilistic counter $\BZ$ needs $\O(\log(n/\gamma(n)))$ bits.
Hence, a compact summary can be stored in $\O(\log(n/\gamma(n)))$ bits.
In contrast to the deterministic sliding window tester, we maintain a set of compact summaries
which represent all runs of $\B$ on the {\em complete} stream read so far
(not only on the active window) with high probability.\footnote{This is similar to the randomized SW-algorithm from \Cref{{lemma-threshold}}
that stores in the probabilistic counters $\BZ_q$ information about the complete input stream.}

\begin{proposition}
\label{prop:ps-alg}
For a given input stream $w \in \Sigma^*$, we can maintain a set of compact summaries $S = \{\kappa_w(q) \mid q \in Q \}$
such that for all $q \in Q$,
\begin{itemize}
    \item $\kappa_w(q)$ starts in $q$, and 
    \item $\Pr[\text{run $\pi_{w,q}$ is represented by $\kappa_w(q)$}] \geq 2/3$.
\end{itemize}
\end{proposition}

\begin{proof}
We maintain for the input word $w \in \Sigma^*$ a set of
random compact summaries $S = \{\kappa_w(q) \mid q \in Q \}$ as follows.

For $w = \eps$, we initialize $S = \{\kappa_{\eps}(q) \mid q \in Q\}$
where $\kappa_{\eps}(q) = (q,0,c_0)$ for $q \in Q$.
If $a \in \Sigma$ is the next input symbol in the stream, then $S$ is updated to the new set $S'$ of compact summaries
by iterating over all transitions $q \xleftarrow{a} p$ in $\B$ and prolonging the compact summary starting in $q$
by that transition.
To prolong a compact summary 
\begin{equation} \label{eq:kappa-w}
\kappa_{w}(q) = (q_m, r_m, c_m) \cdots (q_1, r_1, c_1) 
\end{equation}
we proceed similarly to Algorithm~\ref{alg:ps}.

If $p$ and $q = q_1$ are not in the same SCC then the new compact summary $\kappa_{wa}(p)$ is
\[
	 (q_m,(r_m+1) \bmod g, c'_m) \cdots (q_1,(r_1+1) \bmod g, c'_1) (p, 0, c_0),
\]
where every counter state $c'_i$ is chosen with probability $\rho(c_i,\mathtt{inc}, c'_i)$.

If $p$ and $q = q_1$ belong to the same SCC, then $\kappa_{wa}(p)$ is
\[
	(q_m,(r_m+1) \bmod g, c'_m) \cdots (q_2,(r_2+1) \bmod g, c'_2) (p,r_1, c_1),
\]
where every counter state $c'_i$ with $2 \le i \le m$ is chosen with probability $\rho(c_i,\mathtt{inc}, c'_i)$.

Note that the right-most triple of $\kappa_{w}(q)$ will be $(q,0,c_0)$ with probability $1$.

Finally we claim that for every $q \in Q$, the compact summary $\kappa_w(q)$ from \eqref{eq:kappa-w} computed by the algorithm
represents $\pi_{w,q}$ with probability $2/3$. 
Properties \ref{C1} and \ref{C2} are satisfied by construction.
Furthermore, since the length of $\kappa_w(q)$ is bounded by $|Q|$
and each instance of $\BZ$ has error probability $1/(3|Q|)$
the probability that property \ref{C3} or \ref{C4} is violated for some $i$
is at most $1/3$ by the union bound.
\end{proof}
For the randomized algorithm from the proof of \Cref{prop:ps-alg} the same comment applies that was made after the proof of \Cref{lemma-threshold}: 
the increments of the probabilistic counters do not have to be independent. 
Hence, in each step, only the random bits for incrementing a single $(\ell, h)$-counter (with the above parameters $\ell$ and $h$)
are needed. These random bits can be used for all counters that have to be incremented.

It remains to define an acceptance condition on compact summaries.
For every $q \in Q$ we define
\[
	\Acc_{\mathit{mod}}(q) = \{ \ell \bmod {g} \mid \ell \in \Acc(q) \text{ and } \ell \ge t\}.
\]
Let $\kappa = (q_m, r_m, c_m) \cdots (q_1, r_1, c_1)$ be a compact summary.
Since $c_1$ is the low initial state of the probabilistic counter, there exists a maximal index
$i \in \{1, \dots, m\}$ such that $c_i$ is low.
We say that $\kappa$ is {\em accepting} if $(n - r_i) \bmod g \in \Acc_{\mathit{mod}}(q_i)$.

\begin{proposition}
\label{prop:correctness}
Let $w \in \Sigma^*$ with $|w| \ge n$ and let $\kappa$ be a compact summary which represents $\pi_{w,q_0}$.
\begin{enumerate}[label=(\roman{*}), ref=(\roman{*})]
\item If $\last_n(w) \in L$, then $\kappa$ is accepting. \label{item:compact-acc}
\item If $\kappa$ is accepting, then $\pdist(\last_n(w),L) \le \gamma(n)$. \label{item:compact-rej}
\end{enumerate}
\end{proposition}

\begin{proof}
Consider the SCC-factorization of $\pi = \pi_{w,q_0} = \pi_m \tau_{m-1} \cdots \tau_1 \pi_1$
and a compact summary $\kappa = (q_m, r_m, c_m) \cdots (q_1, r_1, c_1)$ representing $\pi$.
Thus, $q_1 = q_0$ and $c_1 = c_0$.
Consider the maximal index $1 \le i \le m$ where $c_i$ is low,
which means that $|\tau_{i-1} \pi_{i-1} \cdots \tau_1 \pi_1| < n-t$ by \ref{C4}.
The run of $\B$ on $\last_n(w)$ has the form $\pi_k' \tau_{k-1} \pi_{k-1} \cdots \tau_1 \pi_1$
for some suffix $\pi_k'$ of $\pi_k$ and $k \geq i$.
We have $|\pi_k' \tau_{k-1} \cdots \pi_i| = n - |\tau_{i-1} \pi_{i-1} \cdots \tau_1 \pi_1| > t$.
By \ref{C2} we know that
\[
	r_i = |\tau_{i-1} \pi_{i-1} \cdots \tau_1 \pi_1| \bmod {g} = n - |\pi_k' \tau_{k-1} \cdots \pi_i| \bmod {g}.
\]
For \ref{item:compact-acc}~assume that $\last_n(w) \in L$.
Thus, $\pi_k' \tau_{k-1} \pi_{k-1} \cdots \tau_1 \pi_1$
is an accepting run starting in $q_0$.
By \ref{C1} the run $\pi_k' \tau_{k-1} \cdots \pi_i$ starts in $q_i$.
Hence, $\pi_k' \tau_{k-1} \cdots \pi_i$ is an accepting run from $q_i$ of length at least $t$.
By definition of $\Acc_{\mathit{mod}}(q_i)$ we have
\[
	n - r_i \bmod {g} = |\pi_k' \tau_{k-1} \cdots \pi_i| \bmod {g}  \in \Acc_{\mathit{mod}}(q_i),
\]
and therefore $\kappa$ is accepting.

For \ref{item:compact-rej}~assume that $\kappa$ is accepting,
i.e.
\[
	(n - r_i) \bmod g = |\pi_k' \tau_{k-1} \cdots \pi_i|  \bmod g \in \Acc_{\mathit{mod}}(q_i).
\]
Recall that $|\pi_k' \tau_{k-1} \cdots \pi_i| > t$.
By definition of $\Acc_{\mathit{mod}}(q_i)$ there exists an accepting run from $q_i$
whose length is congruent to $|\pi_k' \tau_{k-1} \cdots \pi_i|$ mod $g$
and at least $t$.
By point \ref{item:gt-i} from \Cref{cor:gt} we derive that $|\pi_k' \tau_{k-1} \cdots \pi_i| \in \Acc(q_i)$.
We show that $|\pi_i \tau_{i-1} \pi_{i-1} \cdots \tau_1 \pi_1| \ge n-\gamma(n)+t$ by a case distinction.
If $i = m$, then clearly $|\pi_i \tau_{i-1} \pi_{i-1} \cdots \tau_1 \pi_1| = |w| \ge n \ge n-\gamma(n)+t$.
The latter inequality follows from our assumption $t+1 \le \gamma(n)/4$.
If $i < m$, then $c_{i+1}$ is high by maximality of $i$,
which implies $|\tau_i \pi_i \cdots \tau_1 \pi_1| > n-\gamma(n) + t+1$ by \ref{C3}.
Since $\tau_i$ has length one, we have
$|\pi_i \tau_{i-1} \pi_{i-1} \cdots \tau_1 \pi_1| > n-\gamma(n)+t$.

Since $|\pi_k' \tau_{k-1} \cdots \pi_i| \in \Acc(q_i)$,
we can apply \Cref{lem:sim} and obtain an accepting run $\rho$
of length $|\pi_k' \tau_{k-1} \cdots \pi_i| \in \Acc(q_i)$ starting in $q_i$
which $t$-simulates the internal run $\pi_i$.
The prefix distance between $\rho$ and $\pi_k' \tau_{k-1} \cdots \pi_i$ (which we define as the prefix
distance between the words read along the two runs) is at most
\[
	|\pi_k' \tau_{k-1} \cdots \pi_{i+1} \tau_i| + t = n - |\pi_i \tau_{i-1} \pi_{i-1} \cdots \tau_1 \pi_1| + t \le n
	- n+\gamma(n) = \gamma(n).
\]
Hence, the prefix distance from the accepting run $\rho \tau_{i-1} \pi_{i-1} \cdots \tau_1 \pi_1$
to the run $\pi_k' \tau_{k-1} \pi_{k-1} \cdots \tau_1 \pi_1$ is also at most $\gamma(n)$.
This implies  $\pdist(\last_n(w),L) \le \gamma(n)$.
\end{proof}

We are now ready to prove \Cref{theorem:two-sided-general-HG}.

\begin{proof}[Proof of Theorem \ref{theorem:two-sided-general-HG}]
Assume that the window size is such that $\gamma(n) \ge 4(t+1)$ (recall that $4(t+1)$ is our constant $c$ from \Cref{theorem:two-sided-general-HG}).
We use the algorithm from \Cref{prop:ps-alg}, which is initialized by reading the initial window $\square^n$.
It maintains a compact summary which represents $\pi_{w,q_0}$ with probability at least 2/3
for the read stream prefix $w$.
The algorithm accepts if that compact summary is accepting.
From \Cref{prop:correctness} 
we get:
\begin{itemize}
\item If $\last_n(w) \in L$, then the algorithm accepts with probability at least $2/3$.
\item If $\pdist(\last_n(w),L) > \gamma(n)$, then the algorithm rejects with probability at least $2/3$.
\end{itemize}
This concludes the proof of \Cref{theorem:two-sided-general-HG}.
\end{proof}
From \Cref{theorem:two-sided-general-HG} we can easily deduce \Cref{theorem:two-sided}: 
Let $\gamma(n) = \epsilon n$ for some $0 < \epsilon < 1$ and let $c$ be the constant from 
\Cref{theorem:two-sided-general-HG}. Then the condition $\gamma(n) \ge c$ becomes $n \geq c/\epsilon$.
Hence, for a window size $n \geq c/\epsilon$, \Cref{theorem:two-sided-general-HG} yields a randomized
sliding window tester with two-sided error that uses space $\O(\log(n/\gamma(n)) = \O(\log(1/\epsilon))$.
For $n < c/\epsilon$ we can use a trivial sliding window tester that stores the window content explicitly using
$\O(1/\epsilon)$ bits.

\subsubsection{Sliding window testers with one-sided error}\label{sec:one-sided_ub}

In the following, we turn to sliding window testers with one-sided error
and prove \Cref{thm:false-biased}, i.e.,~we present a false-biased sliding window tester
for languages in $\bigcup(\Triv, \SF)$ with constant Hamming gap using $\O(\log \log n)$ space.
By the following lemma, it suffices to consider the cases $L \in \Triv$ and $L \in \SF$.

\begin{lemma}
	\label{lem:pt-union}
	Let $\P_1$ and $\P_2$ be randomized false-biased sliding window testers for $L_1$ and $L_2$, respectively,
	for window size $n$ with Hamming gap $\gamma(n)$.
	Then, there exists a randomized false-biased sliding window tester for $L_1 \cup L_2$ for window size $n$
	with Hamming gap $\gamma(n)$ using space $\O(s(\P_1) + s(\P_2))$.
\end{lemma}

\begin{proof}
First we reduce the error probability of $\P_i$ ($i \in \{ 1,2\}$) from $1/3$ to $1/9$ 
by running $2$ independent and parallel
copies of $\P_i$ and reject if and only if one of the copies rejects. 
Then, we run both algorithms in parallel and accept if and only if one of them accepts.
If the window belongs to $L_1 \cup L_2$ then either $\P_1$ or $\P_2$ accepts with probability $1$.
If the window $w$ satisfies $\dist(w,L_1 \cup L_2)  = \min(\dist(w,L_1), \dist(w,L_2)) > \gamma(n)$
then $\dist(w,L_i) > \gamma(n)$ for both $i \in \{1,2\}$.
Hence, both algorithms falsely accept with probability at most $1/9$ and the combined 
algorithm falsely accepts with probability at most $1/9 + 1/9 \leq 1/3$.
\end{proof}
The case of a regular trivial language is covered by the following result:
\begin{theorem}
\label{prop-trivial1}
Let $L$ be a language and $\gamma(n)$ be a function. The following statements are equivalent:
\begin{itemize}
\item $L$ is $(\gamma(n)+c)$-trivial for some number $c \in \N$.
\item There is a deterministic sliding window tester
with Hamming gap $\gamma(n)+c'$ for $L$ which uses constant space for some number $c' \in \N$.
\end{itemize}
\end{theorem}

\begin{proof}
Assume first that $L$ is $(\gamma(n)+c)$-trivial. Let $n \in \N$ be a window size.
If $L\cap \Sigma^n=\emptyset$, then the algorithm always rejects,
which is obviously correct since any active window of size $n$ has infinite Hamming distance to $L$. 
On the other hand, if $L\cap \Sigma^n \neq \emptyset$ then the Hamming distance between an arbitrary active window
of size $n$ and $L$ is at most $\gamma(n)+c$. Hence, the algorithm that always accepts achieves
a Hamming gap of $\gamma(n)+c$.

We now show the converse statement.\footnote{The converse statement is not needed for the proof of \Cref{thm:false-biased} 
but we think it is of independent interest.}
For each window size $n \in \N$ let $\P_n$ be a deterministic sliding window tester for $L$
with Hamming gap $\gamma(n)+c'$ such that the number of states of $\P_n$ is constant.
Assume that $\P_n$ has at most $s$ states for every $n$.
Let $N \subseteq \N$ be the set of all $n$ such that $L \cap \Sigma^n \neq \emptyset$.
Note that every $\P_n$ with $n \in N$ accepts a nonempty language.
Every $\P_n$ is a DFA with a most $s$ states over the fixed alphabet $\Sigma$.
The number of pairwise nonisomorphic DFAs with at most $s$ states
over the input alphabet $\Sigma$ is bounded by a fixed constant $d$.
Hence, at most $d$ nonisomorphic DFAs can appear in the list $(\P_n)_{n \in N}$.
We therefore can choose 
numbers $n_1 < n_2 < \cdots < n_e$ from $N$ with $e \leq d$ such that
for every $n \in N$ there exists a unique $i \in \{1,2,\ldots,e\}$ with $n _i \le n$ and $\P_n = \P_{n_i}$
(here and in the following we do not distinguish between isomorphic DFAs).
Let us choose for every $1 \le i \le e$ some word $u_i \in L$ of length $n_i$.
Now, take any $n \in N$ and assume that $\P_n = \P_{n_i}$ where $n_i \leq n$.
Consider any word $u \in \Sigma^* u_i$.
Since $\last_{n_i}(u) = u_i \in L$, $\P_{n_i}$ has to accept $u$.
Hence, $\P_n$ accepts all words from $\Sigma^* u_i$.
In particular, for every word $x$ of length $n-n_i$, $\P_n$ accepts $xu_i$.
This implies that $\dist(x u_i,L) \leq \gamma(n)+c'$ for all $x \in \Sigma^{n-n_i}$.
Recall that this holds for all $n \in N$ and that $N$ is the set of all lengths realized by~$L$.
Hence, if we define $c'' := \max\{n_1, \ldots, n_e\}$,
then every word $w$ of length $n \in N$ has Hamming distance at most $\gamma(n)+c'+c''$ from a word in $L$.
Therefore, $L$ is $(\gamma(n)+c)$-trivial with $c=c'+c''$. 
\end{proof}
Let us now turn to the case of a regular suffix-free language $L$.
We again consider an rDFA $\B = (Q,\Sigma,F,\delta,q_0)$
for $L$ whose nontransient SCCs have uniform period $g \ge 1$.
Since $L$ is suffix-free, $\B$ has the property that no final state can be reached from a
final state by a nonempty run.

We adapt the definition of a path description from \Cref{sec:char-log}.
In the following, a path description is a sequence
\begin{equation}
	\label{eq:pd}
	P = (q_k,a_k,p_{k-1}),C_{k-1},\dots ,(q_2,a_2,p_1),C_1,(q_1,a_1,p_0),C_0,q_0.
\end{equation}
where $C_{k-1}, \dots, C_0$ is a chain (from right to left) in the SCC-ordering of $\B$,
$p_i, q_i \in C_i$, $q_{i+1} \xleftarrow{a_{i+1}} p_i$ is a transition in $\B$
for all $0 \le i \le k-1$, and $q_k \in F$.
Each path description defines a {\em partial rDFA} $\B_P = (Q_P,\Sigma,\{q_k\},\delta_P,q_0)$
by restricting $\B$ to the state set $Q_P = \bigcup_{i=0}^{k-1}C_i \cup \{q_k\}$,
restricting the transitions of $\B$ to internal transitions from the SCCs $C_i$
and the transitions $q_{i+1} \xleftarrow{a_{i+1}} p_i$,
and declaring $q_k$ to be the only final state.
This rDFA is partial since for every state $p$ and every symbol $a \in \Sigma$
there exists at most one transition $q \xleftarrow{a} p$ in $\B_P$.
Since the number of path descriptions $P$ is finite and $\L(\B) = \bigcup_P \L(\B_P)$,
we can fix a single path description $P$ and provide a sliding window tester for $\L(\B_P)$
(we again use \Cref{lem:pt-union} here).

From now on, we fix a path description $P$ as in \eqref{eq:pd}.
The acceptance sets $\Acc_P(q)$ are defined with respect to the restricted automaton $\B_P$. 
If all $C_i$ are transient, then $\L(\B_P)$ is a singleton and we can use a trivial sliding window tester with space complexity $\O(1)$.
Now, assume the contrary and let $0 \le e \le k-1$ be maximal such that $C_e$ is nontransient. 

\begin{lemma} \label{lemma-si-ri}
There exist numbers $r_0, \ldots, r_{k-1}, s_0, \ldots, s_e \in \N$ such that the following holds:
\begin{enumerate}[label=(\roman{*}), ref=(\roman{*})]
\item For all $e+1 \le i \le k$, the set $\Acc_P(q_i)$ is a singleton. \label{item:siri1}
\item For all $0 \le i \le k-1$, every run from $q_i$ to $q_{i+1}$ has length $r_i \pmod g$. \label{item:siri2}
\item For all $0 \le i \le e$, $\Acc_P(q_i) =_{s_i} \sum_{j=i}^{k-1} r_j + g \N$.\label{item:siri3}
\end{enumerate}
\end{lemma}

\begin{proof}
Point~\ref{item:siri1} follows immediately from the definition of transient SCCs.
Let us now show \ref{item:siri2} and \ref{item:siri3}.
Let $0 \le i \le k-1$ and let $N_i$ be the set of lengths of runs of the form $q_{i+1} \xleftarrow{a_{i+1}} p_i \xleftarrow{w} q_i$ in~$\B_P$.
If $C_i$ is transient, then $N_i = \{1\}$.
Otherwise, by \Cref{lemma-alon} there exist a number $r_i \in \N$ and a cofinite set $D_i \subseteq \N$ such that $N_i = r_i + gD_i$.
We can summarize both cases by saying that there exist a number $r_i \in \N$
and a set $D_i \subseteq \N$
which is either cofinite or $D_i = \{0\}$ such that $N_i = r_i + gD_i$.
This implies point~\ref{item:siri2}.
Moreover, the acceptance sets in $\B_P$ satisfy
\[
	\Acc_P(q_i) = \sum_{j=i}^{k-1} N_j = \sum_{j=i}^{k-1} (r_j + gD_j) = \sum_{j=i}^{k-1} r_j + g \sum_{j=i}^{k-1} D_j.
\]
For all $0 \le i \le e$ we get $\Acc_P(q_i) =_{s_i} \sum_{j=i}^{k-1} r_j + g \N$ for some threshold $s_i \in \N$ (note that a nonempty sum of cofinite subsets of $\N$ is again cofinite).
\end{proof}

Let us fix the numbers $r_i$ and $s_i$ from \Cref{lemma-si-ri}.
Let $p$ be a random prime with $\Theta(\log \log n)$ bits. By choosing the $\Theta$-constant
large enough and using  \Cref{lemma-primes} (where we set $m = n$, $a = \ell$ and $b = n$) we
obtain for every $0 \le \ell < n$ the inequality
$\Pr[\ell \equiv n \pmod p] \le 1/3$.
Define the threshold 
\[ 
s = \max\{k, \sum_{j=0}^{k-1} r_j, s_0, \dots, s_e \}
\]
and for a word $w \in \Sigma^*$ define the function
$\ell_w \colon Q \to \N \cup \{\infty\}$ where
\[
	\ell_w(q) = \inf \{ \ell \in \N \mid \delta_P(\last_\ell(w),q) = q_k  \}
\]
(we set $\inf \emptyset = \infty$).
We now define an acceptance condition on $\ell_w(q)$.
If $n \notin \Acc_P(q_0)$, we always reject.
Otherwise, we accept $w$ if and only if $\ell_w(q_0) \equiv n \pmod p$.

\begin{lemma}
\label{lm:acceptingconditiononesided}
Let $n \in \Acc_P(q_0)$ be a window size with $n \ge s + |Q_P|$ and $w \in \Sigma^*$ with $|w| \ge n$.
There exists a constant $c > 0$ such that:
\begin{enumerate}[label=(\roman{*}), ref=(\roman{*})]
\item if $\last_n(w) \in \L(\B_P)$, then $w$ is accepted (in the above sense) with probability $1$, and
\item if $\pdist(\last_n(w), \L(\B_P)) > c$, then $w$ is rejected with probability at least $2/3$.
\end{enumerate}
\end{lemma}

\begin{proof} Consider a word $w \in \Sigma^*$ with $|w| \ge n$.
We consider several cases.

\medskip\noindent
{\em Case 1:} $\last_n(w) \in \L(\B_P)$. Since $\L(\B_P) \subseteq L$ is suffix-free, we have $\ell_w (q_0) \equiv n \pmod p$ and $w$ is accepted with probability $1$,
which shows statement (i) from the lemma.

\medskip\noindent
{\em Case 2:} $\last_n(w) \notin \L(\B_P)$. We then have $\ell_w (q_0) \neq n$, which yields the following two subcases.

\medskip\noindent
{\em Case 2.1:} $\ell_w(q_0) < n$.
Then, by the choice of $p$ we have $\ell_w(q_0) \not \equiv n \pmod p$ with probability at least $2/3$. 
Hence, $w$ is rejected with probability at least $2/3$.

\medskip\noindent
{\em Case 2.2:}  $\ell_w(q_0) > n$.
We will show that this implies $\pdist(\last_n(w), \L(\B_P)) \le c$ for a constant $c$. 
For this $c$, statement (ii) from the lemma then holds: if $\pdist(\last_n(w), \L(\B_P)) > c$, we must have  
$\ell_w(q_0) < n$, which by Case 2.1 implies that $w$ is rejected with probability at least $2/3$.
 
Let $\pi$ be the run of $\B_P$ on $\last_n(w)$ starting from the initial state,
and let its SCC-factorization be $\pi = \pi_m \tau_{m-1} \pi_{m-1} \cdots \tau_0 \pi_0$. 
We have $|\pi|=n$.
Since $\ell_w(q_0) > n$, the run $\pi$ can be strictly extended to a run to $q_k$ and hence we must have $m < k$. 
For all $0 \leq i \leq m$, the run $\pi_i$ is an internal run in the SCC $C_i$ from $q_i$ to $p_i$.
For all $0 \le i \le m-1$ we have $\tau_i = q_{i+1} \, a_{i+1}\, p_i$  and $|\tau_i \pi_i| \equiv r_i \pmod g$
by point~\ref{item:siri2} from \Cref{lemma-si-ri}.
We claim that there exists an index $0 \le i_0 \le m$ such that the following three properties hold:
\begin{enumerate}[ref=(\arabic{*})]
\item \label{it:i} $q_{i_0}$ is nontransient,
\item \label{it:ii} $|\pi_m \tau_{m-1} \pi_{m-1} \cdots \tau_{i_0} \pi_{i_0}| \ge s$,
\item \label{it:iii} $|\pi_m \tau_{m-1} \pi_{m-1} \cdots \tau_{{i_0}+1} \pi_{{i_0}+1}| \le s + m$ \\
(note that $|\pi_m \tau_{m-1} \pi_{m-1} \cdots \tau_{{i_0}+1} \pi_{{i_0}+1}| = 0$ is possible).
\end{enumerate}
Indeed, let $0 \le i \le m$ be the smallest integer such that $q_i$ is nontransient
(recall that $n \ge |Q_P|$ and hence $\pi$ must traverse a nontransient SCC).
Then, the run $\tau_{i-1} \pi_{i-1} \cdots \tau_0 \pi_0$ only passes transient states except for 
its last state $q_i$ and hence its length is bounded by $|Q_P|$.
Therefore, we have
\[
	|\pi_m \tau_{m-1} \pi_{m-1} \cdots \tau_i \pi_i| = n - |\tau_{i-1} \pi_{i-1} \cdots \tau_0 \pi_0| \ge n-|Q_P| \ge s.
\]
Hence, there exists $i$ satisfying properties~\ref{it:i} and~\ref{it:ii} (with $i_0$ replaced by $i$).
Let $0 \le {i_0} \le m$ be the largest integer satisfying properties~\ref{it:i} and~\ref{it:ii}.
We show that property~\ref{it:iii} holds for $i_0$.

If the run $\pi_m \tau_{m-1} \pi_{m-1} \cdots \tau_{{i_0}+1} \pi_{{i_0}+1}$ only passes transient states,
then its length is bounded by $m-{i_0} \le s + m$,  
and we are done.
Otherwise, let ${i_0}+1 \le j \le m$ be the smallest integer such that $q_j$ is nontransient.
The run $\tau_{j-1} \pi_{j-1} \cdots \tau_{{i_0}+1} \pi_{{i_0}+1}$ only passes transient states except for 
its last state $q_j$ 
and therefore it has length $j-{i_0}-1$.
By maximality of ${i_0}$, we have $|\pi_m  \tau_{m-1} \pi_{m-1} \cdots \tau_j \pi_j| < s$
and hence property~\ref{it:iii} holds:
\begin{eqnarray*}
	|\pi_m \tau_{m-1} \pi_{m-1} \cdots \tau_{{i_0}+1} \pi_{{i_0}+1}|  &=& |\pi_m \cdots \tau_j \pi_j| + |\tau_{j-1} \pi_{j-1} \cdots \tau_{{i_0}+1} \pi_{{i_0}+1}|  \\
	&< & s + j - {i_0} \le s + m.
\end{eqnarray*}
In the rest of the proof let $0 \le {i_0} \le m$ be the above index satisfying properties~\ref{it:i}-\ref{it:iii}.
Since $q_{i_0}$ is nontransient, we have $i_0 \le e$
and therefore 
\begin{equation} \label{gl-Acc_P(q_{i_0})-1}
\Acc_P(q_{i_0}) =_s \sum_{j={i_0}}^{k-1} r_j + g \N
\end{equation} 
by \Cref{lemma-si-ri}\ref{item:siri3} and $s \geq s_{i_0}$.
We claim that 
\begin{equation} \label{gl-Acc_P(q_{i_0})-2}
|\pi_m \tau_{m-1} \pi_{m-1} \cdots \tau_{i_0} \pi_{i_0}| \in \Acc_P(q_{i_0}).
\end{equation}
Since $n \ge s$ and $n \in \Acc_P(q_0) =_{s} \sum_{j=0}^{k-1} r_j + g \N$ 
we have $n \in \sum_{j=0}^{k-1} r_j + g \N$. This implies
\begin{align*}
	|\pi_m \tau_{m-1} \pi_{m-1} \cdots \tau_{i_0} \pi_{i_0}| = n - |\tau_{{i_0}-1} \pi_{{i_0}-1} \cdots \tau_0 \pi_0|  
	 \equiv n - \sum_{j=0}^{{i_0}-1} r_j 
	  \equiv \sum_{j={i_0}}^{k-1} r_j \pmod g.
\end{align*}
In addition, we have $|\pi_m \tau_{m-1} \pi_{m-1} \cdots \tau_{i_0} \pi_{i_0}| \geq s$ 
by property~\ref{it:ii}. Since $s \geq \sum_{j={i_0}}^{k-1} r_j$, we get
\[ 
|\pi_m \tau_{m-1} \pi_{m-1} \cdots \tau_{i_0} \pi_{i_0}|  \in \sum_{j={i_0}}^{k-1} r_j + g \N.
\]
Finally, we obtain \eqref{gl-Acc_P(q_{i_0})-2} from \eqref{gl-Acc_P(q_{i_0})-1}.

By \Cref{lem:sim} and \eqref{gl-Acc_P(q_{i_0})-2}, there is an accepting run $\pi'$ from $q_{i_0}$
which $t$-simulates the internal run $\pi_{i_0}$ and has length $|\pi_m \tau_{m-1} \pi_{m-1} \cdots \tau_{i_0} \pi_{i_0}|$.
Here, $t$ is a constant only depending on $\B$.
The prefix distance between the runs
$\pi = \pi_m \tau_{m-1} \pi_{m-1} \cdots \tau_0 \pi_0$ and
$\pi' \tau_{i_0-1} \pi_{{i_0}-1} \cdots \tau_0 \pi_0$ 
is at most $t$ in case $i_0 = m$, and at most
\begin{eqnarray*}
	|\pi_m \tau_{m-1} \pi_{m-1} \cdots \tau_{i_0}| + t &=& |\pi_m \tau_{m-1} \pi_{m-1} \cdots \tau_{{i_0}+1} \pi_{{i_0}+1}| + 1 + t \\
	& \le & 1 + s + m + t =: c
\end{eqnarray*}
in case $i_0 < m$ due to property~\ref{it:iii}:
Hence, the prefix distance between $\last_n(w)$ and $\L(\B_P)$ is bounded by the constant~$c$. As explained before, statement (ii)
of the lemma then holds.
\end{proof}

\begin{proof}[Proof of Theorem \ref{thm:false-biased}]
Let $n \in \N$ be the window size. By the previous discussion,
it suffices to give a randomized sliding window tester (with the properties stated in \Cref{thm:false-biased})
for a fixed partial automaton $\B_P$.
Assume $n \ge s + |Q|$, otherwise a trivial tester can be used.
If $n \notin \Acc_P(q_0)$, the tester always rejects. Otherwise, the tester picks a random prime $p$ with
$\Theta(\log \log n)$ bits and maintains $\ell_w(q) \bmod p$ for all $q \in Q_P$,
where $w$ is the stream read so far, which requires $\O(\log \log n)$ bits.
When a symbol $a \in \Sigma$ is read, we can update $\ell_{wa}$ using $\ell_w$: 
If $q = q_k$, then $\ell_{wa}(q) = 0$, otherwise $\ell_{wa}(q) = 1 + \ell_w(\delta_P(a,q)) \bmod p$
where $1 + \infty = \infty$. The tester accepts if $\ell_w(q_0) \equiv n \pmod p$. 
\Cref{lm:acceptingconditiononesided} guarantees that the tester is false-biased.
\end{proof}
Note that in contrast to the randomized sliding window algorithm from \Cref{sec-random-suffix-free}, the 
randomized sliding window tester from this section only uses the modulo counting technique; Bernoulli counters are not needed.
As a consequence, the tester only has to make a random choice at the beginning (where the prime $p$ is chosen) before the first input symbol arrives. Then it 
continues deterministically.

\subsection{Lower bounds}

In this section we prove our lower bounds. In \Cref{sec-lb-nontriv} we will prove  \Cref{theorem:deterministic_lb} and \Cref{lemma-loglog(n-gamma)}.
\Cref{thm:lb-false-biased} and \Cref{lemma-lb-log(n)-log(gamma(n))} will be shown in \Cref{sec-lb-union-nontriv-sf}.

\subsubsection{Regular nontrivial  languages} \label{sec-lb-nontriv}

In this section, we prove \Cref{theorem:deterministic_lb} and \Cref{lemma-loglog(n-gamma)}.
For this, we first have to study regular trivial languages in more detail.
We will also show a result of independent interest: every regular $o(n)$-trivial language $L$ is already trivial (i.e.,~$\O(1)$-trivial); see \Cref{triv-alon}.

Given $i,j \ge 0$ and a word $w$ of length at least $i+j$ we define
$\cut_{i,j}(w) = y$ such that $w = xyz$, $|x| = i$ and $|z| = j$.
If $|w| < i+j$, then $\cut_{i,j}(w)$ is undefined.
For a language $L$ we define the {\em cut-language} $\cut_{i,j}(L) = \{ \cut_{i,j}(w) \mid w \in L, |w| \ge i+j \}$.

\begin{lemma} \label{lemma-fin-cut-lang}
	If $L$ is regular, then there are finitely many languages $\cut_{i,j}(L)$.
\end{lemma}

\begin{proof}
Let $\A = (Q,\Sigma,q_0,\delta,F)$ be a DFA for $L$.
Given $i,j \ge 0$, let $I$ be the set of states reachable from $q_0$ via $i$ symbols
and let $F'$ be the set of states from which $F$ can be reached via $j$ symbols.
Then, the NFA $\A_{i,j} = (Q,\Sigma,I,\delta,F')$ recognizes $\cut_{i,j}(L)$.
Since there are at most $2^{2|Q|}$ such choices for $I$ and $F'$,
the number of languages of the form $\cut_{i,j}(L)$ must be finite.
\end{proof}

\begin{lemma} \label{cuts}
	If $\cut_{i,j}(L)$ is a length language for some $i,j \ge 0$, then $L$ is trivial.
\end{lemma}

\begin{proof}
	Assume that $\cut_{i,j}(L)$ is a length language.
	Let $n \in \N$ such that $L \cap \Sigma^n \neq \emptyset$.
	We claim that $\dist(w,L) \le i+j$ for all $w \in \Sigma^n$. If $n < i+j$ this is clear.
	So, assume that  $n \ge i+j$. 
	Let $w \in \Sigma^n$ and $w' \in L \cap \Sigma^n$. Then, $\cut_{i,j}(w') \in \cut_{i,j}(L)$
	and hence also $\cut_{i,j}(w) \in \cut_{i,j}(L)$.
	Therefore, there exist $x \in \Sigma^i$ and $z \in \Sigma^j$ such that $x \, \cut_{i,j}(w) \, z \in L$
	satisfies $\dist(w, x \, \cut_{i,j}(w) \, z) \le i+j$.
\end{proof}

The {\em restriction} of a language $L$ to a set of lengths $N \subseteq \N$
is $L|_N = \{ w \in L \mid |w| \in N \}$.
A language~$L$ {\em excludes a word $w$ as a factor} if $w$ is not a factor of any word in $L$.
A simple but important observation is that if $L$ excludes $w$ as a factor
and $v$ contains $k$ disjoint occurrences of $w$,
then $\dist(v,L) \ge k$: If we change at most $k-1$ many symbols in $v$, then the resulting
word $v'$ must still contain $w$ as a factor and hence $v' \notin L$. 

\begin{lemma}
	\label{prop-nontriv}
	Let $L$ be regular. If for all $i, j \ge 0$, $\cut_{i,j}(L)$ is not a length language,
	then there exists an arithmetic progression  $N = d+e\N$  such that 
	the restriction $L|_N$ is infinite and excludes a factor. 
\end{lemma}

\begin{proof}
	First notice that $\cut_{i,j}(L)$ determines $\cut_{i+1,j}(L)$ and $\cut_{i,j+1}(L)$: we have
	$\cut_{i+1,j}(L) = \cut_{1,0}(\cut_{i,j}(L))$ and similarly for $\cut_{i,j+1}(L)$.
	Since the number of cut-languages $\cut_{i,j}(L)$ is finite by \Cref{lemma-fin-cut-lang}, there exist numbers $i \ge 0$ and $d > 0$ such that
	$\cut_{i,0}(L) = \cut_{i+d,0}(L)$. Hence, we have $\cut_{i,j}(L) = \cut_{i+d,j}(L)$ for all $j \ge 0$. 
	By the same argument, there exist numbers $j \ge 0$ and $e > 0$ such that
	$\cut_{i,j}(L) = \cut_{i,j+e}(L) = \cut_{i+d,j}(L) = \cut_{i+d,j+e}(L)$, which implies
	$\cut_{i,j}(L) = \cut_{i,j+h}(L) = \cut_{i+h,j}(L) = \cut_{i+h,j+h}(L)$ for some $h>0$ (we can take $h = ed$).
	This implies that $\cut_{i,j}(L)$ is closed under removing prefixes and suffixes of length $h$.
	
	By assumption $\cut_{i,j}(L)$ is not a length language,
	i.e.,~there exist words $y' \in \cut_{i,j}(L)$ and $y \notin \cut_{i,j}(L)$
	of the same length $k$.
	Let $N = \{ k + i + j + hn \mid n \in \N \}$.
	For any $n \in \N$ the restriction $L|_N$ contains a word of length $k+i+j+hn$
	because $y' \in \cut_{i,j}(L) = \cut_{i+hn,j}(L)$.
	This proves that $L|_N$ is infinite.
	
	Let $u$ be an arbitrary word which contains for every remainder $0 \le r \le h-1$
	an occurrence of $y$ as a factor starting at a position which is congruent to $r$ mod $h$ (these occurrences do not have to be disjoint).
	We claim that $L|_N$ excludes $a^i u a^j$ as a factor where $a$ is an arbitrary symbol.
	Assume that there exists a word $w \in L|_N$ which contains $a^i u a^j$ as a factor.
	Then, $\cut_{i,j}(w)$ contains $u$ as a factor, has length $k+hn$ for some $n \ge 0$,
	and belongs to $\cut_{i,j}(L)$.
	Therefore, $\cut_{i,j}(w)$ also contains $h$ many occurrences of $y$, one per remainder $0 \le r \le h-1$.
	Consider the occurrence of $y$ in $\cut_{i,j}(w)$ which starts at a position that is divisible by $h$,
	i.e.,~we can factorize $\cut_{i,j}(w) = xyz$ such that $|x|$ is a multiple of $h$.
	Since $|\cut_{i,j}(w)|=k+hn$ and $|y|=k$, then $|z|$ is also a multiple of~$h$.
	Therefore, $y \in \cut_{i+|x|,j+|z|}(L) = \cut_{i,j}(L)$, which is a contradiction.
\end{proof}

\begin{lemma} \label{lemma-reg-nontriv}
If $L \in \Reg \setminus \Triv$ then there 
are a restriction $L|_N$ that excludes some factor $w_f$ and words $x,y,z$ such that $|y|>0$ and 
$x y^* z \subseteq L|_N$.
\end{lemma}

\begin{proof}
By \Cref{cuts}, $\cut_{i,j}(L)$ is not a length language for all $i, j \ge 0$.
Let $N$ be the set of lengths from \Cref{prop-nontriv}
such that $L|_N$ is infinite and excludes some factor $w_f$. 
Since $N$ is an arithmetic progression, $L|_N$ is regular. 
Let $\A = (Q,\Sigma,q_0,\delta,F)$ be a DFA for $L|_N$.
Since $\L(\A)$ is infinite, there must exist words $x,y,z$ such that $y \neq \varepsilon$ and 
for $\delta(q_0,x) = q$ we have $\delta(q, y) = q$ and $\delta(q,z) \in F$.
\end{proof}
Before we prove \Cref{theorem:deterministic_lb} let us first show the following result of independent interest:

\begin{theorem} 
	\label{triv-alon}
	For every regular language $L$, the following statements are equivalent:
	\begin{enumerate}[label=(\roman{*}), ref=(\roman{*})]
		\item $L$ is trivial.
		\item $L$ is $o(n)$-trivial. \label{item:eps}
		\item $\cut_{i,j}(L)$ is a length language for some $i, j \ge 0$. \label{item:len}
	\end{enumerate}
\end{theorem}

\begin{proof}
	If $\cut_{i,j}(L)$ is a length language then $L$ is trivial by \Cref{cuts},
	and thus also $o(n)$-trivial. 
	It remains to show the direction \ref{item:eps} to \ref{item:len}.
	Assume that $L$ is $o(n)$-trivial.
	If \ref{item:len} would not hold then
	some infinite restriction $L|_N$ of $L$ excludes a factor $w_f$ by \Cref{prop-nontriv}.
	Hence, if $n \in N$ is a length with $L|_N \cap \Sigma^n \neq \emptyset$,
	then any word $v$ of length $n$
	which contains at least $\lfloor n/|w_f| \rfloor$
	many disjoint occurrences of $w_f$, has distance $\dist(v,L) \ge \lfloor n/|w_f| \rfloor$ to $L$.
	Then, $L$ is not $o(n)$-trivial, which is a contradiction.
\end{proof}

\begin{proof}[Proof of Theorem \ref{theorem:deterministic_lb}]
We will prove the two lower bounds in \Cref{theorem:deterministic_lb}
for the more general class of nondeterministic and co-nondeterministic sliding window testers.
A {\em nondeterministic} sliding window tester for a language $L$
and window size $n \in \N$ with Hamming gap $\gamma(n)$
is a nondeterministic finite automaton $\P_n$ such that for all input words $w \in \Sigma^*$ we have the following (recall from \Cref{sec-automata} that a successful
	run is a run from an initial state to a final state):
\begin{itemize}
	\item If $\last_n(w) \in L$, then there is at least one successful run of $\P_n$ on $w$.
	\item If $\dist(\last_n(w),L) > \gamma(n)$, then there is no successful run of $\P_n$ on $w$.
\end{itemize}
In contrast, $\P_n$ is co-nondeterministic if for all $w \in \Sigma^*$ we have:
\begin{itemize}
	\item If $\last_n(w) \in L$, then all runs of $\P_n$ on $w$ that start in an initial state are successful.
	\item If $\dist(\last_n(w),L) > \gamma(n)$, then there is a nonsuccessful run of $\P_n$ on $w$ that starts in an initial state.
\end{itemize}
The space complexity of $\P_n$ is $\log |\P_n|$.
Clearly, every true-biased (resp., false-biased) sliding window tester
is a nondeterministic (resp., co-nondeterministic) one.

Assume that $L \in \Reg\setminus\Triv$.
We will prove an $\log n - \O(1)$ lower bound for nondeterministic sliding window testers,
and hence also for true-biased and deterministic sliding window testers.
By the power set construction one can transform a co-nondeterministic sliding window tester
with $m$ memory states into an equivalent nondeterministic (and in fact, even deterministic) sliding window tester
with $2^m$ memory states. Hence, an $\log n - \O(1)$ lower bound for nondeterministic sliding window testers
immediately yields an $\log (\log n - \O(1)) \geq \log \log n - \O(1)$ lower bound
for co-nondeterministic sliding window testers,
and hence also for false-biased sliding window testers.

By \Cref{lemma-reg-nontriv} there 
are a restriction $L|_N$ that excludes some factor $w_f$ and words $x,y,z$ such that $|y|>0$ and 
$x y^* z \subseteq L|_N$. Let $c = |w_f| >0$ and choose $0 < \epsilon < 1/c$. Moreover,
let $d = |xz|$ and $e = |y|>0$, which satisfy $d+e\N \subseteq N$.
Recall that every word $v$ that contains $k$ disjoint occurrences of $w_f$ has Hamming distance
at least $k$ from any word in $L|_N$. 

Fix a window size $n\in N$
and consider a nondeterministic sliding window tester $\P_n$ for $L$
and window size $n$ with Hamming gap $\epsilon n$.
Define for $k \geq 0$ the input stream
\[
	v_k = w_f^n x y^k z .
\]
Let $\alpha = c \epsilon < 1$.
If $0 \le k \le \lfloor \frac{(1-\alpha) n - c - d}{e} \rfloor$, then the suffix of $v_k$ of length $n$ 
contains at least
\[
\bigg\lfloor\frac{n - d - e k}{c} \bigg\rfloor \geq 
\bigg\lfloor\frac{n - d - (1-\alpha) n + c + d}{c} \bigg\rfloor = 
\bigg\lfloor\frac{\alpha n + c}{c} \bigg\rfloor = \lfloor \epsilon n + 1 \rfloor > \epsilon n
\]
many disjoint occurrences of $w_f$.
Hence, $\P_n$ has no successful run on an input stream $v_k$ with
$0 \le k \le \lfloor \frac{(1-\alpha) n - c - d}{e} \rfloor$.

Assume now that the window size $n$ satisfies $n \geq d$ and $n \equiv d \pmod e$.
Write $n = d + l e$ for some $l \geq 0$.
We have $l = \frac{n-d}{e} > \lfloor \frac{(1-\alpha) n - c - d}{e} \rfloor$.
The suffix of $v_l = w_f^n x y^l z$ of length $n$ is $x y^l z \in L|_N$.
Therefore, there exists a successful run $\pi$ of $\P_n$ on $v_l$.
Let $m$ be the number of states of $\P_n$.
For $0 \leq i \leq l$ let $p_i$ be the state on the run $\pi$ that is reached
after the prefix $w_f^n x y^i$ of $v_l$.

In order to deduce a contradiction, let us assume that
$m \leq \lfloor \frac{(1-\alpha) n - c - d}{e} \rfloor$.
Then, there must exist numbers $i$ and $j$ with $0 \leq i < j \leq \lfloor \frac{(1-\alpha) n - c - d}{e} \rfloor$
such that $p_i = p_j =: p$. By cutting off cycles at $p$ from the run $\pi$ and repeating this,
we finally obtain a run of $\P_n$ on 
an input stream  $v_k = w_f^n x y^k z$ with $k \leq \lfloor \frac{(1-\alpha) n - c - d}{e} \rfloor$.
This run is still successful. But this contradicts our previous observation that
$\P_n$ has no successful run on an input stream $v_k$ with
$0 \le k \le \lfloor \frac{(1-\alpha) n - c - d}{e} \rfloor$.
We conclude that $\P_n$ must have more than $\lfloor \frac{(1-\alpha) n - c - d}{e} \rfloor$ states.
This implies
\[
s(\P_n) \geq \log \bigg( \frac{(1-\alpha) n - c - d}{e} \bigg) \geq \log n - \O(1) ,
\]
which proves the theorem.
\end{proof}

For the proof of \Cref{lemma-loglog(n-gamma)} we need a promise variant of the communication problem $\mathrm{EQ}_m$ (see \Cref{sec-CC}).
With $\mathrm{EQ}^{\geq}_m$  we denote the following promise communication problem: Alice's (resp., Bob's) input is a number $i \in \{1,\ldots,m\}$ 
(resp., $j \in \{1,\ldots,m\}$) and the promise is that $i \geq j$ (i.e., we do not care about the output of the protocol in case $i < j$). 
If $i = j$ then Bob's final output must be $1$ and if $i > j$ then Bob's final output must be $0$.
We claim that the randomized one-way communication complexity of $\mathrm{EQ}^{\geq}_m$ is $\Omega(\log \log m)$. 

Since the randomized one-way communication complexity of $\mathrm{EQ}_m$ is $\Omega(\log \log m)$ by \Cref{thm:coco}, it suffices to 
show that a randomized one-way protocol for $\mathrm{EQ}^{\geq}_m$ with cost $c(m)$ and error probability $\lambda$ yields a 
randomized one-way protocol for $\mathrm{EQ}_m$ with cost $2 c(m)$ and error probability $2\lambda$. This is easy to see: Assume that 
$P_m$ is a randomized one-way protocol for $\mathrm{EQ}^{\geq}_m$ with cost $c(m)$. To get a 
randomized one-way protocol for $\mathrm{EQ}_m$, Alice and Bob run two copies of $P_m$, one on inputs $i,j$ and the other one on inputs
$m-i, m-j$ in parallel and with independent random bits.
If both copies of $P_m$ yield output $1$ then Bob returns $1$. In all other cases, Bob returns $0$.
If $i=j$ then this combined protocol returns $1$ with probability at least $1-2\lambda$.
On the other hand if $i > j$ or $i < j$ then the protocol returns $0$ with probability at least $1-\lambda$.

\begin{proof}[Proof of Lemma \ref{lemma-loglog(n-gamma)}]
Let $L$ be a language in $L \in \Reg \setminus \Triv$.
By \Cref{lemma-reg-nontriv} there 
are a restriction $L|_N$ that excludes some factor $w_f$ and words $x,y,z$ such that $|y|>0$ and 
$x y^* z \subseteq L|_N$.

Let $b = |y|$ and $c = |xz|$. Note in the following that $b$, $c$, and $|w_f|$ are constants.
Fix a window size $n\in N$ such that 
that $n-c$ is a multiple of $b$. Since $x y^*z \subseteq L|_N$, we have $n \in N$.
Define the word $v = u w_f^k$ where $k = \lfloor \gamma(n) \rfloor + 1 > \gamma(n)$ 
and $u$ is a word of length at most $b-1$ such that $|v|$ is a multiple of $b$. Let $l \in \N$ be such that $|v| = b \cdot l$. Since $b$ divides $n-c$ we can write
$n-c = (m \cdot l + r) \cdot b$ for $m \in \N$ and $0 \le r \le l-1$.
Choose the constant $\epsilon$ from the theorem statement such that
$0 < \epsilon < \frac{1}{|w_f|}$ and hence $\gamma(n) \leq \epsilon n$. 
 Assuming $n$ is large enough, we obtain 
$k =  \lfloor \gamma(n) \rfloor + 1 \leq \frac{1}{|w_f|} (n-b-c)$, i.e., 
$b \cdot l = |v| \leq b+ k \cdot |w_f|  \leq n-c$ and hence $m \geq 1$.
Moreover,  we have $l = |v|/b = \Theta(\gamma(n))$ and therefore 
$$
m = \frac{n-c}{b \cdot l} - \frac{r}{l} = \Theta(n/\gamma(n)).
$$
Consider now a randomized sliding window tester $\P_n$ with two-sided error for $L$ and window size $n$
with Hamming gap $\gamma(n)$.
We show that from $\P_n$ we can obtain a randomized one-way protocol for $\mathrm{EQ}^{\geq}_m$.

Alice produces from her input $i \in \{1,\ldots,m\}$ the word 
$v x y^{r+ (m-i) l}$.
She then runs $\P_n$ on this word and sends the memory state to Bob.
Bob continues the run of the randomized sliding window tester, starting from the transferred memory state,
with the input stream $y^{j l} z$, where $j \in \{1,\ldots,m\}$ is his input.
He obtains the memory state reached after the input $v x y^{r+(m-i+j) l} z$.
Finally, Bob outputs the  answer given by the randomized sliding window tester.
If $i = j$, then $\last_n(v x y^{r+(m-i+j) l} z) = x y^{m \cdot l + r} z  \in L$ and Bob accepts with high probability.
On the other hand, if $i > j$, then $\last_n(v x y^{r+ (m-i+j) l} z)$ contains $v$ (recall that $|v| = b \cdot l$ and hence 
$|x y^{r+(m-i+j) l} z| \leq n-|v|$). Since $v$ contains strictly more than $\gamma(n)$ disjoint occurrences of $w_f$ (an excluded factor of $L|_N$),
we have $\dist(\last_n(w),L) > \gamma(n)$ (here it is important that $n \in N$).
Thus, Bob rejects with high probability. We therefore have a correct protocol for $\mathrm{EQ}^{\geq}_m$.

Since the randomized one-way communication complexity of $\mathrm{EQ}^{\geq}_m$ is $\Omega(\log \log m)$ we finally obtain
$s(\P_n) = \Omega(\log \log m) = \Omega(\log\log(n/\gamma(n))) $.
\end{proof}

\subsubsection{Regular languages that are not finite unions of suffix-free and trivial languages}
\label{sec-lb-union-nontriv-sf}

In this section, we show the lower bounds from \Cref{thm:lb-false-biased} and \Cref{lemma-lb-log(n)-log(gamma(n))}.
We start with the following observation.

\begin{lemma}
	\label{lem:sf-ex}
	Every  suffix-free  language excludes a factor.
\end{lemma}

\begin{proof}
Let $\B=(Q,\Sigma,F,\delta,q_0)$ be an rDFA for $L$. Since $L$ is suffix-free,
we can assume that there exists a unique sink state $q_{\mathit{fail}} \notin F$,
i.e.,~$\delta(a, q_{\mathit{fail}}) = q_{\mathit{fail}}$ for all $a \in \Sigma$,
which is reachable from all states.
We construct a word $w_f \in \Sigma^*$ such that $\delta(p, w_f) = q_{\mathit{fail}}$ for all $p \in Q$.
Let $p_1, \ldots, p_m$ be an enumeration of all states in $Q \setminus \{q_{\mathit{fail}}\}$.
We then construct inductively words $w_0,w_1, \ldots, w_m \in \Sigma^*$ such that 
for all $0 \leq i \leq m$ and $1 \leq j \leq i$:  $\delta(w_i, p_j) = q_{\mathit{fail}}$.
We start with $w_0 = \eps$.
Assume that $w_i$ has been constructed for some $i < m$.
There is a word $x$ such that that $\delta( x, \delta(w_i, p_{i+1})) = q_{\mathit{fail}}$. We set $w_{i+1} = xw_i$.
Then, $\delta(w_{i+1}, p_{i+1}) = \delta(xw_i, p_{i+1})= q_{\mathit{fail}}$ and 
$\delta(w_{i+1},p_j) = \delta(xw_i,p_j)= \delta(x,q_{\mathit{fail}}) = q_{\mathit{fail}}$ for $1 \leq j \leq i$.
Finally, we define $w_f = w_m$.
\end{proof}

\begin{lemma}
\label{lem:not-sf}
Every regular language $L$ satisfies one of the following properties:\footnote{It is not hard to see that the two properties exclude each other, but this is not needed for our further
consideration.}
\begin{itemize}
\item $L \in \bigcup(\Triv, \SF)$ 
\item $L$ has a restriction $L|_N$ which excludes some factor
and contains $y^*z$ for some $y,z \in \Sigma^*$, $|y| > 0$.
\end{itemize}
\end{lemma}

\begin{proof}
	Let $\B = (Q,\Sigma,F,\delta,q_0)$ be an rDFA for $L$.
	Let $\B_r = (Q,\Sigma,F_r,\delta,q_0)$ where $F_r$ is the set of nontransient final states
	and $\B_q = (Q,\Sigma,\{q\},\delta,q_0)$ for $q \in Q$.
	We can decompose $L$ as a union of $L_r = \L(\B_r)$ and all languages $\L(\B_q)$
	over all transient states $q \in F$.
	Notice that $\L(\B_q)$ is suffix-free for all transient $q \in F$
	since any run to $q$ cannot be prolonged to another run to $q$.
	If $L_r$ is trivial, then $L$ satisfies the first property.
	If $L_r$ is nontrivial, then by \Cref{cuts} and \Cref{prop-nontriv}
	there exists an arithmetic progression $N = a + b\N$ such that
	$L_r|_N$ is infinite and excludes some word $w \in \Sigma^*$ as a factor.
	Let $z \in L_r|_N$ be any word. Since some nontransient final state $p$
	is reached in $\B_r$ on input $z$,
	there exists a word $y$ which leads from $p$ back to $p$.
	We can ensure that $|y|$ is a multiple of $b$ by replacing $y$ by $y^b$.
	Then, $y^*z \subseteq L_r|_N \subseteq L|_N$.
	Furthermore, since each language $\L(\B_q)$ excludes some factor $w_q$ by \Cref{lem:sf-ex},
	the language $L|_N \subseteq L_r|_N \cup \bigcup_q \L(\B_q)$
	excludes any concatenation of $w$ and all words $w_q$ as a factor.
\end{proof}

\begin{proof}[Proof of Theorem \ref{thm:lb-false-biased}]
Let $L \in \Reg \setminus \bigcup(\Triv, \SF)$.
By \Cref{lem:not-sf}, $L$ has a restriction $L|_N$ which excludes some factor $w_f$
and contains $y^*z$ for some $y,z \in \Sigma^*$, $|y| > 0$.
Let $c = |w_f| \geq 1$. We choose $0 < \epsilon < 1/c$.
Let $d = |z|$ and $e = |y|$.
Fix a window size $n\in N$ and 
define for $k \geq 0$ the input stream 
$v_k = w_f^n y^k z$.

We show the  lower bound of $\log n - \O(1)$  for co-nondeterministic sliding window testers.
Consider a co-nondeterministic sliding window tester $\P_n$ for $L$ and window size $n$ with Hamming gap $\epsilon n$.
Let $\alpha = c \epsilon < 1$ and $r= \lfloor \frac{(1-\alpha) n - c - d}{e} \rfloor$.
If $0 \le k \le r$, then the suffix of $v_k$ of length $n$  contains at least
\[
\bigg\lfloor\frac{n - d - e k}{c} \bigg\rfloor \geq 
\bigg\lfloor\frac{n - d - (1-\alpha) n + c + d}{c} \bigg\rfloor = 
\bigg\lfloor\frac{\alpha n + c}{c} \bigg\rfloor = \lfloor \epsilon n + 1 \rfloor  > \epsilon n
\]
many disjoint occurrences of $w_f$.
Hence, $\P_n$ must reject the input stream $v_k$ for $0 \le k \le r$,
i.e.,~there is a run of $\P_n$ on $v_k$ 
that starts in an initial state and ends in a nonfinal state.
Consider such a run $\pi$ for $v_r$.
For $0 \leq i \leq r$ let $p_i$ be the state in $\pi$ that is reached after the prefix $w_f^n y^i$ of $v_r$.
Let now $m$ be the number of states of $\P_n$ and assume $m\le r$.
There must exist numbers $i$ and $j$ with $0 \leq i < j \leq r$
such that $p_i = p_j =: p$.
It follows that there is a $\P_n$-run on $y^{j-i}$ that starts and ends in state $p$.
Using that cycle we can now prolong the run $\pi$,
i.e.,~for all $t\ge 0$ there is a run of $\P_n$ on $v_{r+(j-i)\cdot t}=w_f^n y^{r+(j-i)\cdot t}z$ 
that starts in an initial state and ends in a nonfinal state.

Assume now that the window size satisfies $n \geq d$ and $n \equiv d \pmod e$. Write $n = d + l e$ for some $l \geq 0$.
Each $n$ with this property satisfies $n\in N$ since the word $y^lz$ belongs to $L|_N$.
We have $l  = \frac{n-d}{e} > \lfloor \frac{(1-\alpha) n - c - d}{e} \rfloor = r$. For every $k \ge l$,
the suffix of $v_k = w_f^n y^k z$ of length $n$ is $y^l z \in L$.
Therefore, $\P_n$ accepts $v_k$, i.e.,~for all $k\ge l$,
every run of $\P_n$ on $v_k$ that starts in an initial state has to end in a final state.
This contradicts our observation that for all $t\ge 0$ there is a run of $\P_n$ on $v_{r+(j-i)\cdot t}$ that goes from an initial state to a 
nonfinal state. We conclude that $\P_n$ has at least $r+1 \geq \frac{(1-\alpha) n - c - d}{e}$ states.
It follows that 
\[
s(\P_n) \geq  \log \bigg( \frac{(1-\alpha) n - c - d}{e} \bigg) \geq \log n - \O(1).
\]
This proves the theorem.
\end{proof}
Finally, we prove \Cref{lemma-lb-log(n)-log(gamma(n))}.

\begin{proof}[Proof of Lemma \ref{lemma-lb-log(n)-log(gamma(n))}]
Let $L$ be a language in $\Reg \setminus  \bigcup(\Triv, \SF)$. By \Cref{lem:not-sf},
$L$ has a restriction $L|_N$ which excludes some factor $w_f$
and contains $y^*z$ for some $y,z \in \Sigma^*$, $|y| > 0$.
Let $b = |y|$ and $c = |z|$. Note in the following that $b$, $c$, and $|w_f|$ are constants.
Choose a window size $n \ge c$ such that $n-c$ is a multiple of $b$. Since $y^*z \subseteq L|_N$, we have $n \in N$.
Define the word $v = u w_f^k$ where $k = \lfloor \gamma(n) \rfloor + 1 > \gamma(n)$ 
and $u$ is any word of length at most $b-1$ such that $|v|$ is a multiple of $b$. Let $l \in \N$ be such that $|v| = b \cdot l$. Since $b$ divides $n-c$ we can write
$n-c = (m \cdot l + r) \cdot b$ for $m \in \N$ and $0 \le r \le l-1$.
Choose the constant $\epsilon$ from the theorem statement such that
$0 < \epsilon < \frac{1}{|w_f|}$ and hence $\gamma(n) \leq \epsilon n$. 
 Assuming $n$ is large enough, we obtain 
$k =  \lfloor \gamma(n) \rfloor + 1 \leq \frac{1}{|w_f|} (n-b-c)$, i.e., 
$b \cdot l = |v| \leq b+ k \cdot |w_f|  \leq n-c$ and hence $m \geq 1$.
Moreover,  $l = |v|/b = \Theta(\gamma(n))$ ($b$ and $|w_f|$ are constants) and therefore 
$$
m = \frac{n-c}{b \cdot l} - \frac{r}{l} = \Theta(n/\gamma(n)).
$$
Consider now a randomized sliding window tester $\P_n$ with two-sided error for $L$ and window size $n$
with Hamming gap $\gamma(n)$.
We show that from $\P_n$ we can obtain a randomized one-way protocol for $\mathrm{GT}_m$ (the greater-than-function on the interval $\{1,\ldots,m\}$). Recall that $m \geq 1$.

Alice produces from her input $i \in \{1,\ldots,m\}$ the word 
$v y^{r+ (m-i) l}$.
She then runs $\P_n$ on this word and sends the memory state to Bob.
Bob continues the run of the randomized sliding window tester, starting from the transferred memory state,
with the input stream $y^{j l} z$, where $j \in \{1,\ldots,m\}$ is his input.
He obtains the memory state reached after the input $v y^{r+(m-i+j) l} z$.
Finally, Bob outputs the negated answer given by the randomized sliding window tester.
If $i \le j$, then $\last_n(v y^{r+(m-i+j) l} z) = y^{m \cdot l + r} z  \in L$ and Bob rejects with high probability.
On the other hand, if $i > j$, then $\last_n(v y^{r+ (m-i+j) l} z)$ contains $v$ (recall that $|v| = b \cdot l$ and hence 
$|y^{r+(m-i+j) l} z| \leq n-|v|$). Since $v$ contains strictly more than $\gamma(n)$ disjoint occurrences of $w_f$ (an excluded factor of $L|_N$),
we have $\dist(\last_n(w),L) > \gamma(n)$ (here it is important that $n \in N$).
Thus, Bob accepts with high probability. We therefore have a correct protocol for $\mathrm{GT}_m$.

Since the randomized one-way communication complexity of $\mathrm{GT}_m$ is $\Omega(\log m)$ (\Cref{thm:coco}) we finally obtain
$s(\P_n) = \Omega(\log m) = \Omega(\log(n/\gamma(n)))$.
\end{proof}
For example, if $\gamma(n) \le n^c$ for some $0 < c < 1$ then the lower bound in \Cref{lemma-lb-log(n)-log(gamma(n))} is $\Omega(\log n)$.
Note that if $L \in \bigcup(\Triv,\SF)$ then the lower bound of \Cref{lemma-lb-log(n)-log(gamma(n))} does not hold anymore;
this follows from \Cref{thm:false-biased}.

Note the similarity between the proofs of  \Cref{lemma-lb-log(n)-log(gamma(n))} and \Cref{lemma-loglog(n-gamma)}. The difference
is that the word $x$ in the proof of \Cref{lemma-loglog(n-gamma)} is not present in the proof of \Cref{lemma-lb-log(n)-log(gamma(n))}.
The possibly non-empty $x$ in the proof of \Cref{lemma-loglog(n-gamma)} only allows a reduction from $\mathrm{EQ}^{\geq}_m$,
whereas the empty $x$ in the proof of \Cref{lemma-lb-log(n)-log(gamma(n))} allows a reduction from $\mathrm{GT}_m$, which
yields a larger lower bound.

\section{Conclusion and future work}

In this paper we precisely determined the space complexity of regular languages in the sliding window model
in the following settings: deterministic, randomized, deterministic property testing, and randomized property testing.
Two important restrictions that made our results possible but that also
 limit their applicability are the following:
\begin{itemize}
\item Our sliding window algorithms only answer Boolean queries (does the window content belong to a language or not?).
In many applications one wants to compute a certain non-Boolean value, e.g.~the number of $1$'s in the window.
This leads to the question whether our automata theoretic framework for sliding window problems can be extended
to non-Boolean queries. Weighted automata \cite{DrosteKV09} or cost register automata \cite{AlurDDRY13} could be a suitable framework for such an endeavor. 
Another interesting problem in this context is to maintain the distance (e.g. the Hamming distance or edit distance) between the sliding window and a fixed language $L$.
\item The incoming data values in our model are from a fixed finite alphabet. In many practical situations the incoming
data values are from an infinite domain (at least on an abstract level) like the natural numbers or real numbers.
Again, the question arises, whether our automata theoretic approach can be extended to such a setting. 
A popular automata model for words over an infinite alphabet (which are known as data words in this context)
are {\em register automata}, which are also known as finite memory automata \cite{KaminskiF94,NevenSV04}.
In the context of sliding window streaming, deterministic register automata (DRA for short) \cite{FrancezK03} 
might be a good starting point. Benedikt, Ley, and Puppis
\cite{BenediktLP10} proved a Myhill-Nerode-like theorem that characterizes
the class of data languages recognized by DRA for the case that the underlying relational structure $\mathcal{A}$ 
on the data values is either $(D,=)$ (where $=$ denotes the equality relation) or $(D,<)$ for a strict linear order $<$.
As a byproduct of this characterization, they obtain a minimal DRA for any DRA-recognizable language. This DRA is minimal in a very strong 
sense: at the same time it has the minimal number of states and the minimal number of registers
among all equivalent DRA. Using these minimal DRA, one can define space-optimal streaming algorithms
for data languages analogously to the case of words over a finite alphabet. This yields a starting point for studying
space complexity classes for streaming algorithms over infinite domains. 
\end{itemize}

\subsubsection*{Acknowledgement.} We thank the two referees for their helpful comments. The first and third author were partly supported
by the DFG project LO 748/13-1/2 (Streaming Automata Theory).

\newpage
\printbibliography

@phdthesis{GanardiPhD,
	author = {Moses Ganardi},
	bibsource = {dblp computer science bibliography, https://dblp.org},
	school = {University of Siegen, Germany},
	title = {Language recognition in the sliding window model},
	url = {https://dspace.ub.uni-siegen.de/handle/ubsi/1523},
	urn = {urn:nbn:de:hbz:467-15234},
	year = {2019}}

@inproceedings{DBLP:conf/focs/KociumakaPS21,
	author = {Tomasz Kociumaka and Ely Porat and Tatiana Starikovskaya},
	bibsource = {dblp computer science bibliography, https://dblp.org},
	booktitle = {Proceedings of the 62nd {IEEE} Annual Symposium on Foundations of Computer Science, {FOCS} 2021},
	doi = {10.1109/FOCS52979.2021.00090},
	pages = {885--896},
	publisher = {{IEEE}},
	title = {Small-space and streaming pattern matching with k edits},
	url = {https://doi.org/10.1109/FOCS52979.2021.00090},
	year = {2021}}

@inproceedings{Ergun:10,
	author = {Funda Erg{\"{u}}n and Hossein Jowhari and Mert Saglam},
	booktitle = {Proceeding of the 14th International Workshop on Approximation, Randomization, and Combinatorial Optimization. Algorithms and Techniques, {APPROX-RANDOM} 2010},
	doi = {10.1007/978-3-642-15369-3\_41},
	pages = {545--559},
	publisher = {Springer},
	series = {Lecture Notes in Computer Science},
	title = {Periodicity in Streams},
	url = {https://doi.org/10.1007/978-3-642-15369-3\_41},
	volume = {6302},
	year = {2010}}

@inproceedings{stream-periodicity-mismatches,
	author = {Funda Erg{\"{u}}n and Elena Grigorescu and Erfan Sadeqi Azer and Samson Zhou},
	booktitle = {Proceedings of Approximation, Randomization, and Combinatorial Optimization. Algorithms and Techniques, {APPROX/RANDOM} 2017},
	doi = {10.4230/LIPIcs.APPROX-RANDOM.2017.42},
	pages = {42:1--42:21},
	publisher = {Schloss Dagstuhl - Leibniz-Zentrum f{\"{u}}r Informatik},
	series = {LIPIcs},
	title = {Streaming Periodicity with Mismatches},
	url = {https://doi.org/10.4230/LIPIcs.APPROX-RANDOM.2017.42},
	volume = {81},
	year = {2017}}

@inproceedings{stream-periodicity-wildcards,
	author = {Funda Erg{\"{u}}n and Elena Grigorescu and Erfan Sadeqi Azer and Samson Zhou},
	booktitle = {Proceedings of the 13th International Computer Science Symposium in Russia, {CSR} 2018},
	doi = {10.1007/978-3-319-90530-3\_9},
	editor = {Fedor V. Fomin and Vladimir V. Podolskii},
	pages = {90--105},
	publisher = {Springer},
	series = {Lecture Notes in Computer Science},
	title = {Periodicity in Data Streams with Wildcards},
	url = {https://doi.org/10.1007/978-3-319-90530-3\_9},
	volume = {10846},
	year = {2018}}

@inproceedings{DBLP:conf/cpm/GawrychowskiRS19,
	author = {Pawe\l{} Gawrychowski and Jakub Radoszewski and Tatiana Starikovskaya},
	booktitle = {Proceedings of the 30th Annual Symposium on Combinatorial Pattern Matching, {CPM} 2019},
	doi = {10.4230/LIPIcs.CPM.2019.22},
	pages = {22:1--22:14},
	publisher = {Schloss Dagstuhl - Leibniz-Zentrum f{\"{u}}r Informatik},
	series = {LIPIcs},
	title = {Quasi-Periodicity in Streams},
	url = {https://doi.org/10.4230/LIPIcs.CPM.2019.22},
	volume = {128},
	year = {2019}}

@article{DBLP:journals/algorithmica/GawrychowskiMSU19,
	author = {Pawe\l{} Gawrychowski and Oleg Merkurev and Arseny M. Shur and Przemyslaw Uzna\'{n}ski},
	doi = {10.1007/s00453-019-00591-8},
	journal = {Algorithmica},
	number = {9},
	pages = {3630--3654},
	title = {Tight Tradeoffs for Real-Time Approximation of Longest Palindromes in Streams},
	volume = {81},
	year = {2019}}

@inproceedings{DBLP:conf/cpm/MerkurevS19,
	author = {Oleg Merkurev and Arseny M. Shur},
	booktitle = {Proceedings of the 30th Annual Symposium on Combinatorial Pattern Matching, {CPM} 2019},
	pages = {31:1--31:14},
	publisher = {Schloss Dagstuhl - Leibniz-Zentrum f{\"{u}}r Informatik},
	series = {LIPIcs},
	title = {Searching Long Repeats in Streams},
	volume = {128},
	year = {2019}}

@inproceedings{DBLP:conf/spire/MerkurevS19,
	author = {Oleg Merkurev and Arseny M. Shur},
	booktitle = {Proceedings of the 26th International Symposium on String Processing and Information Retrieval, {SPIRE} 2019},
	doi = {10.1007/978-3-030-32686-9\_15},
	pages = {203--220},
	publisher = {Springer},
	series = {Lecture Notes in Computer Science},
	title = {Searching Runs in Streams},
	url = {https://doi.org/10.1007/978-3-030-32686-9\_15},
	volume = {11811},
	year = {2019}}

@article{DBLP:journals/iandc/RadoszewskiS20,
	author = {Jakub Radoszewski and Tatiana Starikovskaya},
	doi = {10.1016/j.ic.2019.104513},
	journal = {Information and Computation},
	pages = {104513},
	title = {Streaming \emph{k}-mismatch with error correcting and applications},
	url = {https://doi.org/10.1016/j.ic.2019.104513},
	volume = {271},
	year = {2020}}

@inproceedings{DBLP:conf/cpm/GolanKKP20,
	author = {Shay Golan and Tomasz Kociumaka and Tsvi Kopelowitz and Ely Porat},
	booktitle = {Proceedings of the 31st Annual Symposium on Combinatorial Pattern Matching, {CPM} 2020},
	doi = {10.4230/LIPIcs.CPM.2020.15},
	pages = {15:1--15:15},
	publisher = {Schloss Dagstuhl - Leibniz-Zentrum f{\"{u}}r Informatik},
	series = {LIPIcs},
	title = {The Streaming k-Mismatch Problem: Tradeoffs Between Space and Total Time},
	url = {https://doi.org/10.4230/LIPIcs.CPM.2020.15},
	volume = {161},
	year = {2020}}

@article{DBLP:journals/algorithmica/GawrychowskiS22,
	author = {Pawel Gawrychowski and Tatiana Starikovskaya},
	doi = {10.1007/s00453-021-00876-x},
	journal = {Algorithmica},
	number = {4},
	pages = {896--916},
	title = {Streaming Dictionary Matching with Mismatches},
	url = {https://doi.org/10.1007/s00453-021-00876-x},
	volume = {84},
	year = {2022}}

@inproceedings{GanardiHL16,
	author = {Moses Ganardi and Danny Hucke and Markus Lohrey},
	bibsource = {dblp computer science bibliography, https://dblp.org},
	booktitle = {Proceedings of the 36th {IARCS} Annual Conference on Foundations of Software Technology and Theoretical Computer Science, {FSTTCS} 2016},
	doi = {10.4230/LIPIcs.FSTTCS.2016.18},
	pages = {18:1--18:14},
	publisher = {Schloss Dagstuhl - Leibniz-Zentrum f\"ur Informatik},
	series = {LIPIcs},
	title = {Querying Regular Languages over Sliding Windows},
	url = {https://doi.org/10.4230/LIPIcs.FSTTCS.2016.18},
	volume = {65},
	year = {2016}}

@inproceedings{GanardiHKLM18,
	author = {Moses Ganardi and Danny Hucke and Daniel K{\"{o}}nig and Markus Lohrey and Konstantinos Mamouras},
	bibsource = {dblp computer science bibliography, https://dblp.org},
	booktitle = {Proceedings of the 35th Symposium on Theoretical Aspects of Computer Science, {STACS} 2018},
	doi = {10.4230/LIPIcs.STACS.2018.31},
	pages = {31:1--31:14},
	publisher = {Schloss Dagstuhl - Leibniz-Zentrum f\"ur Informatik},
	series = {LIPIcs},
	title = {Automata Theory on Sliding Windows},
	url = {https://doi.org/10.4230/LIPIcs.STACS.2018.31},
	volume = {96},
	year = {2018}}

@inproceedings{GanardiHL18,
	author = {Moses Ganardi and Danny Hucke and Markus Lohrey},
	bibsource = {dblp computer science bibliography, https://dblp.org},
	booktitle = {Proceedings of the 45th International Colloquium on Automata, Languages, and Programming, {ICALP} 2018},
	doi = {10.4230/LIPIcs.ICALP.2018.127},
	pages = {127:1--127:13},
	publisher = {Schloss Dagstuhl - Leibniz-Zentrum f\"ur Informatik},
	series = {LIPIcs},
	title = {Randomized Sliding Window Algorithms for Regular Languages},
	url = {https://doi.org/10.4230/LIPIcs.ICALP.2018.127},
	volume = {107},
	year = {2018}}

@inproceedings{GanardiJL18,
	author = {Moses Ganardi and Artur Je\.z and Markus Lohrey},
	bibsource = {dblp computer science bibliography, https://dblp.org},
	booktitle = {Proceedings of the 43rd International Symposium on Mathematical Foundations of Computer Science, {MFCS} 2018},
	doi = {10.4230/LIPIcs.MFCS.2018.15},
	pages = {15:1--15:15},
	publisher = {Schloss Dagstuhl - Leibniz-Zentrum f\"ur Informatik},
	series = {LIPIcs},
	title = {Sliding Windows over Context-Free Languages},
	url = {https://doi.org/10.4230/LIPIcs.MFCS.2018.15},
	volume = {117},
	year = {2018}}

@inproceedings{Ganardi19,
	author = {Moses Ganardi},
	bibsource = {dblp computer science bibliography, https://dblp.org},
	booktitle = {Proceedings of the 36th International Symposium on Theoretical Aspects of Computer Science, {STACS} 2019},
	doi = {10.4230/LIPIcs.STACS.2019.29},
	pages = {29:1--29:17},
	publisher = {Schloss Dagstuhl - Leibniz-Zentrum f\"ur Informatik},
	series = {LIPIcs},
	title = {Visibly Pushdown Languages over Sliding Windows},
	url = {https://doi.org/10.4230/LIPIcs.STACS.2019.29},
	volume = {126},
	year = {2019}}

@inproceedings{GanardiHLS19,
	author = {Moses Ganardi and Danny Hucke and Markus Lohrey and Tatiana Starikovskaya},
	bibsource = {dblp computer science bibliography, https://dblp.org},
	booktitle = {Proceedings of the 30th International Symposium on Algorithms and Computation, {ISAAC} 2019},
	doi = {10.4230/LIPIcs.ISAAC.2019.6},
	pages = {6:1--6:13},
	publisher = {Schloss Dagstuhl - Leibniz-Zentrum f{\"{u}}r Informatik},
	series = {LIPIcs},
	title = {Sliding Window Property Testing for Regular Languages},
	url = {https://doi.org/10.4230/LIPIcs.ISAAC.2019.6},
	volume = {149},
	year = {2019}}

@article{KaminskiF94,
	author = {Michael Kaminski and Nissim Francez},
	bibsource = {dblp computer science bibliography, https://dblp.org},
	doi = {10.1016/0304-3975(94)90242-9},
	journal = {Theor. Comput. Sci.},
	number = {2},
	pages = {329--363},
	title = {Finite-Memory Automata},
	url = {https://doi.org/10.1016/0304-3975(94)90242-9},
	volume = {134},
	year = {1994}}

@article{NevenSV04,
	author = {Frank Neven and Thomas Schwentick and Victor Vianu},
	bibsource = {dblp computer science bibliography, https://dblp.org},
	doi = {10.1145/1013560.1013562},
	journal = {{ACM} Trans. Comput. Log.},
	number = {3},
	pages = {403--435},
	title = {Finite state machines for strings over infinite alphabets},
	url = {https://doi.org/10.1145/1013560.1013562},
	volume = {5},
	year = {2004}}

@inproceedings{AlurDDRY13,
	author = {Rajeev Alur and Loris D'Antoni and Jyotirmoy V. Deshmukh and Mukund Raghothaman and Yifei Yuan},
	bibsource = {dblp computer science bibliography, https://dblp.org},
	booktitle = {Proceedings of the 28th Annual {ACM/IEEE} Symposium on Logic in Computer Science, {LICS} 2013},
	doi = {10.1109/LICS.2013.65},
	pages = {13--22},
	publisher = {{IEEE} Computer Society},
	title = {Regular Functions and Cost Register Automata},
	url = {https://doi.org/10.1109/LICS.2013.65},
	year = {2013}}

@book{DrosteKV09,
	author = {Droste, Manfred and Kuich, Werner and Vogler, Heiko},
	publisher = {Springer},
	title = {Handbook of Weighted Automata},
	year = {2009}}

@article{Morris78a,
	author = {Robert H. {Morris Sr.}},
	bibsource = {dblp computer science bibliography, https://dblp.org},
	doi = {10.1145/359619.359627},
	journal = {Commun. {ACM}},
	number = {10},
	pages = {840--842},
	title = {Counting Large Numbers of Events in Small Registers},
	url = {https://doi.org/10.1145/359619.359627},
	volume = {21},
	year = {1978}}

@article{Flajolet85,
	author = {Philippe Flajolet},
	bibsource = {dblp computer science bibliography, https://dblp.org},
	journal = {{BIT}},
	number = {1},
	pages = {113--134},
	title = {Approximate Counting: {A} Detailed Analysis},
	volume = {25},
	year = {1985}}

@article{FeigenbaumKZ04,
	author = {Joan Feigenbaum and Sampath Kannan and Jian Zhang},
	bibsource = {dblp computer science bibliography, https://dblp.org},
	doi = {10.1007/s00453-004-1105-2},
	journal = {Algorithmica},
	number = {1},
	pages = {25--41},
	title = {Computing Diameter in the Streaming and Sliding-Window Models},
	url = {https://doi.org/10.1007/s00453-004-1105-2},
	volume = {41},
	year = {2005}}

@inproceedings{DatarM02,
	author = {Mayur Datar and Shan Muthukrishnan},
	bibsource = {dblp computer science bibliography, https://dblp.org},
	booktitle = {Proceedings of the 10th European Symposium on Algorithms, {ESA} 2002},
	doi = {10.1007/3-540-45749-6_31},
	editor = {Rolf H. M{\"{o}}hring and Rajeev Raman},
	pages = {323--334},
	publisher = {Springer},
	series = {Lecture Notes in Computer Science},
	title = {Estimating Rarity and Similarity over Data Stream Windows},
	url = {https://doi.org/10.1007/3-540-45749-6_31},
	volume = {2461},
	year = {2002}}

@inproceedings{GolabDDLM03,
	author = {Lukasz Golab and David DeHaan and Erik D. Demaine and Alejandro L{\'{o}}pez{-}Ortiz and J. Ian Munro},
	bibsource = {dblp computer science bibliography, https://dblp.org},
	booktitle = {Proceedings of the 3rd {ACM} {SIGCOMM} Internet Measurement Conference, {IMC} 2003},
	doi = {10.1145/948205.948227},
	pages = {173--178},
	publisher = {{ACM}},
	title = {Identifying frequent items in sliding windows over on-line packet streams},
	url = {https://doi.org/10.1145/948205.948227},
	year = {2003}}

@article{GibbonsT04,
	author = {Phillip B. Gibbons and Srikanta Tirthapura},
	bibsource = {dblp computer science bibliography, https://dblp.org},
	doi = {10.1007/s00224-004-1156-4},
	journal = {Theory Comput. Syst.},
	number = {3},
	pages = {457--478},
	title = {Distributed Streams Algorithms for Sliding Windows},
	url = {https://doi.org/10.1007/s00224-004-1156-4},
	volume = {37},
	year = {2004}}

@book{Kozen97,
	author = {Dexter Kozen},
	bibsource = {dblp computer science bibliography, https://dblp.org},
	isbn = {978-0-387-94907-9},
	publisher = {Springer},
	series = {Undergraduate texts in computer science},
	title = {Automata and computability},
	year = {1997}}

@inproceedings{BravermanGLWZ18,
	author = {Vladimir Braverman and Elena Grigorescu and Harry Lang and David P. Woodruff and Samson Zhou},
	bibsource = {dblp computer science bibliography, https://dblp.org},
	booktitle = {Proceedings of the 21st International Conference on Approximation Algorithms for Combinatorial Optimization Problems, and the 22nd International Conference on Randomization and Computation, {APPROX/RANDOM} 2018},
	doi = {10.4230/LIPIcs.APPROX-RANDOM.2018.7},
	editor = {Eric Blais and Klaus Jansen and Jos{\'{e}} D. P. Rolim and David Steurer},
	pages = {7:1--7:22},
	publisher = {Schloss Dagstuhl - Leibniz-Zentrum f\"ur Informatik},
	series = {LIPIcs},
	title = {Nearly Optimal Distinct Elements and Heavy Hitters on Sliding Windows},
	url = {https://doi.org/10.4230/LIPIcs.APPROX-RANDOM.2018.7},
	volume = {116},
	year = {2018}}

@inproceedings{GolanKP18,
	author = {Shay Golan and Tsvi Kopelowitz and Ely Porat},
	bibsource = {dblp computer science bibliography, https://dblp.org},
	booktitle = {Proceedings of the 45th International Colloquium on Automata, Languages, and Programming, {ICALP} 2018},
	doi = {10.4230/LIPIcs.ICALP.2018.65},
	editor = {Ioannis Chatzigiannakis and Christos Kaklamanis and D{\'{a}}niel Marx and Donald Sannella},
	pages = {65:1--65:16},
	publisher = {Schloss Dagstuhl - Leibniz-Zentrum f\"ur Informatik},
	series = {LIPIcs},
	title = {Towards Optimal Approximate Streaming Pattern Matching by Matching Multiple Patterns in Multiple Streams},
	url = {https://doi.org/10.4230/LIPIcs.ICALP.2018.65},
	volume = {107},
	year = {2018}}

@inproceedings{Starikovskaya17,
	author = {Tatiana Starikovskaya},
	bibsource = {dblp computer science bibliography, https://dblp.org},
	booktitle = {Proceedings of the 28th Annual Symposium on Combinatorial Pattern Matching, {CPM} 2017},
	doi = {10.4230/LIPIcs.CPM.2017.13},
	editor = {Juha K{\"{a}}rkk{\"{a}}inen and Jakub Radoszewski and Wojciech Rytter},
	pages = {13:1--13:11},
	publisher = {Schloss Dagstuhl - Leibniz-Zentrum f\"ur Informatik},
	series = {LIPIcs},
	title = {Communication and Streaming Complexity of Approximate Pattern Matching},
	url = {https://doi.org/10.4230/LIPIcs.CPM.2017.13},
	volume = {78},
	year = {2017}}

@inproceedings{CliffordKP19,
	author = {Rapha{\"{e}}l Clifford and Tomasz Kociumaka and Ely Porat},
	bibsource = {dblp computer science bibliography, https://dblp.org},
	booktitle = {Proceedings of the 30th Annual {ACM-SIAM} Symposium on Discrete Algorithms, {SODA} 2019},
	doi = {10.1137/1.9781611975482.68},
	editor = {Timothy M. Chan},
	pages = {1106--1125},
	publisher = {{SIAM}},
	title = {The streaming k-mismatch problem},
	url = {https://doi.org/10.1137/1.9781611975482.68},
	year = {2019}}

@article{GolanKP16,
	  author       = {Shay Golan and
	                  Tsvi Kopelowitz and
	                  Ely Porat},
	  title        = {Streaming Pattern Matching with d Wildcards},
	  journal      = {Algorithmica},
	  volume       = {81},
	  number       = {5},
	  pages        = {1988--2015},
	  year         = {2019},
	  url          = {https://doi.org/10.1007/s00453-018-0521-7},
	  doi          = {10.1007/S00453-018-0521-7},
	  bibsource    = {dblp computer science bibliography, https://dblp.org}
	}

@inproceedings{GolanP17,
	author = {Shay Golan and Ely Porat},
	bibsource = {dblp computer science bibliography, https://dblp.org},
	booktitle = {Proceedings of the 25th Annual European Symposium on Algorithms, {ESA} 2017},
	doi = {10.4230/LIPIcs.ESA.2017.41},
	editor = {Kirk Pruhs and Christian Sohler},
	pages = {41:1--41:15},
	publisher = {Schloss Dagstuhl - Leibniz-Zentrum f\"ur Informatik},
	series = {LIPIcs},
	title = {Real-Time Streaming Multi-Pattern Search for Constant Alphabet},
	url = {https://doi.org/10.4230/LIPIcs.ESA.2017.41},
	volume = {87},
	year = {2017}}

@article{CohenS06,
	author = {Edith Cohen and Martin J. Strauss},
	bibsource = {dblp computer science bibliography, https://dblp.org},
	doi = {10.1016/j.jalgor.2005.01.006},
	journal = {J. Algorithms},
	number = {1},
	pages = {19--36},
	title = {Maintaining time-decaying stream aggregates},
	url = {https://doi.org/10.1016/j.jalgor.2005.01.006},
	volume = {59},
	year = {2006}}

@inproceedings{BabcockDM02,
	author = {Brian Babcock and Mayur Datar and Rajeev Motwani},
	bibsource = {dblp computer science bibliography, https://dblp.org},
	booktitle = {Proceedings of the 13th Annual {ACM-SIAM} Symposium on Discrete Algorithms, {SODA} 2002},
	editor = {David Eppstein},
	pages = {633--634},
	publisher = {{ACM/SIAM}},
	title = {Sampling from a moving window over streaming data},
	url = {http://dl.acm.org/citation.cfm?id=545381.545465},
	year = {2002}}

@article{Ben-BasatEF18,
	  author       = {Ran Ben Basat and
	                  Gil Einziger and
	                  Roy Friedman},
	  title        = {Give me some slack: Efficient network measurements},
	  journal      = {Theor. Comput. Sci.},
	  volume       = {791},
	  pages        = {87--108},
	  year         = {2019},
	  url          = {https://doi.org/10.1016/j.tcs.2019.05.012},
	  doi          = {10.1016/J.TCS.2019.05.012},
	  bibsource    = {dblp computer science bibliography, https://dblp.org}
	}

@inproceedings{TangwongsanH017,
	author = {Kanat Tangwongsan and Martin Hirzel and Scott Schneider},
	bibsource = {dblp computer science bibliography, https://dblp.org},
	booktitle = {Proceedings of the 11th {ACM} International Conference on Distributed and Event-based Systems, {DEBS} 2017},
	doi = {10.1145/3093742.3093925},
	pages = {66--77},
	publisher = {{ACM}},
	title = {Low-Latency Sliding-Window Aggregation in Worst-Case Constant Time},
	url = {https://doi.org/10.1145/3093742.3093925},
	year = {2017}}

@article{FeigenbaumKSV02,
	author = {Joan Feigenbaum and Sampath Kannan and Martin Strauss and Mahesh Viswanathan},
	bibsource = {dblp computer science bibliography, https://dblp.org},
	doi = {10.1007/s00453-002-0959-4},
	journal = {Algorithmica},
	number = {1},
	pages = {67--80},
	title = {Testing and Spot-Checking of Data Streams},
	url = {https://doi.org/10.1007/s00453-002-0959-4},
	volume = {34},
	year = {2002}}

@article{AlonKNS00,
	author = {Noga Alon and Michael Krivelevich and Ilan Newman and Mario Szegedy},
	bibsource = {dblp computer science bibliography, https://dblp.org},
	doi = {10.1137/S0097539700366528},
	journal = {{SIAM} J. Comput.},
	number = {6},
	pages = {1842--1862},
	title = {Regular Languages are Testable with a Constant Number of Queries},
	url = {https://doi.org/10.1137/S0097539700366528},
	volume = {30},
	year = {2000}}

@article{GoldreichGR98,
	author = {Oded Goldreich and Shafi Goldwasser and Dana Ron},
	bibsource = {dblp computer science bibliography, https://dblp.org},
	doi = {10.1145/285055.285060},
	journal = {J. {ACM}},
	number = {4},
	pages = {653--750},
	title = {Property Testing and its Connection to Learning and Approximation},
	url = {https://doi.org/10.1145/285055.285060},
	volume = {45},
	year = {1998}}

@article{AlonMS99,
	author = {Noga Alon and Yossi Matias and Mario Szegedy},
	bibsource = {dblp computer science bibliography, https://dblp.org},
	doi = {10.1006/jcss.1997.1545},
	journal = {J. Comput. Syst. Sci.},
	number = {1},
	pages = {137--147},
	title = {The Space Complexity of Approximating the Frequency Moments},
	url = {https://doi.org/10.1006/jcss.1997.1545},
	volume = {58},
	year = {1999}}

@inproceedings{Gawrychowski11,
	author = {Pawel Gawrychowski},
	bibsource = {dblp computer science bibliography, https://dblp.org},
	booktitle = {Proceedings of the 16th International Conference on Implementation and Application of Automata, {CIAA} 2011},
	doi = {10.1007/978-3-642-22256-6_14},
	editor = {B{\'{e}}atrice Bouchou{-}Markhoff and Pascal Caron and Jean{-}Marc Champarnaud and Denis Maurel},
	pages = {142--153},
	publisher = {Springer},
	series = {Lecture Notes in Computer Science},
	title = {Chrobak Normal Form Revisited, with Applications},
	url = {https://doi.org/10.1007/978-3-642-22256-6_14},
	volume = {6807},
	year = {2011}}

@inproceedings{JiraskovaM14,
	author = {Galina Jir{\'{a}}skov{\'{a}} and Peter Mlyn{\'{a}}rcik},
	bibsource = {dblp computer science bibliography, https://dblp.org},
	booktitle = {Proceedings of the 16th International Workshop on Descriptional Complexity of Formal Systems, {DCFS} 2014},
	doi = {10.1007/978-3-319-09704-6_20},
	editor = {Helmut J{\"{u}}rgensen and Juhani Karhum{\"{a}}ki and Alexander Okhotin},
	pages = {222--233},
	publisher = {Springer},
	series = {Lecture Notes in Computer Science},
	title = {Complement on Prefix-Free, Suffix-Free, and Non-Returning {NFA} Languages},
	url = {https://doi.org/10.1007/978-3-319-09704-6_20},
	volume = {8614},
	year = {2014}}

@article{KremerNR99,
	author = {Ilan Kremer and Noam Nisan and Dana Ron},
	bibsource = {dblp computer science bibliography, https://dblp.org},
	doi = {10.1007/s000370050018},
	journal = {Comput. Complex.},
	number = {1},
	pages = {21--49},
	title = {On Randomized One-Round Communication Complexity},
	url = {https://doi.org/10.1007/s000370050018},
	volume = {8},
	year = {1999}}

@book{KushilevitzN97,
	author = {Eyal Kushilevitz and Noam Nisan},
	bibsource = {dblp computer science bibliography, https://dblp.org},
	isbn = {978-0-521-56067-2},
	publisher = {Cambridge University Press},
	title = {Communication complexity},
	year = {1997}}

@article{RosserS62,
	author = {J.~Barkley Rosser and Lowell Schoenfeld},
	journal = {Illinois J. Math.},
	number = {1},
	pages = {64--94},
	title = {Approximate formulas for some functions of prime numbers},
	volume = {6},
	year = {1962}}

@book{MitzenmacherU17,
	address = {New York, NY, USA},
	author = {Mitzenmacher, Michael and Upfal, Eli},
	edition = {2nd},
	publisher = {Cambridge University Press},
	title = {Probability and Computing: Randomization and Probabilistic Techniques in Algorithms and Data Analysis},
	year = {2017}}

@article{Rabin63,
	author = {Michael O. Rabin},
	bibsource = {dblp computer science bibliography, https://dblp.org},
	doi = {10.1016/S0019-9958(63)90290-0},
	journal = {Inform. Control},
	number = {3},
	pages = {230--245},
	title = {Probabilistic Automata},
	url = {https://doi.org/10.1016/S0019-9958(63)90290-0},
	volume = {6},
	year = {1963}}

@book{Paz71,
	address = {Orlando, FL, USA},
	author = {Paz, Azaria},
	isbn = {0125476507},
	publisher = {Academic Press, Inc.},
	title = {Introduction to Probabilistic Automata (Computer Science and Applied Mathematics)},
	year = {1971}}

@inproceedings{Ben-BasatEFK16,
	author = {Ran Ben{-}Basat and Gil Einziger and Roy Friedman and Yaron Kassner},
	bibsource = {dblp computer science bibliography, https://dblp.org},
	booktitle = {Proceedings of the 15th Scandinavian Symposium and Workshops on Algorithm Theory, {SWAT} 2016},
	doi = {10.4230/LIPIcs.SWAT.2016.11},
	editor = {Rasmus Pagh},
	pages = {11:1--11:14},
	publisher = {Schloss Dagstuhl - Leibniz-Zentrum f\"ur Informatik},
	series = {LIPIcs},
	title = {Efficient Summing over Sliding Windows},
	url = {https://doi.org/10.4230/LIPIcs.SWAT.2016.11},
	volume = {53},
	year = {2016}}

@inproceedings{ArasuM04,
	author = {Arvind Arasu and Gurmeet Singh Manku},
	bibsource = {dblp computer science bibliography, https://dblp.org},
	booktitle = {Proceedings of the 23rd {ACM} {SIGACT-SIGMOD-SIGART} Symposium on Principles of Database Systems, {PODS} 2004},
	doi = {10.1145/1055558.1055598},
	editor = {Catriel Beeri and Alin Deutsch},
	pages = {286--296},
	publisher = {{ACM}},
	title = {Approximate Counts and Quantiles over Sliding Windows},
	url = {https://doi.org/10.1145/1055558.1055598},
	year = {2004}}

@article{Straubing85,
	author = {Howard Straubing},
	doi = {10.1016/0022-4049(85)90062-3},
	journal = {J. Pure Appl. Algebra},
	pages = {53--94},
	title = {Finite semigroup varieties of the form {$V * D$}},
	url = {http://www.sciencedirect.com/science/article/pii/0022404985900623},
	volume = {36},
	year = {1985}}

@inproceedings{SegoufinV02,
	author = {Luc Segoufin and Victor Vianu},
	bibsource = {dblp computer science bibliography, https://dblp.org},
	booktitle = {Proceedings of the 21st {ACM} {SIGACT-SIGMOD-SIGART} Symposium on Principles of Database Systems, {PODS} 2002},
	doi = {10.1145/543613.543622},
	editor = {Lucian Popa and Serge Abiteboul and Phokion G. Kolaitis},
	pages = {53--64},
	publisher = {{ACM}},
	title = {Validating Streaming {XML} Documents},
	url = {https://doi.org/10.1145/543613.543622},
	year = {2002}}

@inproceedings{PoratP09,
	author = {Benny Porat and Ely Porat},
	bibsource = {dblp computer science bibliography, https://dblp.org},
	booktitle = {Proceedings of the 50th Annual {IEEE} Symposium on Foundations of Computer Science, {FOCS} 2009},
	doi = {10.1109/FOCS.2009.11},
	pages = {315--323},
	publisher = {{IEEE} Computer Society},
	title = {Exact and Approximate Pattern Matching in the Streaming Model},
	url = {https://doi.org/10.1109/FOCS.2009.11},
	year = {2009}}

@article{MagniezMN14,
	author = {Fr{\'{e}}d{\'{e}}ric Magniez and Claire Mathieu and Ashwin Nayak},
	bibsource = {dblp computer science bibliography, https://dblp.org},
	doi = {10.1137/130926122},
	journal = {{SIAM} J. Comput.},
	number = {6},
	pages = {1880--1905},
	title = {Recognizing Well-Parenthesized Expressions in the Streaming Model},
	url = {https://doi.org/10.1137/130926122},
	volume = {43},
	year = {2014}}

@inproceedings{KrebsLS11,
	author = {Andreas Krebs and Nutan Limaye and Srikanth Srinivasan},
	bibsource = {dblp computer science bibliography, https://dblp.org},
	booktitle = {Proceedings of the 36th International Symposium on Mathematical Foundations of Computer Science, {MFCS} 2011},
	doi = {10.1007/978-3-642-22993-0_38},
	editor = {Filip Murlak and Piotr Sankowski},
	pages = {412--423},
	publisher = {Springer},
	series = {Lecture Notes in Computer Science},
	title = {Streaming Algorithms for Recognizing Nearly Well-Parenthesized Expressions},
	url = {https://doi.org/10.1007/978-3-642-22993-0_38},
	volume = {6907},
	year = {2011}}

@article{KonradM13,
	author = {Christian Konrad and Fr{\'{e}}d{\'{e}}ric Magniez},
	bibsource = {dblp computer science bibliography, https://dblp.org},
	doi = {10.1145/2504590},
	journal = {{ACM} Trans. Database Syst.},
	number = {4},
	pages = {27:1--27:36},
	title = {Validating {XML} documents in the streaming model with external memory},
	url = {https://doi.org/10.1145/2504590},
	volume = {38},
	year = {2013}}

@book{HopcroftU79,
	author = {John E. Hopcroft and Jeffrey D. Ullman},
	bibsource = {dblp computer science bibliography, https://dblp.org},
	isbn = {0-201-02988-X},
	publisher = {Addison-Wesley},
	title = {Introduction to Automata Theory, Languages and Computation},
	year = {1979}}

@inproceedings{GawrychowskiJ09,
	author = {Pawel Gawrychowski and Artur Je\.z},
	bibsource = {dblp computer science bibliography, https://dblp.org},
	booktitle = {Proceedings of the 34th International Symposium on Mathematical Foundations of Computer Science 2009, {MFCS} 2009},
	doi = {10.1007/978-3-642-03816-7_31},
	editor = {Rastislav Kr{\'{a}}lovic and Damian Niwinski},
	pages = {356--368},
	publisher = {Springer},
	series = {Lecture Notes in Computer Science},
	title = {Hyper-minimisation Made Efficient},
	url = {https://doi.org/10.1007/978-3-642-03816-7_31},
	volume = {5734},
	year = {2009}}

@article{FrandsenMS97,
	author = {Gudmund Skovbjerg Frandsen and Peter Bro Miltersen and Sven Skyum},
	bibsource = {dblp computer science bibliography, https://dblp.org},
	doi = {10.1145/256303.256309},
	journal = {J. {ACM}},
	number = {2},
	pages = {257--271},
	title = {Dynamic word problems},
	url = {https://doi.org/10.1145/256303.256309},
	volume = {44},
	year = {1997}}

@inproceedings{FrancoisMRS16,
	author = {Nathana{\"{e}}l Fran{\c{c}}ois and Fr{\'{e}}d{\'{e}}ric Magniez and Michel de Rougemont and Olivier Serre},
	bibsource = {dblp computer science bibliography, https://dblp.org},
	booktitle = {Proceedings of the 24th Annual European Symposium on Algorithms, {ESA} 2016},
	doi = {10.4230/LIPIcs.ESA.2016.43},
	editor = {Piotr Sankowski and Christos D. Zaroliagis},
	pages = {43:1--43:17},
	publisher = {Schloss Dagstuhl - Leibniz-Zentrum f\"ur Informatik},
	series = {LIPIcs},
	title = {Streaming Property Testing of Visibly Pushdown Languages},
	url = {https://doi.org/10.4230/LIPIcs.ESA.2016.43},
	volume = {57},
	year = {2016}}

@book{Eilenberg74,
	author = {Samuel Eilenberg},
	bibsource = {dblp computer science bibliography, https://dblp.org},
	isbn = {0122340019},
	publisher = {Academic Press},
	series = {Pure and applied mathematics},
	title = {Automata, languages, and machines},
	volume = {A},
	year = {1974}}

@article{deBr46,
	author = {Nicolaas G.~de Bruijn},
	journal = {Nederl. Akad. Wetensch. Proc.},
	language = {English},
	number = {7},
	pages = {758--764},
	title = {A combinatorial problem},
	volume = {49},
	year = {1946}}

@article{DatarGIM02,
	author = {Mayur Datar and Aristides Gionis and Piotr Indyk and Rajeev Motwani},
	bibsource = {dblp computer science bibliography, https://dblp.org},
	doi = {10.1137/S0097539701398363},
	journal = {{SIAM} J. Comput.},
	number = {6},
	pages = {1794--1813},
	title = {Maintaining Stream Statistics over Sliding Windows},
	url = {https://doi.org/10.1137/S0097539701398363},
	volume = {31},
	year = {2002}}

@inproceedings{CliffordS16,
	author = {Rapha{\"{e}}l Clifford and Tatiana Starikovskaya},
	bibsource = {dblp computer science bibliography, https://dblp.org},
	booktitle = {Proceedings of the 43rd International Colloquium on Automata, Languages, and Programming, {ICALP} 2016},
	doi = {10.4230/LIPIcs.ICALP.2016.20},
	editor = {Ioannis Chatzigiannakis and Michael Mitzenmacher and Yuval Rabani and Davide Sangiorgi},
	pages = {20:1--20:14},
	publisher = {Schloss Dagstuhl - Leibniz-Zentrum f\"ur Informatik},
	series = {LIPIcs},
	title = {Approximate {H}amming Distance in a Stream},
	url = {https://doi.org/10.4230/LIPIcs.ICALP.2016.20},
	volume = {55},
	year = {2016}}

@inproceedings{CliffordFPSS16,
	author = {Rapha{\"{e}}l Clifford and Allyx Fontaine and Ely Porat and Benjamin Sach and Tatiana Starikovskaya},
	bibsource = {dblp computer science bibliography, https://dblp.org},
	booktitle = {Proceedings of the 27th Annual {ACM-SIAM} Symposium on Discrete Algorithms, {SODA} 2016},
	doi = {10.1137/1.9781611974331.ch142},
	editor = {Robert Krauthgamer},
	pages = {2039--2052},
	publisher = {{SIAM}},
	title = {The \emph{k}-mismatch problem revisited},
	url = {https://doi.org/10.1137/1.9781611974331.ch142},
	year = {2016}}

@inproceedings{CliffordFPSS15,
	author = {Rapha{\"{e}}l Clifford and Allyx Fontaine and Ely Porat and Benjamin Sach and Tatiana Starikovskaya},
	bibsource = {dblp computer science bibliography, https://dblp.org},
	booktitle = {Proceedings of the 23rd Annual European Symposium, {ESA} 2015},
	doi = {10.1007/978-3-662-48350-3_31},
	editor = {Nikhil Bansal and Irene Finocchi},
	pages = {361--372},
	publisher = {Springer},
	series = {Lecture Notes in Computer Science},
	title = {Dictionary Matching in a Stream},
	url = {https://doi.org/10.1007/978-3-662-48350-3_31},
	volume = {9294},
	year = {2015}}

@article{BreslauerG14,
	author = {Dany Breslauer and Zvi Galil},
	bibsource = {dblp computer science bibliography, https://dblp.org},
	doi = {10.1145/2635814},
	journal = {{ACM} Trans. Algorithms},
	number = {4},
	pages = {22:1--22:12},
	title = {Real-Time Streaming String-Matching},
	url = {https://doi.org/10.1145/2635814},
	volume = {10},
	year = {2014}}

@article{BravermanOZ12,
	author = {Vladimir Braverman and Rafail Ostrovsky and Carlo Zaniolo},
	bibsource = {dblp computer science bibliography, https://dblp.org},
	doi = {10.1016/j.jcss.2011.04.004},
	journal = {J. Comput. Syst. Sci.},
	number = {1},
	pages = {260--272},
	title = {Optimal sampling from sliding windows},
	url = {https://doi.org/10.1016/j.jcss.2011.04.004},
	volume = {78},
	year = {2012}}

@inproceedings{BravermanO07,
	author = {Vladimir Braverman and Rafail Ostrovsky},
	bibsource = {dblp computer science bibliography, https://dblp.org},
	booktitle = {Proceedings of the 48th Annual {IEEE} Symposium on Foundations of Computer Science, {FOCS} 2007},
	doi = {10.1109/FOCS.2007.55},
	pages = {283--293},
	publisher = {{IEEE} Computer Society},
	title = {Smooth Histograms for Sliding Windows},
	url = {https://doi.org/10.1109/FOCS.2007.55},
	year = {2007}}

@book{Berstel79,
	author = {Jean Berstel},
	bibsource = {dblp computer science bibliography, https://dblp.org},
	isbn = {3519023407},
	publisher = {Teubner},
	series = {Teubner Studienb{\"{u}}cher : Informatik},
	title = {Transductions and context-free languages},
	url = {http://www.worldcat.org/oclc/06364613},
	volume = {38},
	year = {1979}}

@article{BabuLRV13,
	author = {Ajesh Babu and Nutan Limaye and Jaikumar Radhakrishnan and Girish Varma},
	bibsource = {dblp computer science bibliography, https://dblp.org},
	doi = {10.1016/j.tcs.2012.12.028},
	journal = {Theor. Comput. Sci.},
	pages = {13--23},
	title = {Streaming algorithms for language recognition problems},
	url = {https://doi.org/10.1016/j.tcs.2012.12.028},
	volume = {494},
	year = {2013}}

@inproceedings{BabcockDMO03,
	author = {Brian Babcock and Mayur Datar and Rajeev Motwani and Liadan O'Callaghan},
	bibsource = {dblp computer science bibliography, https://dblp.org},
	booktitle = {Proceedings of the 22nd {ACM} {SIGACT-SIGMOD-SIGART} Symposium on Principles of Database Systems, {PODS} 2003},
	doi = {10.1145/773153.773176},
	editor = {Frank Neven and Catriel Beeri and Tova Milo},
	pages = {234--243},
	publisher = {{ACM}},
	title = {Maintaining variance and k-medians over data stream windows},
	url = {https://doi.org/10.1145/773153.773176},
	year = {2003}}

@article{AlurM04,
	  author       = {Rajeev Alur and
	                  P. Madhusudan},
	  title        = {Adding nesting structure to words},
	  journal      = {J. {ACM}},
	  volume       = {56},
	  number       = {3},
	  pages        = {16:1--16:43},
	  year         = {2009},
	  url          = {https://doi.org/10.1145/1516512.1516518},
	  doi          = {10.1145/1516512.1516518},
	  bibsource    = {dblp computer science bibliography, https://dblp.org}
	}

@book{Aggarwal07,
	bibsource = {dblp computer science bibliography, https://dblp.org},
	doi = {10.1007/978-0-387-47534-9},
	editor = {Charu C. Aggarwal},
	isbn = {978-0-387-28759-1},
	publisher = {Springer},
	series = {Advances in Database Systems},
	title = {Data Streams - Models and Algorithms},
	url = {https://doi.org/10.1007/978-0-387-47534-9},
	volume = {31},
	year = {2007}}

@inproceedings{BenediktLP10,
	author = {Michael Benedikt and Clemens Ley and Gabriele Puppis},
	booktitle = {Proceedings of the 4th Alberto Mendelzon International Workshop on Foundations of Data Management},
	publisher = {CEUR-WS.org},
	series = {{CEUR} Workshop Proceedings},
	title = {What You Must Remember When Processing Data Words},
	volume = {619},
	year = {2010}}

@article{FrancezK03,
	author = {Nissim Francez and Michael Kaminski},
	journal = {Theor. Comput. Sci.},
	number = {1-3},
	pages = {155--175},
	title = {An algebraic characterization of deterministic regular languages over infinite alphabets},
	volume = {306},
	year = {2003}}

@inproceedings{BarloyMP21,
	author = {Corentin Barloy and Filip Murlak and Charles Paperman},
	booktitle = {Proceedings of the 40th {ACM} {SIGMOD-SIGACT-SIGAI} Symposium on Principles of Database Systems, PODS'21},
	doi = {10.1145/3452021.3458320},
	pages = {109--125},
	publisher = {{ACM}},
	title = {Stackless Processing of Streamed Trees},
	url = {https://doi.org/10.1145/3452021.3458320},
	year = {2021}}

@article{CuMa12,
	author = {Cugola, Gianpaolo and Margara, Alessandro},
	doi = {10.1145/2187671.2187677},
	journal = {ACM Computing Surveys},
	number = {3},
	title = {Processing Flows of Information: From Data Stream to Complex Event Processing},
	url = {https://doi.org/10.1145/2187671.2187677},
	volume = {44},
	year = {2012}}

@inproceedings{ZhangDI14,
	author = {Haopeng Zhang and Yanlei Diao and Neil Immerman},
	booktitle = {Proceedings of the International Conference on Management of Data, {SIGMOD} 2014},
	doi = {10.1145/2588555.2593671},
	pages = {217--228},
	publisher = {{ACM}},
	title = {On complexity and optimization of expensive queries in complex event processing},
	url = {https://doi.org/10.1145/2588555.2593671},
	year = {2014}}

@article{JN14,
	author = {Rahul {Jain} and Ashwin {Nayak}},
	doi = {10.1109/TIT.2014.2339859},
	issn = {0018-9448},
	journal = {IEEE Transactions on Information Theory},
	month = {Oct},
	number = {10},
	pages = {6646-6668},
	title = {The Space Complexity of Recognizing Well-Parenthesized Expressions in the Streaming Model: The Index Function Revisited},
	volume = {60},
	year = {2014}}

@article{Roughgarden16,
	author = {Tim Roughgarden},
	doi = {10.1561/0400000076},
	journal = {Found. Trends Theor. Comput. Sci.},
	number = {3-4},
	pages = {217--404},
	title = {Communication Complexity (for Algorithm Designers)},
	url = {https://doi.org/10.1561/0400000076},
	volume = {11},
	year = {2016}}

@article{GanardiHLTOCS21,
	  author       = {Moses Ganardi and
	                  Danny Hucke and
	                  Markus Lohrey},
	  title        = {Derandomization for Sliding Window Algorithms with Strict Correctness$^*$},
	  journal      = {Theory Comput. Syst.},
	  volume       = {65},
	  number       = {3},
	  pages        = {1--18},
	  year         = {2021},
	  url          = {https://doi.org/10.1007/s00224-020-10000-1},
	  doi          = {10.1007/S00224-020-10000-1},
	  bibsource    = {dblp computer science bibliography, https://dblp.org}
	}

@inproceedings{AmarilliJP21,
	author = {Antoine Amarilli and Louis Jachiet and Charles Paperman},
	booktitle = {Proceedings of the 48th International Colloquium on Automata, Languages, and Programming, {ICALP} 2021},
	doi = {10.4230/LIPICS.ICALP.2021.116},
	editor = {Nikhil Bansal and Emanuela Merelli and James Worrell},
	pages = {116:1--116:17},
	publisher = {Schloss Dagstuhl - Leibniz-Zentrum f{\"{u}}r Informatik},
	series = {LIPIcs},
	title = {Dynamic Membership for Regular Languages},
	volume = {198},
	year = {2021}}

@inproceedings{FrandsenHMRS95,
	author = {Gudmund Skovbjerg Frandsen and Thore Husfeldt and Peter Bro Miltersen and Theis Rauhe and S{\o}ren Skyum},
	booktitle = {Proceedings of the 4th International Workshop on Algorithms and Data Structures, {WADS} '95},
	doi = {10.1007/3-540-60220-8\_54},
	pages = {98--108},
	publisher = {Springer},
	series = {Lecture Notes in Computer Science},
	title = {Dynamic Algorithms for the {Dyck} Languages},
	volume = {955},
	year = {1995}}

@inproceedings{GanardiJLS22,
	author = {Moses Ganardi and Louis Jachiet and Markus Lohrey and Thomas Schwentick},
	booktitle = {Proceedings of the 42nd {IARCS} Annual Conference on Foundations of Software Technology and Theoretical Computer Science, {FSTTCS} 2022},
	doi = {10.4230/LIPICS.FSTTCS.2022.38},
	pages = {38:1--38:23},
	publisher = {Schloss Dagstuhl - Leibniz-Zentrum f{\"{u}}r Informatik},
	series = {LIPIcs},
	title = {Low-Latency Sliding Window Algorithms for Formal Languages},
	volume = {250},
	year = {2022}}

@incollection{Pin97,
  author       = {Jean{-}Eric Pin},
  editor       = {Grzegorz Rozenberg and
                  Arto Salomaa},
  title        = {Syntactic Semigroups},
  booktitle    = {Handbook of Formal Languages, Volume 1: Word, Language, Grammar},
  pages        = {679--746},
  publisher    = {Springer},
  year         = {1997},
  url          = {https://doi.org/10.1007/978-3-642-59136-5\_10},
  doi          = {10.1007/978-3-642-59136-5\_10},
  bibsource    = {dblp computer science bibliography, https://dblp.org}
}
	

\end{document}